\documentclass[12pt,letterpaper,twoside]{article}
\usepackage{color,graphicx,natbib,lscape,here,geometry,setspace,multirow,enumitem}
\usepackage[dvipsnames]{xcolor}
\usepackage{graphicx}
\usepackage{authblk}
\usepackage{silence}
\usepackage{rotating}
\usepackage{multirow}

\usepackage{booktabs}
\usepackage{color}
\usepackage{colortbl}
\definecolor{greyc}{rgb}{0.9,0.9,0.9}
\providecommand{\shadeRow}{\rowcolor[rgb]{0.9,0.9,0.9}}
\usepackage{setspace}
\usepackage{algorithm2e}
\usepackage{enumitem}
\usepackage{tikz}
\usepackage{standalone}
\usepackage{longtable}

\usepackage{newtxtext,newtxmath}
\newcommand*{\scfont}{\fontfamily{ptm}\selectfont}

\definecolor{nblue}{HTML}{000660}
\usepackage[colorlinks=true,urlcolor=nblue,linkcolor=nblue,citecolor=nblue]{hyperref}
\usepackage{bigstrut}

\usepackage{tabularx}
\usepackage{threeparttable}
\usepackage{dcolumn}

\usepackage{etoolbox} 
\makeatletter
\patchcmd{\BR@backref}{\newblock}{\newblock[}{}{}
\patchcmd{\BR@backref}{\par}{]\par}{}{}
\makeatother

\usepackage{pdflscape}
\usepackage{afterpage}

\usepackage{longtable}
\usepackage{array}
\newcolumntype{C}[1]{>{\centering\arraybackslash}p{#1}}

\usepackage{comment}
\usepackage{mathtools}

\usepackage[hang,flushmargin]{footmisc} 
\usepackage[english]{babel}

\pdfoutput=1
 \usepackage{float}
 \restylefloat{table}

\newcommand*\xbar[1]{%
   \hbox{%
     \vbox{%
       \hrule height 0.7 pt 
       \kern 0.2ex
       \hbox{%
         \kern-0.05em
         \ensuremath{#1}%
         \kern-0.05em
       }%
     }%
   }%
} 

\newcommand*\xAbar[1]{%
   \hbox{%
     \vbox{%
       \hrule height 0.7 pt 
       \kern 0.16ex
       \hbox{%
         \kern-0.05em
         \ensuremath{#1}%
         \kern-0.05em
       }%
     }%
   }%
} 

\newcommand*\uxbar[1]{%
   \hbox{%
     \vbox{%
       \hrule height 0.7 pt 
       \kern -1.8ex
       \hbox{%
         \kern-0.05em
         \ensuremath{#1}%
         \kern-0.05em
       }%
     }%
   }%
} 

\usepackage[title,titletoc]{appendix}
\makeatletter

\renewenvironment{appendices}{%
    \begin{oldappendices}%
    \renewcommand{\thefigure}{\ifnum \c@section>\z@ \thesection.\fi\@arabic\c@figure}%
    \@addtoreset{figure}{section}%
    \renewcommand{\thetable}{\ifnum \c@section>\z@ \thesection.\fi\@arabic\c@table}%
    \@addtoreset{table}{section}}{%
    \end{oldappendices}%
}\makeatother

\usepackage{titlesec} 
\titleformat{\section}[block]{\large}{\thesection. }{0em}{\MakeUppercase} 
\titleformat{\subsection}[block]{\large}{\thesubsection. }{0em}{\itshape} 
\titleformat{\subsubsection}[block]{\large}{}{0em}{\itshape} 

\let\natbibcitet\citet
\renewcommand\citet{\bibpunct{(}{)}{,}{a}{,}{,}\natbibcitet}
\let\natbibcitep\citep
\renewcommand\citep{\bibpunct{(}{)}{;}{a}{,}{;}\natbibcitep}
\newcommand{\bi}{\begin{itemize}}
\newcommand{\ei}{\end{itemize}}
\newcommand{\be}{\begin{equation}}
\newcommand{\ee}{\end{equation}}
\defcitealias{ieo14}{IEO, 2014}

\long\def\symbolfootnote[#1]#2{\begingroup%
\def\thefootnote{\fnsymbol{footnote}}\footnote[#1]{#2}\endgroup}

\makeatletter
\def\ubar#1{\underline{\sbox\tw@{$#1$}\dp\tw@\z@\box\tw@}}
\def\obar#1{\overline{\sbox\tw@{$#1$}\dp\tw@\z@\box\tw@}}
\makeatother

\usepackage{bm}
\usepackage{caption}
\usepackage{subcaption}

\makeatletter\let\p@subfigure\thefigure\makeatother

\floatstyle{plaintop}
\restylefloat{table}

\usepackage{fancyhdr} 
\usepackage{cleveref}
\crefname{chapter}{Chapter}{Chapters}
\crefname{section}{Section}{Sections}
\crefname{subsection}{Section}{Sections}
\crefname{subsubsection}{Section}{Sections}
\crefname{figure}{Figure}{Figures}
\crefname{table}{Table}{Tables}
\crefname{equation}{Equation}{Equations}
\crefname{appendix}{Appendix}{Appendices}
\crefname{appendices}{Appendix}{Appendices}

\crefname{appsec}{Appendix}{Appendices}

\usepackage{catoptions}
\makeatletter

\def\Autoref#1{%
  \begingroup
  \edef\reserved@a{\cpttrimspaces{#1}}%
  \ifcsndefTF{r@#1}{%
    \xaftercsname{\expandafter\testreftype\@fourthoffive}
      {r@\reserved@a}.\\{#1}%
  }{%
    \ref{#1}%
  }%
  \endgroup
}
\def\testreftype#1.#2\\#3{%
  \ifcsndefTF{#1autorefname}{%
    \def\reserved@a##1##2\@nil{%
      \uppercase{\def\ref@name{##1}}%
      \csn@edef{#1autorefname}{\ref@name##2}%
      \autoref{#3}%
    }%
    \reserved@a#1\@nil
  }{%
    \autoref{#3}%
  }%
}
\makeatother

\newcolumntype{d}[1]{D{.}{.}{#1}}

\title{\LARGE{Combining Shrinkage and Sparsity in Conjugate Vector Autoregressive Models}\vspace*{0.75cm}}
\author{\large{\uppercase{Niko Hauzenberger$^{1}$,~ Florian Huber$^{1}$,~and Luca Onorante$^2$}\thanks{\noindent Corresponding author: Florian Huber. Department of Economics and Salzburg Centre of European Union Studies, University of Salzburg. Address: M\"{o}nchsberg 2a, 5020 Salzburg, Austria. Email: \href{mailto:florian.huber@sbg.ac.at}{florian.huber@sbg.ac.at}. The first two authors gratefully acknowledge financial support by the Austrian Science Fund (FWF): ZK $35$ and by funds of the Oesterreichische Nationalbank (Austrian Central Bank, Anniversary Fund, project number: $18127$). We would like to thank Michael Pfarrhofer and Gary Koop for helpful comments and suggestions.}\\ \textit{$^{1}$Department of Economics and Salzburg Centre of European Union Studies,\\ University of Salzburg}\\ \textit{$^{2}$Joint Research Centre \\ European Commission}}}
\date{}


\def\equationautorefname~#1\null{%
  Eq.~(#1)\null
}
\def\equationautorefname~#1\null{
Eq.~(#1)\null
}

         \addto\captionsenglish{%
}
\addto\extrasenglish{%
}  

\begin{document}
\maketitle\thispagestyle{empty}\normalsize\small
\vspace*{1cm}
\begin{center}
\begin{minipage}{0.8\textwidth}
\noindent\small Conjugate priors allow for fast inference in large dimensional vector autoregressive (VAR) models but, at the same time, introduce the restriction that each equation features the same set of explanatory variables. This paper proposes a straightforward means of post-processing posterior estimates of a conjugate Bayesian VAR to effectively perform equation-specific covariate selection. Compared to existing techniques using shrinkage alone, our approach combines shrinkage and sparsity in both the VAR coefficients and the error variance-covariance matrices, greatly reducing estimation uncertainty in large dimensions while maintaining computational tractability. We illustrate our approach by means of two applications. The first application uses synthetic data to investigate the properties of the model across different data-generating processes, the second application analyzes the predictive gains from sparsification in a  forecasting exercise for US data. \\
\MakeUppercase{\textit{JEL Codes}}: C11, C30, E3, E37\\[0.5em] \MakeUppercase{\textit{keywords}}: Shrinkage, sparsity, conjugate BVAR, density forecasting\\[2em]
\begin{center}\today\end{center}
\end{minipage}
\end{center}
\normalsize

\renewcommand{\thepage}{\arabic{page}}
\setcounter{page}{1}

\newpage\section{Introduction}\label{sec:intro}
This paper deals with estimating vector autoregressive (VAR) models of the following form,
\begin{equation}
\bm y_t = \bm A_1 \bm y_{t-1}+\dots+ \bm A_p \bm y_{t-p} +\bm C +\bm \varepsilon_t, \label{eq:VAR}
\end{equation}
where $\bm y_t = (y_{1t}, \dots, y_{mt})'$ denotes an $m$-dimensional vector of time series measured in time $t=1,\dots,T$, $\bm A_j~(j=1,\dots,p)$ is an $(m \times m)$-dimensional matrix of coefficients associated with the $j^{th}$ lag of $\bm y_t$, $\bm C$ is an $m$-dimensional intercept vector, and $\bm \varepsilon_t \sim \mathcal{N}(\bm 0, \bm \Sigma)$ is a Gaussian shock vector with zero mean and an $(m \times m)$-dimensional variance-covariance matrix $\bm \Sigma$.  For further convenience, let $\bm a = \text{vec}\{(\bm A_1, \dots, \bm A_p , \bm C)' \}$ denote a vector of dimension $k=m(mp+1)$ of vectorized coefficients with $a_i ~(i=1,\dots, k)$ denoting its $i^{th}$ element. This model class has been extensively used for forecasting and  policy analysis in central banks \citep[see][]{alessi2014central} as well as a natural starting point for unveiling stylized time series facts in order to estimate theoretical models \citep[see, inter alia,][]{hall2012information}. 

Conditional on the first $p$ observations, estimation of the model in \autoref{eq:VAR} can be carried out using ordinary least squares (OLS). In this case, however, overfitting issues arise, translating into imprecise out-of-sample forecasts. As a potential solution, the Bayesian literature uses informative priors to push the system towards a prior model. For instance, the widely used Minnesota prior assumes that the elements in $\bm y_t$ follow a random walk a priori \citep{doan1984forecasting, litterman1986forecasting, giannone2015prior}. Theoretically inspired restrictions stemming from structural models can also be used to inform parameter estimates and thus improve inference \citep{ingram1994supplanting, del2004priors}. The key feature of these priors is that they are conjugate, implying that the likelihood and the prior feature the same distributional from. This yields closed-form solutions for key posterior quantities and, if simulation-based techniques are necessary, greatly improves estimation speed. 

In VARs conjugacy requires that each equation in the system features the same set of predictors, potentially leading to model misspecification \citep[see, for example,][]{george2008bayesian,koop2013forecasting}. This translates into a Kronecker structure in the likelihood, prior, and the resulting posterior distribution, implying that inversion of the posterior variance-covariance matrix of the coefficients is computationally cheap. In contrast, models based on non-conjugate priors allow for different predictors across equations by specifying the prior on the VAR coefficients independently of $\bm \Sigma$. This, however, is computationally much more demanding, since the convenient Kronecker structure is lost.\footnote{For recent solutions that allow for estimating large-scale VARs under non-conjugate priors, see \cite{carriero2019large}.} 

Apart from reduced flexibility in terms of covariate selection across equations, typical shrinkage priors push many VAR coefficients towards zero. Under continuous  shrinkage priors, however, this implies that the probability of observing a coefficient that exactly equals zero is zero \citep[see, for example,][]{griffin2010inference,carvalho2010horseshoe, bhattacharya2015dirichlet, huber2017adaptive, huber2019inducing}. Spike and slab priors allow for shrinking coefficients exactly to zero. These priors rely on an additional set of auxiliary binary indicators that determine whether a coefficient is zero or non-zero. In large models with $k$ covariates (such as the VAR models we consider), estimating these indicators is cumbersome since the number of potential models is $2^k$. In such a situation, Markov chain Monte Carlo (MCMC) techniques often fail to explore this vast model space and convergence issues arise \citep[see][]{polson2010shrink}.

In recent contributions, \citet{hahncarvalho2015dss} and \citet{bhattacharya2018signal} propose a way to circumvent insufficient zero shrinkage/variable selection in light of an increasing amount of predictors (i.e. the curse of dimensionality problem). They estimate a large-scale regression model under a suitable shrinkage prior and then post-process a point estimator (the posterior mean) such that the distance between the fit of the model based on the shrinkage prior and a model based on a sparse estimator (i.e. with coefficients set equal to zero) is minimized while accounting for a penalty term that depends on the $L1$-norm of the coefficients. This approach, labeled decoupled shrinkage and selection (DSS), yields a sparse estimator and is analogous to solving a LASSO-type problem. One key disadvantage, however, is the non-automatic nature of this approach. A semi-automatic approach that is similar in nature is described in \citet{bhattacharya2018signal}. In this framework, an  optimization problem to efficiently set coefficients associated with irrelevant predictors to zero is solved. But instead of performing cross validation, this task depends only on a single tuning parameter that needs to be chosen by the researcher. These techniques that combine shrinkage and sparsity have been shown to work well in a wide range of applications ranging from finance \citep{puelz2017variable, puelz2019portfolio} to macroeconomics \citep{huber2019inducing}.

In this paper, we deal with both issues discussed above by proposing a fully conjugate VAR model coupled with the prior proposed in \citet{kadiyala1997numerical} and \citet{koop2013forecasting}, that allows for different covariates across equations, sparsity in terms of the VAR coefficients, and data-based zero restrictions on the covariance parameters in $\bm \Sigma$. Instead of post-processing posterior mean/median estimates, we follow \cite{huber2019inducing} in sparsifying each draw from the posterior distribution. This yields an approximate posterior distribution for a  sparse vector of coefficients which can be used for uncertainty quantification. The key advantage is that this significantly reduces estimation uncertainty if the data generating process is sparse. For example, if there is strong evidence that  $a_{i}$ equals zero, our proposed framework is capable of selecting this restriction consistently across different draws from the posterior distribution of $a_i$. This implies that the posterior variance of $a_i$, in the limiting case that each draw of $a_i$ is set to zero, also equals zero. In terms of forecasting, the reduced estimation uncertainty could then lead to more precise predictions, especially in situations where $k$ is large.

The merits of our proposed approach are illustrated by means of two applications. In the first application, we use synthetic data obtained from a set of different data generating processes (DGPs) that differ in terms of sparsity, model size, and number of observations. Across DGPs, we find that i.) our framework successfully detects zero values in both the VAR coefficients and the error variance-covariance matrices and ii.) it outperforms other Bayesian VARs (BVARs) in terms of root mean square errors (RMSEs). In the second application, we forecast US output, inflation, and short-term interest rates using the dataset compiled in \cite{mccracken2016fred}. We find that applying the additional sparsification step often improves point and density forecasts. In turbulent times (such as the period of the global financial crisis), however, our results also show that using sparsification could harm the accuracy of density forecasts by underestimating the predictive variance. Nevertheless, these accuracy losses are never substantial and forecasts are still competitive.

The remainder of the paper is structured as follows.
\autoref{sec:econometrics} introduces the conjugate Bayesian VAR while \autoref{sec:DSS} discusses the techniques to achieve sparsity in this model. 
 \autoref{sec:sim} documents model features when applied to synthetic data.  \autoref{sec:forc} summarizes the results of the forecast exercise with real data. Finally, \autoref{sec:concl} concludes the findings of the paper and an appendix provides details on data. 
\clearpage
\section{Bayesian Vector Autoregressions}\label{sec:econometrics}
\subsection{The Natural Conjugate Prior}\label{sec:model}
Before discussing prior implementation, it is worth noting that \autoref{eq:VAR} can be rewritten as a standard regression model,
\begin{equation}
\bm y_t = (\bm I_m \otimes \bm x_t') \bm a + \bm \varepsilon_t,  \label{eq: reg1}
\end{equation}
with $\bm x_t=(\bm y_{t-1}, \dots, \bm y_{t-p}, 1)'$ denoting an $n(=pm+1)$-dimensional vector of explanatory variables. In terms of full-data matrices $\bm Y$ (with $t^{th}$ row $\bm y'_t$) and $\bm X$ (with $t^{th}$ row $\bm x'_t$), the model reads
\begin{equation}
\bm Y = \bm X \bm A + \bm E, \label{eq: reg2}
\end{equation}
where $\bm A = (\bm A_1, \dots, \bm A_p , \bm C)'$ and $\bm E$ is a $(T \times m)$-dimensional matrix of stacked shocks with $t^{th}$ row given by $\bm \varepsilon'_t$. 

The model in \autoref{eq: reg2} features $k$ parameters in $\bm a$ and $w =m (m+1)/2$ free parameters in $\bm \Sigma$. If $m$ and $p$ become large, the number of parameters sharply increases, making precise estimation almost impossible. To deal with this issue, Bayesian econometricians rely on informative priors. This implies that more weight is placed on the prior and the resulting posterior distribution of $\bm a$ and $\bm \Sigma$ will be strongly influenced by the prior model (such as the random walk).

The general form of the conjugate prior in VARs assumes dependence between $\bm a$ and $\bm \Sigma$ and is given by
\begin{equation}
\bm a | \bm \Sigma \sim \mathcal{N}\left(\bm a_0,\bm \Sigma \otimes \bm V_0(\bm \delta)\right), \label{eq:priorA}
\end{equation}
where $\bm a_0$ denotes a $k$-dimensional prior mean vector and $\bm V_0(\bm \delta)$ is a prior variance-covariance matrix that depends on a lower dimensional set of $q$ hyperparameters in $\bm \delta$. In what follows, we assume that $\bm y_t$ is stationary and thus a common choice for the prior mean would be $\bm a_0 = \bm 0$.   

For $\bm V_0(\bm \delta)$,  we use a variant of the conjugate Minnesota prior  \citep{kadiyala1997numerical, koop2013forecasting} that can be implemented using a set of dummy observations \citep[][]{banbura2010large} that are then concatenated to $\bm Y$ and $\bm X$:
\begin{align*}
\uxbar{\bm Y} = \begin{pmatrix}\label{eq:MINNES1}
 \text{diag}(\phi_1 \hat{\sigma}_1,\dots,\phi_m \hat{\sigma}_m)/\theta_1\\\\
\boldsymbol{0}_{m (p-1) \times m}\\ \\
{diag}(\hat{\sigma}_1,\dots,\hat{\sigma}_m)\\ \\
\boldsymbol{0}_{1 \times m}
\end{pmatrix}, \quad \uxbar{\bm X} = \begin{pmatrix}
\boldsymbol{J}_p \otimes \text{diag}(\hat{\sigma}_1,\dots,\hat{\sigma}_m)/\theta_1 ~~ \boldsymbol{0}_{mp \times 1}\\ \\
\boldsymbol{0}_{m \times m p}~~~~~~~~~~~~~~~~~~~~~~~~~ \boldsymbol{0}_{m \times 1}\\ \\
\boldsymbol{0}_{1 \times m p} ~~~~~~~~~~~~~~~~~~~~~~~~~~\pi^{-1/2}
\end{pmatrix},
\end{align*}
with $\bm V_0(\bm \delta) = (\uxbar{\bm X}'\uxbar{\bm X})^{-1}$. Here, $\boldsymbol{J}_p = (1,\dots,p)'$, $\pi$ is a hyperparameter that determines the prior variance on the intercepts, and $\phi_i~(i=1,\dots,m)$ represents the prior mean associated with the coefficient on the first, own lag of a given variable (which is consequently set equal to zero). In addition, we let $\hat{\sigma}^2_i$ denote the OLS variances obtained by estimating $m$ univariate AR($p$) models for each element in $\bm y_t$. Finally,  $\theta_1$ is a hyperparameter that controls the overall tightness of the prior.  Lower values of $\theta_1$ imply a stronger prior belief, effectively pushing the elements in $\bm a$ towards $\bm a_0$. For $\pi$, we simply set it equal to a large value to render the prior on the intercept weakly informative. This prior setup implies that $\bm \delta = (\theta_1, \pi)'$ is a $2$-dimensional vector.

The final ingredient is a conjugate prior on $\bm \Sigma$. Here, conjugacy implies a prior on $\bm \Sigma$ that does not depend on $\bm a$ and follows an inverted Wishart distribution:
\begin{equation}
\bm \Sigma \sim \mathcal{W}^{-1}(s_0, \bm S_0). \label{eq: priorWish}
\end{equation}
We let $s_0$ denote the prior degrees of freedom and $\bm S_0$ a prior scaling matrix.  The main shortcoming of this prior choice  is that shrinking \textit{specific} covariances in $\bm \Sigma$ to zero is impossible. For instance, even if there exists significant evidence that contemporaneous relations across elements in $\bm y_t$ equal zero, this prior is not capable of selecting such restrictions and the resulting posterior estimate of $\bm \Sigma$ (and of its inverse) will be non-sparse.

In the general case (i.e. with any form of the prior hyperparameters), one can show that the conditional posterior of $\bm a$ is given by
\begin{equation}
\bm a|\bm \Sigma, \bm Y, \bm X \sim \mathcal{N}(\text{vec}(\xbar{\bm A}), \bm \Sigma \otimes \xbar{\bm V}), \label{eq:postA}
\end{equation}
with
\begin{align}
\xbar{\bm V} &= [\bm X' \bm X+\bm V_0 (\bm \delta)^{-1}]^{-1},\\
\xbar{\bm A} &= \xbar{\bm V} (\bm X' \bm Y+\bm  V_0(\bm \delta)^{-1} \bm A_0).
\end{align}
Here, we let $\bm A_0$ denote an $(n \times m)$-dimensional matrix reshaped such that $\bm a_0 = \text{vec}(\bm A_0)$. 

Using the Minnesota dummies, the posterior moments can be obtained by applying Theil-Goldberger mixed estimation \citep{theil1961pure}:
\begin{align*}
\xbar{\bm V} =\left(\xbar{\bm X}'\xbar{\bm X}\right)^{-1},\quad
\xAbar{\bm A} = \xbar{\bm V} \hspace{1pt} \xbar{\bm X}'\xbar{\bm Y},
\end{align*}
where  $\xbar{\bm Y} =(\bm Y',  \uxbar{\bm Y})'$ and $\xbar{\bm X}= (\bm X, \uxbar{\bm X})'$ denote  full-data matrices augmented with dummy observations.

Under the prior in \autoref{eq: priorWish}, the posterior distribution also follows an inverted Wishart distribution,
\begin{equation}
\bm \Sigma|\bm Y, \bm X \sim \mathcal{W}^{-1}(s_1, \bm S_1). \label{eq: postWish}
\end{equation}
The posterior degrees of freedom are denoted by $s_1= T + s_0$ and $\bm S_1$ represents the $(m \times m)$-dimensional posterior scaling matrix, obtained by using the Minnesota-specific dummy observations:
\begin{align*}
\bm S_1 = (\xbar{\bm Y} -  \xbar{\bm X} \hspace{1pt} \xAbar{\bm A})'(\xbar{\bm Y} -  \xbar{\bm X} \hspace{1pt} \xAbar{\bm A}),\quad \bm s_1 = T +  s_0.
\end{align*} 

A key advantage of conjugacy is the Kronecker structure in \autoref{eq:postA}, which implies that $\mathcal{\bm  V} = \bm \Sigma \otimes \xbar{\bm V}$ is a block-diagonal matrix and computing the inverse or the Cholesky factor is computationally cheap. By contrast, if $\mathcal{\bm V}$ were a full $(k \times k)$ matrix, computation would quickly become cumbersome and impossible even for moderate values of $m$ and $p$. One further advantage of the conjugate prior is that the one-step-ahead predictive density and the marginal likelihood (ML) are available in closed form \citep[see, for instance,][]{zellner1985bayesian}. This implies that if interest centers on one-step-ahead forecasts, no posterior simulation is required.\footnote{For higher-order forecasts or other quantities such as impulse responses, Monte Carlo simulation is necessary.}  

Unfortunately, the conjugate prior has  two important shortcomings. First, each equation must include the same set of covariates (see \autoref{eq: reg1}), a feature  that could be unappealing if the researcher wishes to introduce theoretically motivated restrictions across equations. Second, the structure of the prior variance-covariance matrix implies that for each equation $j=1,\dots,m$, the prior variance is given by $\sigma^2_{jj} \bm V_0(\bm \delta)$, with $\sigma^2_{jj}$ denoting the $(j,j)^{th}$ element of $\bm \Sigma$. One consequence of this is that across equations, the prior variances are proportional to each other. This implies that it is not possible to discriminate between coefficients on own  (which we define as lags of the $j^{th}$ endogenous variable in equation $j$) and other (defined as the lags of the $i^{th}$ endogenous variable for $i \neq j$ within equation $j$) lags. The methods we discuss in the next section allow for different treatment of own and other lags in a simple way.

\section{Achieving Sparsity in VAR Models}\label{sec:DSS}
From a forecaster's perspective, heavily parameterized models, such as large-scale VARs, have another important shortcoming. The continuous shrinkage prior described in \autoref{sec:model} implies that the probability of observing exact zeros in $\bm a$ equals zero. One could ask whether it makes a big difference to zero out different $a_i$'s  as opposed to setting them close to zero. Setting them close, but not exactly to zero essentially implies that there exists a lower bound of accuracy one can achieve under the specific prior distribution \citep{huber2019inducing}. For small-scale systems, this has negligible implications on predictive accuracy. However, if $k$ is large (i.e. of order $1,000$ or $10,000$), parameter uncertainty adds up and potentially dominates the predictive variance. To see this point, notice that under the conjugate prior, the one-step-ahead predictive density follows a multivariate $t$-distribution \citep[see][]{koop2013forecasting} with predictive variance given by
\begin{equation}
\text{Var}(\bm y_{T+1}|\bm Y, \bm X) = \frac{1}{s_1-2} \left(1 +  \sum_{i=1}^n \sum_{j=1}^n \left(x_{i T+1} x_{j T+1} v_{ij} \right)\right) \bm S_1,\label{eq: varPost1}
\end{equation}
with $\text{Var}( \bullet )$  denoting the variance operator, $x_{i T+1}$ is the $i^{th}$ element of $\bm x_{T+1}$  and $v_{ij}$ referring to the $(i,j)^{th}$ element in $\xbar{\bm V}$. 

Here, it can be seen that the predictive variance depends on the variance of the reduced-form shocks in $\bm \varepsilon_{T+1}$ and parameter uncertainty arising from the term in the parenthesis.  Equation (\ref{eq: varPost1}) indicates that if $n$ increases, posterior uncertainty rises even if the $v_{ij}$'s associated with irrelevant predictors are small. This point clearly highlights the difference between sparsity and shrinkage, namely the fact that under a sparse model, $v_{ij}$ would be equal to zero if and only if the relevant predictor is excluded from the model.  In the next subsections, we will show how this lower bound on accuracy (determined by small but non-zero values of $v_{ij}$) can be removed.

Equation (\ref{eq: varPost1}), moreover, highlights that the uncertainty associated with coefficients potentially adds up in large-scale models and the variance implied by the reduced-form shocks further influences the predictive variance. Without additional restrictions, these two sources  can act in opposite directions. In the case of a large-scale model, the variance-covariance matrix might be underestimated due to overfitting while uncertainty surrounding parameter estimates is too large. The second effect is mainly driven by the fact that in VAR models and with standard macroeconomic datasets, covariates are often highly correlated and this, in combination with insufficient shrinkage, inflates variance estimates of the regression coefficients.  Since these two sources play a crucial role in forming accurate forecasts, it is imperative to treat both of those carefully.


\subsection{Achieving Sparsity on the VAR Coefficients}
Since obtaining a sparse representation of $\bm a$ is unfeasible in high dimensions due to the necessity to explore a model space of cardinality $2^k$, we follow a different route that combines shrinkage and sparsity. Our approach follows \cite{hahncarvalho2015dss} and \cite{bhattacharya2018signal} and  is based on manipulating an estimator $\hat{\bm a}$  ex-post by solving the following optimization problem,
\begin{equation}
\bm {\hat a}^* =  \underset{ {\bm \alpha}}{\text{arg min}}\left\lbrace \frac{1}{2}\left\|(\bm Z  \hat{\bm a} - \bm Z \bm {\alpha} )\right\|^2_2 + \sum_{j = 1}^{k} \kappa_j |{\alpha}_j|\right\rbrace,\label{eq: loss}
\end{equation}
with $\bm Z = (\bm I_m \otimes \bm X)$, $\bm \alpha$ being a sparse $k$-dimensional vector and $\left\| \bm m \right\|_2$ denoting the Euclidean norm of a vector $\bm m$. Equation (\ref{eq: loss}) consists of two components. The first part measures the Euclidean distance between the fit of an unrestricted model, estimated using the shrinkage prior described in \autoref{sec:model}, and a sparse model determined by $\bm \alpha$.  The second part is a penalty term that penalizes non-zero values in $\bm \alpha$, with $\kappa_j$ denoting variable-specific penalties.  In light of large $k$ (which is almost always the case in moderately-sized VARs), choosing the tuning parameters $\kappa_j$ by means of cross-validation becomes computational prohibitive.

To circumvent the necessity to employ cross-validation, we adopt  the signal adaptive variable selection (SAVS) estimator proposed in \cite{bhattacharya2018signal}. We rewrite \autoref{eq: loss} in terms of the $j^{th}$ column of $\bm Z$, $\bm Z_j$, and solve the optimization problem in \autoref{eq: loss}  for each covariate individually, adopting the coordinate descent algorithm \citep{friedman2007pathwise}. This yields the following solution to the optimization problem in \autoref{eq: loss},\footnote{Strictly speaking, this is the solution obtained after the first iteration of the optimization algorithm, which, conditional on initializing the algorithm at the posterior mean, already indicates convergence at this stage.}
\begin{equation}
\hat{a}_j^* = \text{sign}(\hat{a}_j)~ ||\bm Z_j||^{-2} \left( |\hat{a}_j|  ~  ||\bm Z_j||^2 - \kappa_j\right)_+, \label{eq: SAVS}
\end{equation}
for $j=1,\dots,k$, with $\text{sign}(c)$ returning the sign of a real number $c$ and $c_+ = \text{max}\{c, 0\}$. 

We set the penalty term as follows:
\begin{equation}
\kappa_j = \frac{\lambda}{|\hat{a}_j|^{\zeta}},
\label{eq: penalty}
\end{equation}
which depends on the non-sparse estimate $\hat{a}_j$ and two hyperparameters $\lambda>0$ and $\zeta\ge 1$.   Setting $\zeta \ge 1$  implies that smaller values of $\hat{a}_j$ receive a larger penalty and are likely to be zeroed out by the SAVS algorithm. 

A typical choice, proposed in \cite{bhattacharya2018signal}, sets  $\lambda=1$ and $\zeta=2$. We show below that, in simulations, this choice works well. The approach stipulated in \cite{hahncarvalho2015dss} is obtained by setting  $\zeta = 1$ while inferring $\lambda$ by visually inspecting the posterior output. More specifically, \cite{hahncarvalho2015dss} suggest choosing $\lambda$  such that the variation-explained by a sparsified linear predictor (which is akin to a standard R$^2$) statistically equals that of the non-sparsified model. For carrying out structural analysis, this poses no problem since it needs to be done once. However, if the researcher is interested in assessing forecasting accuracy, this procedure has to be repeated sequentially over a hold-out period. This makes the non-automatic nature of the approach problematic.

As stated above, the natural conjugate Minnesota prior is not capable of discriminating between own and  lags other variables. In this paper, we modify the SAVS estimator accordingly. In what follows, we replace $\lambda$ with a lag-wise parameter that increases the weight associated with coefficients on higher order lags of $\bm y_t$ and impose a stronger penalty on coefficients related to the other lags within a given equation. Moreover, we do not sparsify the diagonal elements of $\bm A_1$. These  parameters are specified for each $\bm A_l$ such that
\begin{equation}\label{eq:lambdalag}
\lambda_{l,ij} =\begin{cases}
	\lambda ~(l-1)^2 & \text{if $i = j$ } \\
    \lambda ~l^2  & \text{if $i \neq j$},
  \end{cases}
\end{equation}
for   $l =1, \dots, p; i=1,\dots,m; j=1,\dots,m$. Here, we assume that $\lambda$ is some lag-invariant scaling parameter and $\lambda_{l,ij}$ increases quadratically with the lag order. Note that for the first, own lag of a given equation, we set the penalty equal to zero, capturing the notion that this covariate is crucial and should never be set equal to zero \citep{banbura2010large}. For coefficients on lags of other variables we increase the penalty slightly by multiplying $\lambda$ with $l^2$ instead of $(l-1)^2$.\footnote{In the empirical application, we specify the penalty on the intercept term equal to zero.}
\subsection{Sparsification of the Variance-Covariance Matrix}
Up to this point, we have focused attention on obtaining a sparse representation of the VAR coefficients. 
In large dimensions, $\bm \Sigma$ also contains $w$ free elements and, without using more sophisticated shrinkage techniques, the existing estimate would be prone to overfitting. As a potential remedy, we propose post-processing the estimates of the precision matrix $\bm \Sigma^{-1}$ (i.e. the inverse of $\bm \Sigma$). \cite{friedman2008sparse} and, more recently, \cite{bashir2018post} propose methods to ex-post sparsify precision matrices using the graphical lasso. We follow this literature and specify a loss function similar to \autoref{eq: loss} that aims to strike a balance between model fit and parsimony.

Let $\bm \Omega$ be a sparse estimate of $\bm \Sigma^{-1}$ with elements given by $\omega_{ij}$. The loss function is then given by
\begin{equation}
\hat{\bm  \Omega}^{*} =  \underset{\bm \Omega}{\text{arg min}} \left\lbrace \text{tr}\left( \bm \Omega \hat{\bm S} \right) - \log~ \det \left(\bm \Omega \right) + \sum_{i \neq j} \rho_{ij} |{\omega}_{ij}|\right\rbrace, \label{eq:lossdet}
\end{equation}
with $\hat{\bm S}$ denoting an estimate of the variance-covariance matrix, $\rho_{ij}$ referring to a parameter-specific penalty and $\log \det(\bullet)$ being the log-determinant while $\text{tr}(\bullet)$ denotes the trace of a square matrix. The term $\text{tr}\left( \bm \Omega \hat{\bm S} \right) - \log \left(\det \left(\bm \Omega \right)\right)$ measures the (negative) expected fit whereas $\sum_{i \neq j} \rho_{ij} |{\omega}_{ij}|$ constitutes a penalty term that penalizes non-zero precision parameters in $\bm \Omega$.  Similarly to \autoref{eq: loss}, \autoref{eq:lossdet} aims to find a sparse precision matrix that describes the data well while being parsimonious.\footnote{Note that, if $\hat{\omega}^*_{ij}$ with $i \neq j$, the $(i,j)^{th}$ element of $\hat{\bm \Omega}^*$, is set to zero, then $y_{it}$ and $y_{jt}$ exhibit no contemporaneous relationship.} 

Optimizing \autoref{eq:lossdet} is challenging and suitable penalty parameters need to be defined. We follow \cite{friedman2008sparse}  in adopting the coordinate descent algorithm and state \autoref{eq:lossdet} as a set of independent soft-threshold problems which can be solved for each off-diagonal element, respectively.

To determine the penalty parameter, we follow \cite{glasso} and use:
\begin{equation}
\rho_{ij} = \frac{\varpi}{|\hat{s}^*_{ij}|^{\frac{\kappa}{2}}},
\label{eq: penalty_sig}
\end{equation}
with $|\hat{s}^*_{ij}|$ denoting the absolute size of the $(i,j)^{th}$ element of $\hat{\bm S}^{-1}$ and $\varpi$ is a scalar penalty parameter while $\kappa \ge 1$ controls the penalty on small precision parameters. \autoref{eq: penalty_sig} nests the specification stipulated in \cite{bashir2018post} if we set $\kappa  = 1$, $\hat{s}_{ij}$ to an initial estimate of the $(i,j)^{th}$ element of the precision matrix, and cross-validate $\varpi$.  

It is worth discussing a promising alternative approach to regularization of precision matrices. One could also regularize $\bm \Sigma^{-1}$ by stating it as a set of nodewise regressions \citep{meinshausen2006high}. Exploiting the triangular decomposition of the precision matrix one can treat each node as an independent lasso problem and use the other endogenous variables as covariates. This strategy would imply that one replaces the optimization problem in \autoref{eq:lossdet} by a set of $m$ independent (node-specific) problems as outlined in \autoref{eq: loss}. As noted by \cite{friedman2008sparse} and \cite{banerjee2008model}, however, this approach is a special case of the graphical lasso and thus closely related.

\subsection{Posterior Inference}
Before discussing our posterior simulation algorithm, it is worth noting that up to this point, the different sparsification techniques have been proposed such that some estimate (i.e. the posterior mean/median) is used and then ex-post sparsified. This technique provides a sparse point estimator of $\bm a$ and $\bm \Sigma$ but is not capable of controlling for posterior uncertainty conditional on zeroing out the $v_{ij}$'s. 

Following \cite{huber2019inducing}, we sparsify \textit{each draw} from the joint posterior distribution of $\bm a$ and $\bm \Sigma$. Drawing from the joint posterior is easily achieved. Let $\bm a^{(r)}$ and $\bm \Sigma^{(r)}$ denote the $r^{th}$ draw from the posterior, then we first sample $\bm \Sigma^{(r)}$ from its marginal posterior distribution (\ref{eq: postWish}) and, conditional on this draw, we sample $\bm a^{(r)}$ from (\ref{eq:postA}). 
Given this pair of draws, the corresponding loss functions becomes:
\begin{align}
    \bm {\hat a}^{* (r)} &=  \underset{ {\bm \alpha}}{\text{arg min}}\left\lbrace \frac{1}{2}\left\|(\bm Z  \bm a^{(r)} - \bm Z \bm {\alpha} )\right\|^2_2 + \sum_{j = 1}^{k} \kappa^{(r)}_j |{\alpha}_j|\right\rbrace,\label{eq:lossfun_draw2} \\
\hat{\bm  \Omega}^{* (r)} &=  \underset{\bm \Omega}{\text{arg min}} \left\lbrace \text{tr}\left( \bm \Omega \bm \Sigma^{(r)} \right) - \log~ \det \left(\bm \Omega \right) + \sum_{i \neq j} \rho^{(r)}_{ij} |{\omega}_{ij}|\right\rbrace. \label{eq:lossdet2}
\end{align}

Equations (\ref{eq:lossfun_draw2}) and (\ref{eq:lossdet2}) indicate that   we search for an optimal action that minimizes the loss (instead of the expected loss) \textit{for each draw}. This guarantees that the corresponding sparse estimates associated with the $r^{th}$ draw of $\bm {\hat a}^{* (r)}$ and $\hat{\bm  \Omega}^{* (r)}$  are optimal. To be consistent with the definition of the (variable-specific) penalty parameters in (\ref{eq: penalty}) and (\ref{eq: penalty_sig}), we replace the corresponding point estimators with the draws of $a_j$ and $s_{ij}$.

This approach is similar in nature to \cite{woody2019model} who  perform (approximate) uncertainty quantification around sparse estimators. As opposed to our approach, \cite{woody2019model} estimate a regression model using MCMC and then project each draw into the sparse posterior for the optimal model. This is very similar to our strategy with the main exception that we base our inferences on all sparsified MCMC draws. More precisely, while \cite{woody2019model} rely on a single optimal model (selected using the posterior mean) to project the non-sparse posterior draws into the sparse regression with $q$ selected covariates, our approach allows for uncertainty about this set of $q$ regressors. In the simulation exercise, we will show that the corresponding (sparse) point estimate is close to the one of the traditional approach and thus works  well empirically.

Our approach can be viewed as an approximate algorithm to  draw from the joint posterior of sparsified coefficients $p(\hat{\bm a}^*, \hat{\bm  \Omega}^{*}| \bm Y, \bm X )$. The  optimization problem in (\ref{eq:lossfun_draw2}) is solved using the SAVS estimator. This approach, however, has been developed under the assumption that the point estimate used is the posterior mean/median. \cite{bhattacharya2018signal} show that, using any of these implies that the gradient descent algorithm quickly (after one iteration) converges. Using draws from the posterior of $\bm a$ instead yields similar favorable properties of the optimization algorithm, leading to convergence after one iteration.\footnote{More precise results based on averages of the loss functions of the optimization routine are available from the first author upon request.}

As opposed to the approach proposed in \cite{hahncarvalho2015dss}, our method allows for uncertainty quantification and computation of non-linear functions of the parameters such as impulse responses or higher order predictive distributions. Moreover, it allows for selecting appropriate submodels (defined through inclusion/exclusion of covariates and/or covariance relations in $\bm \Sigma$). Applying sparsification to each draw implicitly yields a sparse estimator of $\bm a$ and $\bm \Sigma$ which can be viewed as a specific restricted version of the non-sparsified model. Since we average across these different sparse estimators, we effectively average across different models. Doing this is similar to Bayesian Model Averaging. In contrast, the traditional method can be viewed as approximate Bayesian Model Selection with the shortcoming that uncertainty across models is not taken into consideration.

 As mentioned in the introductory section, applying the SAVS algorithm to sparsify draws from the joint posterior during MCMC  potentially implies that point estimators such as the posterior mean of $\bm \hat{\bm a}^*$  and $\hat{\bm \Sigma}^*$ are non-sparse. However,  this strongly depends on the information contained in the posterior distribution; if there is significant information that a given coefficient is equal to zero, the corresponding point estimator of the sparsified coefficient could also be exactly zero.

\section{Simulation-based Evidence}\label{sec:sim}
We use a set of different data generating processes (DGPs) that vary in terms of dimension ($m \in \{3, 10, 30\}$), length of the time series ($T \in \{80, 240\})$  and whether the model is sparse, moderately sparse or dense to analyze whether sparsification improves estimation accuracy. All simulated VAR models feature five lags ($p = 5$), with the coefficient matrix $\bm A_j$ (for $j = \{1, \dots, 5\}$) being drawn from $\mathcal{N}(0, (\xi/j)^2)$. In the case of $m = 3$ we set $\xi = 0.3$ and, for stability reasons, we define $\xi = 0.2$ for $m=10$ and $\xi = 0.1$ for $m=30$. Moreover, we add $0.25$ to the diagonal elements of $\bm A_1$.  To capture that higher lag orders become less important, we rescale the variance of the Gaussian by $1/j^2$ (for $j=2, \dots, 5$). Similarly, the non-zero off-diagonal elements of the lower Cholesky factor of $\bm \Sigma$  are sampled from $\mathcal{N}(0, \xi^2)$, while the elements of diag~($\bm \Sigma$) are all non-zero and set to $0.25$. 

We obtain DGPs that feature different levels of sparsity by randomly setting off-diagonal elements in  $\bm A_j$ (for $j = \{1, \dots, 5\}$) and in the lower Cholesky factor of $\bm \Sigma$ to zero. As stated above, we consider three levels of sparsity. The  dense model features around $10\%$ zeroes in the coefficients  while the moderately sparse model features around $60\%$ zeroes.  Finally, we also consider an extremely sparse DGP with approximately $90\%$ zeroes. The dense DGP turns out to be a challenging case for our model. This is because it features a large number of non-zero but small coefficients (especially for $\bm A_j$ with $j>1$) which might be erroneously set equal to zero. 

To assess the sensitivity of the results with respect to different choices of $\lambda$ and $\varpi$, we compute a range of sparse models and benchmark it to the non-sparse competitor. This non-sparse competitor is a Minnesota-prior BVAR with hyperparameters obtained by optimizing the marginal likelihood of the model over a grid \citep[see, for example,][]{carriero2019large}. Moreover, we consider the stochastic search variable selection (SSVS) prior as competitor \citep{george1993variable, george1997approaches}. This model assumes a mixture of Gaussians prior to introduce sparsity but has the severe drawback of being non-conjugate and thus challenging to estimate in large dimensions. For the SSVS we follow \cite{george2008bayesian} and re-scale the spike and slab component with the OLS coefficient variances denoted by $\hat{v}_{ii}$. The corresponding spike variance is then given by $0.01 \times \hat{v}_{ii}$ while the slab variance is considerably larger with $100 \times \hat{v}_{ii}$. 

Finally, we add two additional competing specifications. The first one (labeled SAVS-Median) is the traditional SAVS approach stipulated in \cite{bhattacharya2018signal} which sparsifies the posterior median. This model allows us to assess whether sparsifying the posterior median yields similar insights (in terms of point estimates) as our approach. The second one, labeled CDA, sparsifies each draw from the joint posterior using a standard coordinate descent algorithm (i.e. without stopping after the first iteration). This specification serves to illustrate whether using more iterations yields similar insights compared to stopping after the first iteration.

\begin{table}[!tbp]
{\tiny
\begin{center}
\begin{tabular}{llllclllllcll}
\toprule
\multicolumn{1}{l}{\bfseries }&\multicolumn{3}{c}{\bfseries DGP}&\multicolumn{1}{c}{\bfseries }&\multicolumn{5}{c}{\bfseries Specification}&\multicolumn{1}{c}{\bfseries }&\multicolumn{2}{c}{\bfseries Alternatives with $\lambda = 1$}\tabularnewline
\cline{2-4} \cline{6-10} \cline{12-13}
\multicolumn{1}{l}{}&\multicolumn{1}{c}{m}&\multicolumn{1}{c}{T}&\multicolumn{1}{c}{Sparsity}&\multicolumn{1}{c}{}&\multicolumn{1}{c}{MIN - $\lambda = 0.01$}&\multicolumn{1}{c}{MIN - $\lambda = 0.1 $}&\multicolumn{1}{c}{MIN - $\lambda = 0.5 $}&\multicolumn{1}{c}{MIN - $\lambda = 1 $}&\multicolumn{1}{c}{SSVS}&\multicolumn{1}{c}{}&\multicolumn{1}{c}{SAVS - Median}&\multicolumn{1}{c}{CDA}\tabularnewline
\midrule
{\scshape COEFFICIENTS}&&&&&&&&&&&&\tabularnewline
~~&S&80&Sparse&&0.654&0.504&0.451&\textbf{0.446}&0.546&&0.446&0.446\tabularnewline
~~&&&Moderate&&0.727&0.636&\textbf{0.623}&0.629&0.669&&0.629&0.629\tabularnewline
~~&&&Dense&&0.847&\textbf{0.818}&0.830&0.846&0.860&&0.846&0.846\tabularnewline
~~&&240&Sparse&&0.618&0.454&0.414&\textbf{0.413}&0.505&&0.413&0.413\tabularnewline
~~&&&Moderate&&0.717&0.612&\textbf{0.604}&0.611&0.651&&0.611&0.611\tabularnewline
~~&&&Dense&&0.844&\textbf{0.819}&0.846&0.865&0.865&&0.865&0.865\tabularnewline
\midrule
{\scshape }&&&&&&&&&&&&\tabularnewline
~~&M&80&Sparse&&0.637&0.491&0.436&\textbf{0.436}&0.632&&0.436&0.436\tabularnewline
~~&&&Moderate&&0.758&0.697&\textbf{0.696}&0.710&0.797&&0.710&0.710\tabularnewline
~~&&&Dense&&\textbf{0.913}&0.924&0.970&0.999&1.018&&0.999&0.999\tabularnewline
~~&&240&Sparse&&0.588&0.423&\textbf{0.377}&0.382&0.505&&0.382&0.382\tabularnewline
~~&&&Moderate&&0.716&\textbf{0.642}&0.644&0.665&0.692&&0.665&0.665\tabularnewline
~~&&&Dense&&\textbf{0.886}&0.890&0.927&0.960&0.963&&0.960&0.960\tabularnewline
\midrule
{\scshape }&&&&&&&&&&&&\tabularnewline
~~&L&80&Sparse&&0.579&0.507&\textbf{0.507}&0.507&1.572&&0.507&0.507\tabularnewline
~~&&&Moderate&&0.800&\textbf{0.794}&0.798&0.798&1.513&&0.798&0.798\tabularnewline
~~&&&Dense&&\textbf{0.984}&1.023&1.032&1.032&1.432&&1.032&1.032\tabularnewline
~~&&240&Sparse&&0.577&0.441&\textbf{0.439}&0.442&0.704&&0.442&0.442\tabularnewline
~~&&&Moderate&&0.760&\textbf{0.735}&0.770&0.779&0.876&&0.779&0.779\tabularnewline
~~&&&Dense&&\textbf{0.953}&1.005&1.070&1.086&1.075&&1.086&1.086\tabularnewline
\midrule
{\scshape COVARIANCES}&&&&&&&&&&&&\tabularnewline
~~&S&80&Sparse&&0.996&0.971&0.903&0.847&\textbf{0.647}&&0.847&0.847\tabularnewline
~~&&&Moderate&&0.998&0.981&0.935&0.899&\textbf{0.792}&&0.899&0.899\tabularnewline
~~&&&Dense&&1.000&\textbf{1.000}&1.001&1.007&1.090&&1.007&1.007\tabularnewline
~~&&240&Sparse&&0.987&0.919&0.771&\textbf{0.694}&0.700&&0.694&0.694\tabularnewline
~~&&&Moderate&&0.994&0.955&0.871&0.835&\textbf{0.829}&&0.835&0.835\tabularnewline
~~&&&Dense&&0.999&\textbf{0.999}&1.010&1.039&1.153&&1.039&1.039\tabularnewline
\midrule
{\scshape }&&&&&&&&&&&&\tabularnewline
~~&M&80&Sparse&&0.998&0.980&0.918&0.859&\textbf{0.594}&&0.859&0.859\tabularnewline
~~&&&Moderate&&0.999&0.988&0.953&0.922&\textbf{0.826}&&0.922&0.922\tabularnewline
~~&&&Dense&&1.000&\textbf{0.999}&1.000&1.007&1.174&&1.007&1.007\tabularnewline
~~&&240&Sparse&&0.992&0.936&0.788&0.687&\textbf{0.639}&&0.687&0.686\tabularnewline
~~&&&Moderate&&0.996&0.973&0.915&0.889&\textbf{0.863}&&0.889&0.889\tabularnewline
~~&&&Dense&&\textbf{1.000}&1.003&1.032&1.081&1.248&&1.081&1.082\tabularnewline
\midrule
{\scshape }&&&&&&&&&&&&\tabularnewline
~~&L&80&Sparse&&0.998&0.985&0.934&0.882&\textbf{0.632}&&0.882&0.881\tabularnewline
~~&&&Moderate&&0.999&0.988&0.950&0.913&\textbf{0.809}&&0.913&0.912\tabularnewline
~~&&&Dense&&0.999&0.992&0.968&\textbf{0.950}&1.004&&0.950&0.950\tabularnewline
~~&&240&Sparse&&0.993&0.938&0.780&0.658&\textbf{0.604}&&0.658&0.656\tabularnewline
~~&&&Moderate&&0.996&0.966&0.887&0.841&\textbf{0.827}&&0.841&0.840\tabularnewline
~~&&&Dense&&0.999&\textbf{0.993}&0.998&1.026&1.106&&1.026&1.026\tabularnewline
\bottomrule
\end{tabular}
\caption{MAE ratios of coefficients and covariances to non-sparse BVAR estimates.\label{tab:MAD-A}} 
\begin{minipage}{\textwidth}
 \footnotesize\textbf{Notes:} Bold numbers indicate the smallest MAE ratios. We simulate a DGP for a \textit{small-scale} ($m = 3$), \textit{medium-scale} ($m = 10$) and \textit{large-scale} ($m = 30$) VAR for two different number of observations $T$ and for three different degrees of sparsity (zero parameters as percentage of total number of coefficients $k = m(mp + 1)$ and covariances $w = m(m + 1)/2$, ranging from a dense DGP to a fully sparse DGP. \texttt{SAVS-Med.} refers to a sparse estimator, where we minimize the expected loss of the posterior median \cite[see][]{hahncarvalho2015dss}, while for the \texttt{CDA} specification we replace the SAVS estimator with a coordinate descent algorithm, i.e. we do not stop after the first iteration for both coefficients and covariances. 
\end{minipage}\end{center}}
\end{table}

Table~\ref{tab:MAD-A} shows (relative) mean absolute errors (MAE) between the posterior median of the coefficients for the sparsified BVAR and the true parameter values, averaged across $150$ replications per DGP. Note that all numbers in the table feature a numerical standard error which is relatively small. Nevertheless, these findings need to be interpreted with some caution and we aim to focus on results that imply substantial differences relative to the benchmark after taking into account the simulation-induced variation. All MAEs are divided by the MAEs of non-sparse competitor. 

The upper panel  of the table presents the results for the VAR coefficients while the lower panel of \autoref{tab:MAD-A} displays the MAEs associated with the covariance parameters. In order to investigate how differing values of $\lambda$ and $\varpi$ impact estimation accuracy, we also estimate the model over a grid of values for $\lambda \in \{0.01, 0.1, 0.5, 1\}$ and set $\varpi = \lambda/10$. Before proceeding, it is worth noting that we estimate all VAR models with  five lags. 

Considering the upper panel of \autoref{tab:MAD-A}, a few results are worth emphasizing. \textit{First}, we observe that sparsification pays off in terms of achieving lower estimation errors. This improvement strongly depends on the true level of sparsity, with strong accuracy gains if the DGP is very sparse and the data sample is short. Especially when $T$ is small relative to the number of parameters, sparsification improves against the traditional Bayesian VAR model. 

\textit{Second}, the sensitivity of estimation accuracy with respect to $\lambda$ varies with the level of sparsity. For instance, we find slightly more pronounced differences if the DGP is either dense or moderately dense but, as long as $\lambda$ is set greater to $0.01$ we find only small differences in MAEs. It is worth noting, however, that these small differences across different penalty terms are often insignificant. Once we take into account numerical standard errors the specific choice of $\lambda$ (as long as it is not set too small) seems to play a minor role with differences being smaller than ten percent in MAE terms (and thus often within one standard deviation) for most models considered. This also provides some evidence that the specific choice proposed in \cite{bhattacharya2018signal} (i.e. $\lambda =1$) works well in most circumstances.

\textit{Third}, for large models we find that the SAVS estimator yields substantial gains, improving upon the shrinkage-only estimator by large margins. These improvements even arise if the DGP is characterized by relatively few zeros in the VAR coefficients. This finding is not surprising given the fact that the absolute number of zeros increases with the dimension of the parameter space and the small but negligible posterior estimates under the Minnesota BVAR have a detrimental effect on estimation accuracy.

\textit{Fourth}, we find that the conjugate VARs in combination with SAVS very often improve upon the non-conjugate and more flexible VAR coupled with the SSVS prior. For large-scale models, we even find that the SSVS prior performs rather poorly and this might be caused by mixing issues in the indicators that determine the mixture of Gaussian to be chosen. In smaller-sized models, accuracy differences decrease but still favor our proposed sparsified model. 

\textit{Finally, and importantly,} comparing the performance of the SAVS - Median and CDA approaches with the corresponding model based on setting $\lambda=1$ reveals no differences in estimation accuracy. This remarkable result shows that sparsifying each draw from the joint posterior yields (almost) identical sparsified point estimators and provides evidence that using only a single iteration of the coordinate descent algorithm seems to be sufficient when compared to using a stopping rule (and thus potentially many more iterations).

The lower panel of \autoref{tab:MAD-A} provides  similar but more mixed insights. Sparsification of the variance-covariance matrix sometimes yields accuracy improvements over its non-sparsified counterpart. These improvements range from being small (or in some rare cases even negative) to very large (in the case the DGP is sparse and the model is moderately large). We conjecture that the somewhat smaller improvements in predictive accuracy arise from the Wishart-distributed prior imposed on $\bm \Sigma^{-1}$. This prior, by construction, is not capable of discriminating between relevant and irrelevant covariance parameters since it uniformly pushes the posterior estimate of $\bm \Sigma^{-1}$ towards a diagonal matrix. Again, we find no discernible differences between SAVS-median and CDA to the respective model with $\lambda=1$.

Considering the performance of the SSVS prior shows that it provides more accurate estimates of $\bm \Sigma$ but at substantially higher computational costs. In most instances where the SSVS prior improves upon our SAVS-based model, these performance gains are often small (i.e. below ten to 15 percent in MAE terms). So given that the additional costs of applying SAVS to the posterior draws of $\bm \Sigma$ are negligible, we can recommend adding this additional step to further improve estimation accuracy.

\begin{figure}[htbp]
\centering
\begin{minipage}{0.49\textwidth}
\centering
\textit{Mean absolute error}
\end{minipage}
\begin{minipage}{0.49\textwidth}
\centering
\textit{Standard deviation of estimates}
\end{minipage}
\begin{minipage}{1\textwidth}
\centering
\vspace{10pt}
(a) \textit{Coefficients}
\end{minipage}
\begin{minipage}{0.49\textwidth}
\centering
\includegraphics[width=\textwidth]{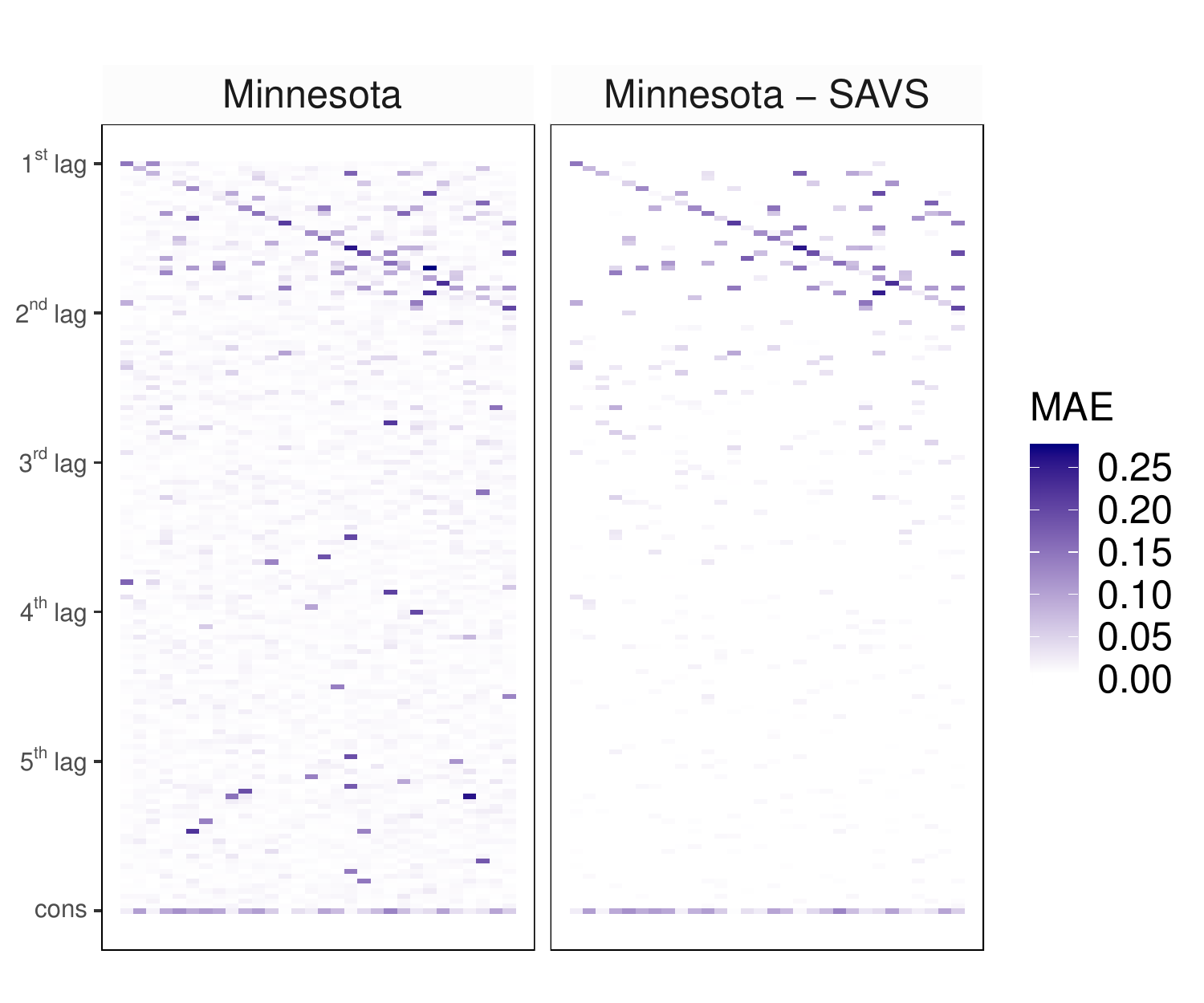}
\end{minipage}
\begin{minipage}{0.49\textwidth}
\centering
\includegraphics[width=\textwidth]{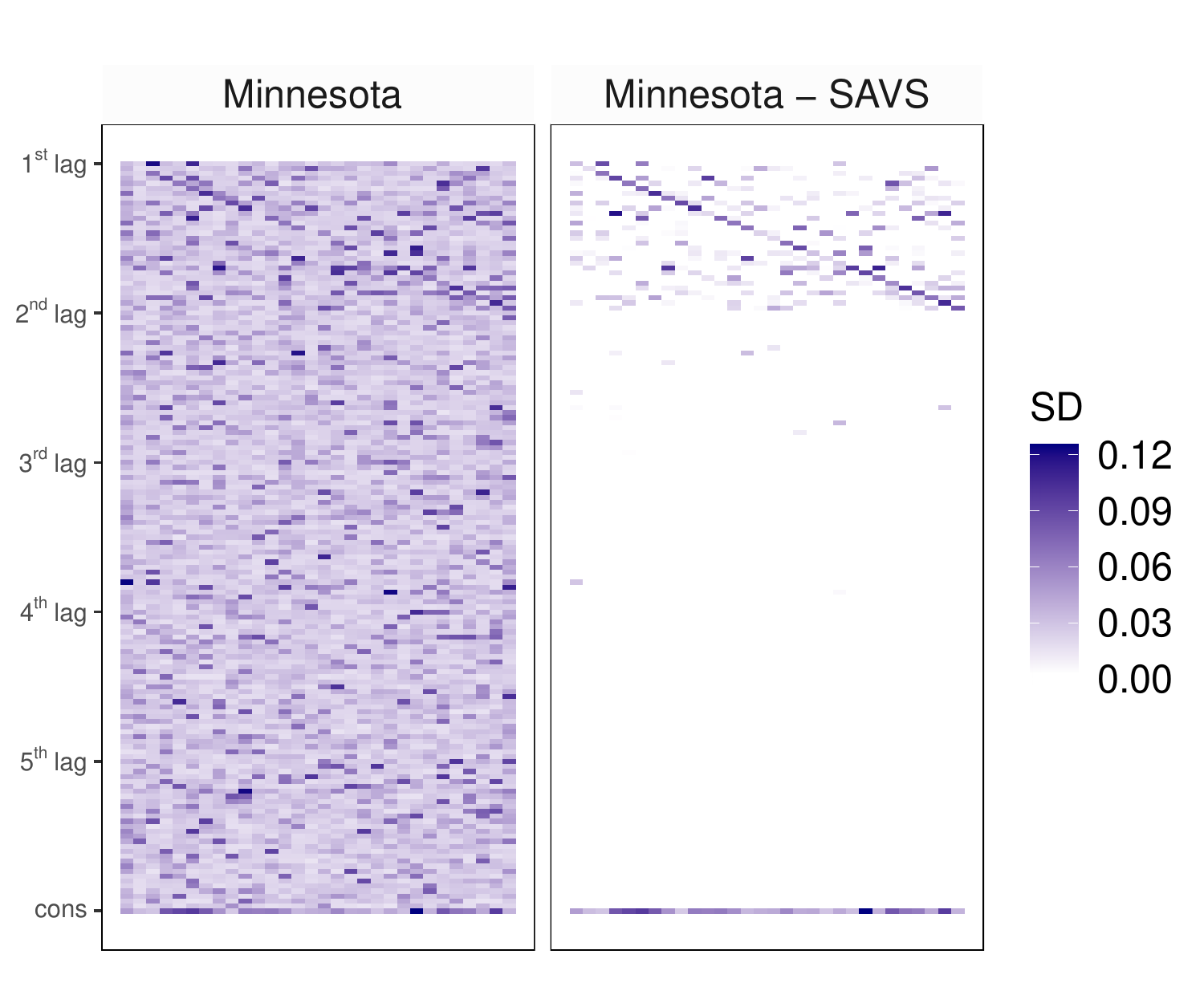}
\end{minipage}
\begin{minipage}{1\textwidth}
\centering
(b) \textit{Lower Cholesky factor of the variance-covariance matrix}
\end{minipage}
\begin{minipage}{0.49\textwidth}
\centering
\includegraphics[width=\textwidth]{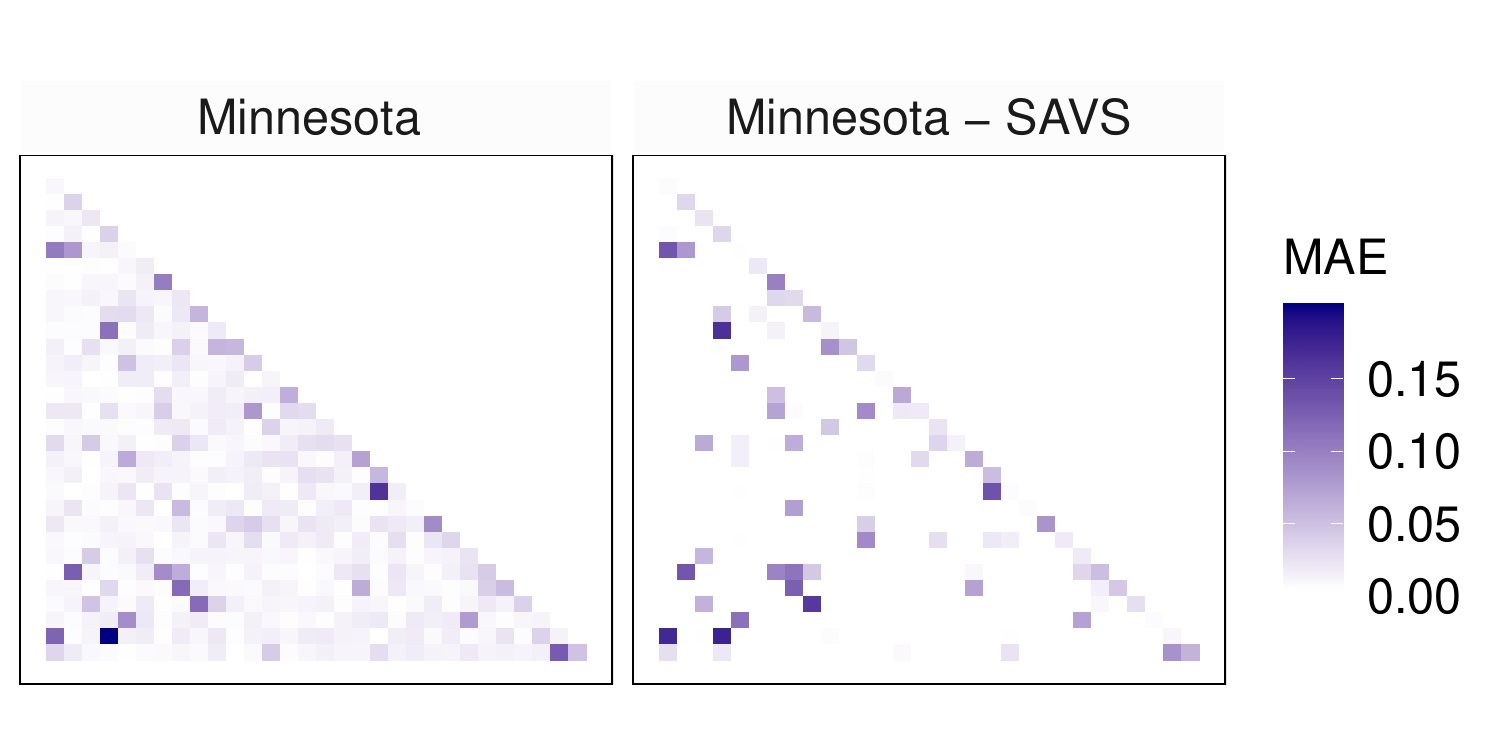}
\end{minipage}
\begin{minipage}{0.49\textwidth}
\centering
\includegraphics[width=\textwidth]{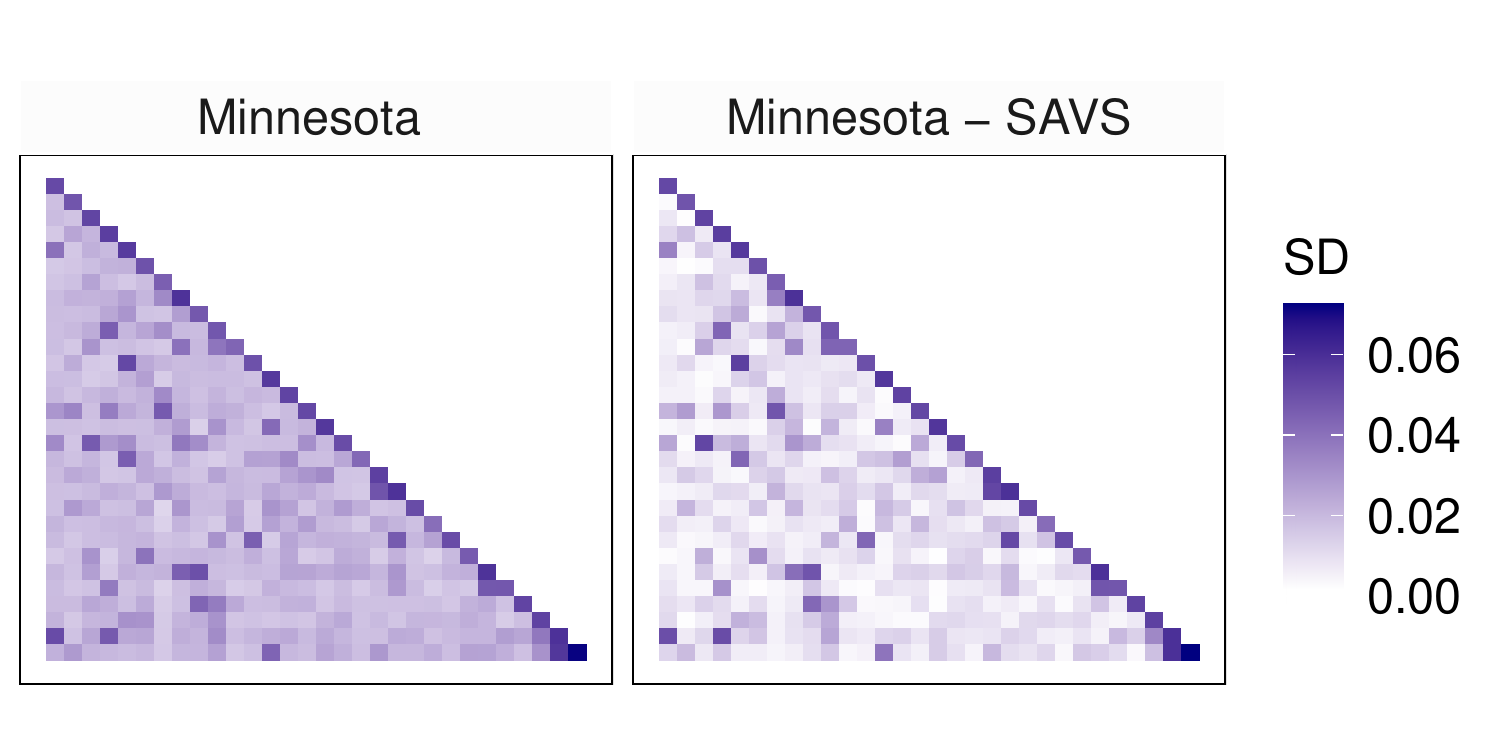}
\end{minipage}
\caption{Heatmaps of coefficients and covariances for $m =30$ endogenous variables, $p = 5$, $T = 240$ and the degree of sparsity is $90\%$. Panel (a) lists the $m$ endogenous variables on the $x$-axis and the $(mp +1)$ regressors for each equation on the $y$-axis. The $i^{th}$ $(m \times m)$ block denotes the $i^{th}$ lag coefficient matrix, while the constant is ordered last (indicated by cons). Moreover, in panel (b) the lower Cholesky factor of the variance-covariance matrix is an $(m \times m)$-dimensional lower triangular matrix.}  
\label{fig:sparse0980}
\end{figure}

We stressed one key advantage in \autoref{sec:DSS}, namely that sparsification reduces estimation uncertainty by zeroing out the coefficient under scrutiny during posterior simulation. Thus, while the discussion in the previous  paragraphs highlights that using sparsification improves estimation performance in terms of point estimators, we now investigate its consequences on the posterior variance of $\bm a$.   \autoref{fig:sparse0980}(a) and (b) are heatmaps that show the absolute distance between the posterior median and the true coefficients (left panel) as well as a corresponding heatmap that presents the corresponding posterior standard deviation of the parameter under scrutiny (right panel). These heatmaps are created for a single realization from the sparse DGP with $T= 240$ and $m = 30$. 

Considering the heatmaps reveals that sparsification improves estimation accuracy and accurately detects zeroes, as evidenced by the abundance of white cells in the figure. The slight bias along the main diagonal (which also exists under the shrinkage-only model) stems from the informative prior that is centered around zero. However, note that even with a high degree of shrinkage, the corresponding estimate of $\bm a$ with the Minnesota prior is quite dense. By contrast, applying SAVS yields a very sparse coefficient matrix and, in addition, a sparsified estimate of the variance-covariance specification.

The left panel of  \autoref{fig:sparse0980} shows that sparsity is, not surprisingly, accompanied by appreciable decreases in posterior variance. White cells imply that the posterior standard deviation is (close to) zero. This is often the case if we post-process the posterior using SAVS. As expected, we see that the coefficients associated with the first lag often feature considerable posterior uncertainty. For higher lag orders, however, SAVS effectively reduces posterior uncertainty and, in light of the strong accuracy gains for the point estimator, leads to a much more favorable bias-variance relationship. For a standard Minnesota prior without SAVS, we observe a more dense heatmap with a great deal of purple shaded areas. We would like to stress that even for the Minnesota prior these standard deviations are often small (especially compared to some weakly informative prior). But these small elements in $v_{ij}$ could add up (see \autoref{eq: varPost1}) and thus be deleterious for predictive accuracy. And it is precisely this problem which we try to circumvent by applying SAVS.

On the variance-covariance matrix we, again, see more white cells under the sparse model. This shows that we reduce estimation uncertainty. Considering the right panel of \autoref{fig:sparse0980}(b), moreover, implies that we also successfully reduce posterior uncertainty around the free elements in $\bm \Sigma$.

To sum up, our simulation exercise shows that using SAVS almost always improves estimates of $\bm a$ and $\bm \Sigma$. These accuracy gains increase strongly with the degree of sparsity in the DGP. At a first glance, we find some differences across the competing values of $\lambda$. As a general recommendation, we advise to stick to the standard setup proposed in \cite{bhattacharya2018signal} and set $\lambda = 1$. This choice works reasonably well and is always close to the optimal value of $\lambda$ (and it often coincides with it).

\section{Forecasting Application}\label{sec:forc}
\subsection{Data Overview, Design of the Forecasting Exercise and Competitors}
In simulations, we have shown that our approach yields more precise parameter estimates if the DGP is sparse. To illustrate the merits of our approach for forecasters in central banks and other policy institutions, we now carry out a US macroeconomic forecasting exercise. Our empirical work relies on the dataset discussed in \cite{mccracken2016fred}. Several recent papers have analyzed this dataset using various shrinkage and sparsification techniques \citep{giannone2017economic, cross2020macroeconomic} and find mixed evidence for sparsity. Nevertheless, using this application we aim to illustrate that even when there is little evidence in favor of sparsity, the proposed framework is still capable of improving upon a model that relies solely on shrinkage priors.

In this application we use the quarterly variant of the \cite{mccracken2016fred} dataset that  spans the period from $1959$:Q$1$ up to $2018$:Q$4$. 
To investigate whether combining shrinkage and sparsity improves  predictive performance, we split the sample in two parts. The first part, called the training sample, runs from $1959$:Q$1$ to $1989$:Q$4$. This initial period is used to compute the $h$-step-ahead predictive distribution (for $h \in \{1, 4, 8\}$).  After obtaining the predictive density for all $h$, we expand the initial estimation period by one quarter (i.e. to $1990:$Q$1$) and repeat this procedure until we reach the penultimate point in the sample (i.e. $2018$:Q$3$).  The period from $1990$:Q$1$ to $2018$:Q$4$ is consequently labeled the hold-out or verification period. For each quarter in the hold-out period, we evaluate the predictive accuracy of the models using root-mean-squared forecast errors (RMSEs), log predictive likelihoods (LPLs), and normalized forecast errors.\footnote{As alternative metric, we also computed compute continuous rank probability scores (CRPS).  Compared to the LPLs, the CRPS yield qualitatively similar insights. For the sake of brevity, we focus on LPLs in the paper. The CRPS are available from the first author upon request.}

The existing literature highlights the necessity to exploit large information sets \citep[see, for instance,][]{banbura2010large, carriero2015bayesian, giannone2015prior, koop2013forecasting}. Building on this evidence, we apply our techniques to a VAR model that features $m=165$ macroeconomic and financial variables. Out of these, we select three traditional target variables, namely output (GDPC1), consumer price inflation (CPIAUCSL) and the Federal Funds Rate (FEDFUNDS). Since the computational burden of using non-conjugate priors increases dramatically with model size, we do not use the SSVS prior for the large dataset.

Apart from this large-scale VAR (L-VAR), we investigate how our techniques perform across different model sizes and dimension-reduction techniques. These competing models range from small and medium-scale VARs to dynamic factor models in the spirit of \cite{bernanke2005measuring}. These competing approaches are:\footnote{In Appendix \ref{app:data} we show a detailed list of the variables included along the transformation codes.}

\begin{itemize}
\item \textbf{S-VAR:} This specification is inspired by the literature using small-scale three equation VAR models that feature the three target variables exclusively.

\item \textbf{M-VAR:} This model extends the small-scale model by additionally including financial market variables. In total, this model includes $m=21$ variables and thus resembles the size of typical reduced-form models employed by the ECB to carry out its short-term inflation projections. 

\item \textbf{FA-VAR:} As a competitor that exploits the full information set but reduces the dimensionality of the estimation problem, we use a factor-augmented VAR (FA-VAR). This model augments the small-scale VAR by including three principal components extracted from the remaining quantities \citep[see][]{banbura2010large, koop2013forecasting}.
\end{itemize}
 All models feature $p=5$ lags of the endogenous variables.
 
\subsection{Choice of hyperparameters}
Since we use VAR models that do not only differ in the size of their information sets, but also how these information is used during estimation, careful choice of the prior hyperparameters is necessary. Using the prior outlined above, we need to set $\theta_1$ as well as the sparsification parameters $\lambda$ and $\varpi$. 

To set the hyperparameters of the Minnesota prior, we follow two different routes.  The first (and simplest) way, is to set $\theta_1$ such that shrinkage increases with the size of the information set \citep[see, for instance,][]{banbura2010large, koop2013forecasting} and select $\theta_1$ on a grid of potential values.

The second approach is based on optimizing the marginal likelihood (which is available in closed form) a priori \citep[see][]{carriero2015bayesian}. One problem with this strategy, however, is that the marginal likelihood might be ill-behaved, which renders optimization difficult.\footnote{To circumvent this issue  \cite{banbura2010large} and \cite{koop2013forecasting}, for example, define a training sample which serves the purpose to calibrate $\theta_1$ by minimizing the distance of the mean square error (MSE) of a large-scale model and a three-variable VAR estimated with OLS. Intuitively, this strategy implies that large dimensional models are shrunk to a larger degree than smaller-scale models \citep{de2008forecasting}.} 

Optimizing the marginal likelihood in huge models is often unfeasible due to numerical reasons. As a solution, we assess the sensitivity of the forecasts with respect to three choices of the hyperparameters $\theta_1 = \{0.025, 0.05, 0.075\}$. These three values all lead to an informative prior but the weight placed on prior information ranges from being large ($\theta_1 = 0.025$) to being more moderate ($\theta_1 = 0.075$).
Using these  values enables us to assess  how shrinkage and sparsification interact. For example, if $\theta_1$is set to $0.025$ 
it is very likely that the SAVS estimator will lead to a sparse model while a shrinkage parameter $\theta_1=0.075$ allows for larger elements in $\bm a$ and thus a more dense model under the SAVS estimator.

For the small- and medium-scale models and the FA-VAR specification, we define a large grid of values for $\theta_1 \in \{0.01, 0.025, 0.050, 0.075, 0.10, 0.125, 0.15, 0.20, 0.25, 0.30, 0.35, 0.40, 0.45, 0.50, 0.75, 1, 2, 5\}$.  Using this grid, we seek the value of $\theta_1$ that maximizes the marginal likelihood. Over the hold-out sample, this procedure yields an average of $\theta_1 = 0.457$ for the small-scale model, $\theta_1 = 0.392$ for the FA-VAR specification and $\theta_1 = 0.149$ for the medium-scale model. These values indicate that the larger the model becomes, the more weight needs to be placed on the prior. Similar to \cite{carriero2015bayesian}, we find that the hyperparameters tend to display little variation over the hold-out period.   For example, in the case of the medium-scale model, we find that $\theta_1$ ranges from $0.125$ to $0.15$.

Finally, we investigate how forecasting performance changes for different values of $\lambda$ and $\varpi$, again, using a grid of candidate values. More precisely, we set $\lambda \in \{0.01, 0.1, 0.5, 1\}$ and $\varpi = \lambda/10$.




\subsection{Point Forecasting Performance}

In this sub-section, we start by considering point forecasting accuracy of the different models and by comparing sparse with non-sparse models.
Table~\ref{tab:RMSE} depicts the relative RMSEs to a small-scale VAR with a Minnesota prior (henceforth called the benchmark model) and without sparsification for the three target variables. The asterisks indicate statistical significance for each model relative to the benchmark as measured by the Diebold-Mariano (DM) test \citep{diebold1995comparing}. The numbers in parentheses refer to the ranking of the three best specifications using the procedure outlined in \cite{hansen2011model}.

\subsection*{One-quarter-ahead Point Forecasts}
We start by considering the average model performance (in terms of computing average RMSEs across the three focus variables) for the one-step-ahead forecasts. In general, there exists no single superior modeling approach that outperforms its competitors in a statistically significant manner. However, when it comes to lowest RMSEs, our results suggest that the most accurate one-step-ahead forecasts can be found for smaller-sized models with sparsification (with the S-VAR with $\lambda = 0.5$ yielding the lowest average RMSEs). Irrespective of $\lambda$, small-scale models seem to perform well and are the only ones that outperform the benchmark Minnesota BVAR.

To analyze point forecasts in more detail, we now discuss the one-step-ahead forecasting performance across the three focus variables. For output and one-step-ahead forecasts, we observe that most large models display relative RMSEs below or close to one, suggesting that their predictions tend to improve forecast accuracy compared to the ones obtained from using the simple small-scale Bayesian VAR. When we compare sparsified large-scale BVARs with 
the non-sparsified counterparts we find limited evidence that sparsification improves point forecasts (but it also never substantially hurts predictive accuracy). Nevertheless, we find that  sparsified large-scale BVARs with $\theta_1 = \{0.05, 0.075\}$ and $\lambda = 0.01$ improve upon the benchmark by around eleven to twelve percent in RMSE terms. 

Turning to the medium-scale models, we find  more pronounced improvements relative to the benchmark model and that using SAVS improves predictive accuracy. These performance gains depend on the specific value of $\lambda$. For example, setting $\lambda = 0.01$ yields a model that improves upon all competitors and produces statistically significant better forecasts than the non-sparsified benchmark model (at the five percent significance level). The second best performing model is the medium-scale VAR that sets $\lambda = 0.1$ (with differences between $\lambda = 0.01$ and $\lambda = 0.1$ being small). Finally, if $\lambda$ is set too large (i.e. equal to $0.5$ or $1$) we see that forecast quality tends to converge towards the benchmark model.

Although the M-VAR with a Minnesota prior tends to outperform the FA-VAR (irrespective of the choice of $\lambda$), a reverse picture emerges when considering the SSVS prior. With the SSVS prior a FA-VAR outperforms its medium-scale counterpart. For the SSVS model we would expect such an outcome, since the forecast performance of the spike-and-slab prior commonly deteriorates with model size \citep[see, for instance,][]{koop2013forecasting}.

Considering the one-step-ahead forecasts of inflation, it appears that none of the larger-scale competitors is capable of improving upon the benchmark small-scale VAR. Notice, however, that for the medium-sized model and the FA-VAR specification, introducing sparsity through our SAVS estimator sometimes yields more accurate short-run predictions and tends to outperform not only the non-sparsified model, but also the SSVS prior. In fact, a FA-VAR and M-VAR equipped with an SSVS prior produces quite poor one-quarter-ahead inflation forecasts. For the S-VAR, some sparsification slightly improves forecast accuracy. Setting $\lambda = 0.01$ and thus introducing a small penalty on non-zero elements in $\hat{\bm a}^*$ yields the most precise inflation forecasts. In general, and similarly to output, we find relative RMSEs close to unity (and in fact often slightly exceeding unity). This is consistent with the literature on inflation and mainly driven by the high persistence of inflation. 

The one-quarter-ahead interest rate forecasts reveals two insights. First, we find that medium- to large-sized models yield forecasts that tend to be more accurate than the ones obtained from the benchmark VAR. This result is particularly pronounced for the moderately-sized VAR and $\lambda = 0.5$. 
Second, comparing sparse with non-sparse models shows that for interest rates, introducing sparsity pays off markedly. This is even visible for the small-scale model and $\lambda=1$ (ranked second), with RMSEs being over $26$ percent lower as compared to the shrinkage-only case. The largest gains, as expected, can be obtained in situations with increased model size. For the medium and large models, the increases in accuracy after sparsification are substantial and statistically significant. And these display little differences across the values of $\lambda$.  We conjecture that some of the increases in predictive performance are driven by the zero lower bound, which is  prominently featured in our hold-out sample. Methods that shrink but do not sparsify yield forecasts that are non-zero. This implies larger forecasts errors as compared to models that set coefficients in the interest rate equation to zero and thus predict that short-term interest rates to be zero in the next period.\footnote{A substantially lower RMSE error of the non-sparsified large-scale model with a relatively tight Minnesota prior ($\theta_1 = 0.025$) compared to the non-sparsified L-VAR with $\theta =\{0.05, 0.075\}$ underpins this observation.} This is the main source of the strong accuracy gains obtained by applying the SAVS step.

\subsection*{Multi-steps-ahead Point Forecasts}
One important limitation of our forecasting exercise is that we rely on using final vintage data. This essentially implies that, for each point in the training sample, we use information that is not available at the time the forecast is produced. This effect is most pronounced in the case of the one-step-ahead forecast. Once we increase the forecast horizon, the corresponding information advantage becomes smaller (since we use information up to time $T$ to predict $\bm y_{T+h}$ for  $h > 1$). This is because data revisions after a few quarters are typically very close to final vintage data. 

The discussion above suggests that studying higher order forecasts in more detail is worthwhile. 
Next, we consider four and eight-step-ahead predictions. In summary, the results differ from the ones discussed above. While we find again mixed evidence that sparsification improves point predictions, one striking difference between one- and multi-step-ahead forecasts is that more information seems to be beneficial for accurately forecasting multiple periods ahead.  Specifically, when interest centers on predicting one-year or two-year-ahead, using more information often improves predictive performance. For the one-year-ahead horizon, our findings suggest that a non-sparsified L-VAR with a tight Minnesota prior ($\theta_1 = 0.025$) yields the most accurate point forecasts. When two-year-ahead forecasts are taken under consideration, the sparse medium-scale model (with $\lambda = 0.5)$ dominates all competitors.

For output we observe small gains for the larger models. For one-year-ahead predictions,  the single best performing model is the L-VAR with $\theta_1 = 0.075$ and $\lambda = 0.01$. But with close to two percent in RMSE terms, these gains are muted and not statistically significant. In general we find that the benchmark is difficult to beat and no model is capable of significantly outperforming it. This does not carry over to two-year-ahead forecasts. Here we find at least some models significantly improving (at the ten percent level) upon the benchmark by around $3.4$ to $3.2$ percent. The best performing model is the L-VAR with $\theta_1 = 0.025$ and $\lambda = 1$, closely followed by two other sparsified L-VARs (with $\theta_1 = 0.025$ and $\lambda = 0.5$ and with $\theta_1 = 0.075$ and $\lambda = 0.01$). For output, it is worth noting that we find larger gains from sparsification when two-year-ahead forecasts are being considered. In most cases, the sparsified models improve upon their dense counterparts.

Analyzing one-year-ahead inflation forecasts, again, provides limited evidence that SAVS improves forecast accuracy. While there are many cases where a given model can be slightly improved by adding the sparsification step, these differences appear to be insignificant and small. When we consider two-year-ahead inflation predictions, we find slightly more predictive evidence for sparsity (with the medium-scale VAR with $\lambda = 0.5$ performing best). For this forecast horizon, we generally observe that sparsification often yields accuracy gains which become more pronounced in larger models. 

For the interest rate, the forecast gains of sparsification slightly diminish for higher-order forecasts. Generally speaking, there are little differences between all specifications, simply due to the fact that the forecasts of every stationary autoregressive model, irrespective of its size and degree of shrinkage, converge to the unconditional mean of the endogenous variables.

Overall, there is only mixed  evidence that point forecasts can be improved by adding the SAVS step. There are cases where gains from SAVS become more pronounced (such as longer-run forecasts for output and inflation or short-term interest rate forecasts) but there also exist several instances where using a sparse model slightly hurts forecast accuracy. One important thing to note is that sparsification only very rarely harms predictive accuracy in a statistically significant manner. 

To further substantiate this observation we consider the model confidence set (MCS) procedure of \cite{hansen2011model} and implemented by \cite{bernardi2018model} to obtain a measure for model uncertainty.  \autoref{tab:MCS_RMSE} shows the cardinality of the superior model set across variables and forecast horizons. Moreover, the table shows the number of sparsified models included in the MCS.  Considering a total number of $33$ models (including the benchmark) and using a mean squared error loss function, the MCS procedure suggests a quite large superior model set, including around $26$ to $33$ models. These numbers strongly depend on the variable and forecast horizon considered. In general, only few specifications are eliminated by the MCS procedure with no single class of models significantly outperforming its alternatives. It is worth emphasizing that in  most instances, the share of models using SAVS is high.  

Notice, however, that focusing exclusively on point forecasts ignores what can be considered the main advantage of SAVS: the corresponding reduction in estimation uncertainty and the potentially positive effect on the full predictive distribution. This theme will be the subject of the next section.

\begin{table}[!tbp]
{\tiny
\begin{center}
\scalebox{0.8}{
\begin{tabular}{llcrrrcrrrcrrrcrrr}
\toprule
\multicolumn{1}{l}{\bfseries }&\multicolumn{1}{c}{\bfseries }&\multicolumn{1}{c}{\bfseries }&\multicolumn{3}{c}{\bfseries Avg.}&\multicolumn{1}{c}{\bfseries }&\multicolumn{3}{c}{\bfseries Marginal One-quarter-ahead}&\multicolumn{1}{c}{\bfseries }&\multicolumn{3}{c}{\bfseries Marginal One-year-ahead}&\multicolumn{1}{c}{\bfseries }&\multicolumn{3}{c}{\bfseries Marginal Two-year-ahead}\tabularnewline
\cline{4-6} \cline{8-10} \cline{12-14} \cline{16-18}
\multicolumn{1}{l}{}&\multicolumn{1}{c}{}&\multicolumn{1}{c}{}&\multicolumn{1}{c}{One-quarter-ahead}&\multicolumn{1}{c}{One-year-ahead}&\multicolumn{1}{c}{Two-year-ahead}&\multicolumn{1}{c}{}&\multicolumn{1}{c}{GDPC1}&\multicolumn{1}{c}{CPIAUCSL}&\multicolumn{1}{c}{FEDFUNDS}&\multicolumn{1}{c}{}&\multicolumn{1}{c}{GDPC1}&\multicolumn{1}{c}{ CPIAUCSL}&\multicolumn{1}{c}{FEDFUNDS}&\multicolumn{1}{c}{}&\multicolumn{1}{c}{GDPC1}&\multicolumn{1}{c}{ CPIAUCSL}&\multicolumn{1}{c}{FEDFUNDS}\tabularnewline
\midrule
{\scshape }&&&&&&&&&&&&&&&&&\tabularnewline
~~&MIN&&$ 6$&$ 6$&$ 6$&&$ 5$&$ 6$&$ 5$&&$ 6$&$ 6$&$ 6$&&$ 5$&$ 6$&$ 6$\tabularnewline
~~&MIN - $\lambda = 0.01$&&$ 6$&$ 6$&$ 6$&&$ 4$&$ 6$&$ 5$&&$ 6$&$ 6$&$ 5$&&$ 6$&$ 6$&$ 6$\tabularnewline
~~&MIN - $\lambda = 0.1$&&$ 5$&$ 6$&$ 6$&&$ 6$&$ 5$&$ 4$&&$ 6$&$ 6$&$ 6$&&$ 6$&$ 6$&$ 5$\tabularnewline
~~&MIN - $\lambda = 0.5$&&$ 6$&$ 6$&$ 4$&&$ 5$&$ 6$&$ 6$&&$ 6$&$ 6$&$ 6$&&$ 6$&$ 4$&$ 5$\tabularnewline
~~&MIN - $\lambda = 1$&&$ 6$&$ 6$&$ 5$&&$ 6$&$ 6$&$ 6$&&$ 6$&$ 6$&$ 6$&&$ 6$&$ 4$&$ 5$\tabularnewline
~~&SSVS&&$ 3$&$ 1$&$ 3$&&$ 3$&$ 3$&$ 0$&&$ 1$&$ 3$&$ 3$&&$ 3$&$ 3$&$ 3$\tabularnewline
\midrule
{\scshape }&&&&&&&&&&&&&&&&&\tabularnewline
~~&Total&&$32$&$31$&$30$&&$29$&$32$&$26$&&$31$&$33$&$32$&&$32$&$29$&$30$\tabularnewline
\bottomrule
\end{tabular}}
\caption{Number of models included in the superior models sets (SMSs) for point forecasts according to different shrink-and-sparsify specifications. \label{tab:MCS_RMSE}
}
\begin{minipage}{\linewidth}
\footnotesize \textbf{Notes}: The SMSs are obtained with the model confidence set (MCS) procedure \citep{hansen2011model} at a $25$ percent significance level across variables and forecast horizons.
The loss function is specified in terms of mean squared errors. For each variable-horizon combination we use $33$ different specifications (including the benchmark). Note that the SSVS prior is only considered for the three smallest information sets. 
\end{minipage}
\end{center}}
\end{table}

\begin{landscape}\begin{table}[!tbp]
{\tiny
\begin{center}
\scalebox{0.9}{
\begin{tabular}{llclllclllclllclll}
\toprule
\multicolumn{1}{l}{\bfseries }&\multicolumn{1}{c}{\bfseries Spec.}&\multicolumn{1}{c}{\bfseries }&\multicolumn{3}{c}{\bfseries Avg.}&\multicolumn{1}{c}{\bfseries }&\multicolumn{3}{c}{\bfseries Marginal One-quarter-ahead}&\multicolumn{1}{c}{\bfseries }&\multicolumn{3}{c}{\bfseries Marginal One-year-ahead}&\multicolumn{1}{c}{\bfseries }&\multicolumn{3}{c}{\bfseries Marginal Two-year-ahead}\tabularnewline
\cline{2-2} \cline{4-6} \cline{8-10} \cline{12-14} \cline{16-18}
\multicolumn{1}{l}{}&\multicolumn{1}{c}{}&\multicolumn{1}{c}{}&\multicolumn{1}{c}{One-quarter-ahead}&\multicolumn{1}{c}{One-year-ahead}&\multicolumn{1}{c}{Two-year-ahead}&\multicolumn{1}{c}{}&\multicolumn{1}{c}{GDPC1}&\multicolumn{1}{c}{CPIAUCSL}&\multicolumn{1}{c}{FEDFUNDS}&\multicolumn{1}{c}{}&\multicolumn{1}{c}{GDPC1}&\multicolumn{1}{c}{ CPIAUCSL}&\multicolumn{1}{c}{FEDFUNDS}&\multicolumn{1}{c}{}&\multicolumn{1}{c}{GDPC1}&\multicolumn{1}{c}{ CPIAUCSL}&\multicolumn{1}{c}{FEDFUNDS}\tabularnewline
\midrule
{\scshape S}&&&&&&&&&&&&&&&&&\tabularnewline
   ~~&   MIN - $\lambda = 0.01$&   &   0.980 (3)&   1.003&   1.000&   &   1.008&   \textbf{0.975 (1)}&   0.951&   &   1.008&   0.996&   1.049&   &   1.010&   0.994&   1.024\tabularnewline
   ~~&   MIN - $\lambda = 0.1$&   &   0.977 (2)&   0.999&   0.994&   &   1.002&   0.992 &   0.833&   &   1.021&   0.995&   0.973&   &   0.979&   0.994***&   1.039\tabularnewline
   ~~&   MIN - $\lambda = 0.5$&   &   \textbf{0.976 (1)}&   1.000&   0.997&   &   1.005&   1.000&   0.756** (3)&   &   1.002&   1.000&   0.995&   &   1.015&   0.984&   1.077\tabularnewline
   ~~&   MIN - $\lambda = 1$&   &   0.981&   0.996&   0.982 &   &   1.004&   1.011&   0.736*** (2)&   &   0.996&   0.998&   0.983&   &   0.984&   0.977&   1.033\tabularnewline
\shadeRow   ~~&   SSVS&   &   0.991&   1.030&   0.987&   &   1.018&   0.989 (2)&   0.948&   &   1.059&   1.021&   1.029&   &   1.017&   0.975&   1.017\tabularnewline
\midrule
{\scshape FA}&&&&&&&&&&&&&&&&&\tabularnewline
\shadeRow   ~~&   MIN&   &   1.041&   1.031**&   1.006&   &   0.937&   1.022*&   1.317&   &   1.083***&   1.020&   0.977&   &   1.083&   0.987&   \textbf{0.943 (1)}\tabularnewline
   ~~&   MIN - $\lambda = 0.01$&   &   1.024&   1.013&   1.008&   &   0.905&   1.015&   1.272&   &   1.043&   1.010&   0.959&   &   1.020&   1.000&   1.053\tabularnewline
   ~~&   MIN - $\lambda = 0.1$&   &   1.025&   1.019&   0.990&   &   0.903&   1.037&   1.185&   &   1.081**&   0.996&   1.042&   &   0.999&   0.983&   1.034\tabularnewline
   ~~&   MIN - $\lambda = 0.5$&   &   1.047&   1.027&   1.001&   &   0.909&   1.066*&   1.201&   &   1.110**&   0.993*&   1.095&   &   1.001&   0.995&   1.066\tabularnewline
   ~~&   MIN - $\lambda = 1$&   &   1.064&   1.014&   0.995&   &   0.903&   1.092*&   1.203&   &   1.055&   1.000&   1.017&   &   0.980&   0.992**&   1.072\tabularnewline
\shadeRow   ~~&   SSVS&   &   1.164***&   1.181***&   1.154***&   &   0.857&   1.213***&   1.411***&   &   1.116***&   1.198***&   1.204**&   &   1.074*&   1.161***&   1.317***\tabularnewline
\midrule
{\scshape M}&&&&&&&&&&&&&&&&&\tabularnewline
\shadeRow   ~~&   MIN&   &   1.067&   0.985&   1.006&   &   0.844 (3)&   1.112**&   1.216&   &   1.011&   0.980 (2)&   0.960*&   &   1.039&   0.996&   1.010\tabularnewline
   ~~&   MIN - $\lambda = 0.01$&   &   1.041&   0.983 (2)&   0.997&   &   \textbf{0.815** (1)}&   1.114**&   1.040&   &   0.996&   0.982 (3)&   0.953&   &   1.007&   0.993&   1.011\tabularnewline
   ~~&   MIN - $\lambda = 0.1$&   &   1.011&   1.008&   0.985&   &   0.837*** (2)&   1.098**&   0.817***&   &   1.020&   1.004&   1.012&   &   1.010&   0.972 (2)&   1.051\tabularnewline
   ~~&   MIN - $\lambda = 0.5$&   &   1.002&   0.999&   \textbf{0.972 (1)}&   &   0.939&   1.065&   \textbf{0.725*** (1)}&   &   1.011&   0.995&   0.994&   &   0.978&   \textbf{0.964 (1)}&   1.027\tabularnewline
   ~~&   MIN - $\lambda = 1$&   &   1.025&   0.993&   0.988&   &   1.013&   1.070&   0.759**&   &   1.004&   0.992&   0.975&   &   0.995&   0.981&   1.037\tabularnewline
\shadeRow   ~~&   SSVS&   &   1.189***&   1.223***&   1.215***&   &   0.938&   1.240***&   1.353***&   &   1.147*&   1.216***&   1.472***&   &   1.129&   1.212***&   1.490***\tabularnewline
\midrule
{\scshape L ($\theta_1 = 0.025$)}&&&&&&&&&&&&&&&&&\tabularnewline
\shadeRow   ~~&   MIN&   &   1.000&   \textbf{0.978 (1)}&   0.995&   &   0.854**&   1.068&   0.876&   &   0.994 (3)&   \textbf{0.975 (1)}&   0.956&   &   1.029&   0.983&   1.010 (3)\tabularnewline
   ~~&   MIN - $\lambda = 0.01$&   &   1.041&   0.994&   0.983&   &   0.996&   1.096*&   0.782**&   &   1.020&   0.989&   0.963&   &   0.990&   0.978&   1.023\tabularnewline
   ~~&   MIN - $\lambda = 0.1$&   &   1.055&   1.003&   0.991&   &   1.005&   1.105*&   0.848**&   &   1.020&   1.002&   0.966&   &   0.988&   0.986&   1.061\tabularnewline
   ~~&   MIN - $\lambda = 0.5$&   &   1.045&   0.990&   0.983&   &   1.021&   1.087&   0.833*&   &   0.992 (2)&   0.992&   0.967&   &   0.969 (2)&   0.982&   1.035\tabularnewline
   ~~&   MIN - $\lambda = 1$&   &   1.046&   0.999&   0.980 &   &   1.025&   1.085*&   0.851*&   &   1.010&   1.001&   0.947 (2)&   &   \textbf{0.966* (1)}&   0.980&   1.026\tabularnewline
\midrule
{\scshape L ($\theta_1 = 0.050$)}&&&&&&&&&&&&&&&&&\tabularnewline
\shadeRow   ~~&   MIN&   &   1.023&   1.000&   0.984 (3)&   &   0.881&   1.055&   1.110&   &   1.010&   1.000&   0.970&   &   1.011&   0.976&   0.974 (2)\tabularnewline
   ~~&   MIN - $\lambda = 0.01$&   &   1.020&   0.994&   0.983&   &   0.891*&   1.091&   0.848**&   &   0.995&   0.995&   0.975&   &   1.007&   0.972 (3)&   1.023\tabularnewline
   ~~&   MIN - $\lambda = 0.1$&   &   1.039&   0.995&   0.984&   &   0.986&   1.092*&   0.823**&   &   1.023&   0.987&   0.992&   &   0.982&   0.982&   1.011\tabularnewline
   ~~&   MIN - $\lambda = 0.5$&   &   1.052&   1.003&   0.983* (2)&   &   1.032&   1.091&   0.854*&   &   1.006&   1.003&   0.987&   &   0.981*&   0.981&   1.013\tabularnewline
   ~~&   MIN - $\lambda = 1$&   &   1.043&   0.993&   0.986&   &   1.020&   1.084*&   0.837*&   &   0.998&   0.993&   0.978&   &   0.990&   0.980&   1.035\tabularnewline
\midrule
{\scshape L ($\theta_1 = 0.075$)}&&&&&&&&&&&&&&&&&\tabularnewline
\shadeRow   ~~&   MIN&   &   1.037&   0.997&   1.011**&   &   0.921&   1.042&   1.222&   &   1.004&   0.997&   0.983&   &   1.059*&   0.995&   1.023\tabularnewline
   ~~&   MIN - $\lambda = 0.01$&   &   1.039&   0.984 (3)&   0.984&   &   0.885&   1.104*&   0.963&   &   \textbf{0.982 (1)}&   0.988&   0.961&   &   0.973* (3)&   0.983&   1.036\tabularnewline
   ~~&   MIN - $\lambda = 0.1$&   &   1.029&   0.993&   0.985&   &   0.981&   1.084*&   0.780**&   &   1.003&   0.996&   \textbf{0.943 (1)}&   &   0.984&   0.979&   1.042\tabularnewline
   ~~&   MIN - $\lambda = 0.5$&   &   1.050&   0.994&   0.984&   &   1.056&   1.087&   0.799**&   &   1.004&   0.996&   0.949* (3)&   &   0.993&   0.980&   0.996 \tabularnewline
   ~~&   MIN - $\lambda = 1$&   &   1.050&   0.993&   0.990&   &   1.036&   1.094*&   0.806**&   &   1.016&   0.989&   0.957&   &   0.990&   0.988&   1.016\tabularnewline
\bottomrule
\end{tabular}}
\caption{RMSE ratios relative to the \textit{small-scale} BVAR with a Minnesota prior. \label{tab:RMSE}}
\begin{minipage}{\linewidth}
\footnotesize \textbf{Notes}: Bold numbers indicate lowest RMSE ratios for each horizon and target variable (and therefore best performing models over the full hold-out sample), while the numbers in parenthesis refer to the ranking of the three best specifications according to the MCS \cite{hansen2011model} procedure (including the benchmark). Gray shaded rows denote non-sparsified models. Asterisks indicate statistical significance for each model relative to the benchmark at the $1$ ($^{***}$), $5$ ($^{**}$) and $10$ ($^{*}$) percent significance levels. 
\end{minipage}\end{center}}
\end{table}\end{landscape}

\subsection{Density Forecasting Performance}
When focusing on point forecasting accuracy we necessarily disregard any information on how well the different models capture higher order moments of the predictive distribution. However, the discussion in \autoref{sec:DSS} suggests that applying SAVS might improve density forecasts by zeroing out irrelevant predictors throughout MCMC sampling, eventually exerting a positive effect on the full predictive distribution by reducing the predictive variance. 

Table~\ref{tab:LPS} shows differences in LPLs relative to the small-scale Minnesota VAR. Apart from focusing on variable-specific relative LPLs, the table also shows joint LPLs over the three target variables. These serve as a general measure  on how well some approach performs in forecasting output, inflation and interest rates jointly. Numbers greater than zero imply that a given model improves upon the benchmark while negative values suggest that the benchmark yields more precise density predictions. Moreover, we compute a two-sided \cite{amisano2007comparing} test for each specification with a null hypothesis of equal average LPLs relative to the benchmark and report the statistical significance based on the obtained $p$-values.\footnote{Note that the test statistics must be interpreted carefully since we use a recursive forecast design.}
\subsection*{One-quarter-ahead Density Forecasts}
We start with the joint predictive LPLs and the one-step-ahead horizon shown in Table~\ref{tab:LPS}. It is worth noting that these numbers can be interpreted as a training sample marginal likelihood \citep{geweke2010comparing}. For this measure, we find that the best performing model is the moderately-sized sparse VAR with $\lambda=0.1$, suggesting that using SAVS appreciably improves density predictions.

The large VARs are generally outperformed by smaller-sized ones according to one-step-ahead joint LPLs. For these models, we also find that sparsification hurts density performance, a finding that is absent when we consider the other models. For instance, in the case of the small and medium-scale VARs, we find statistical significant accuracy gains from using the additional SAVS step. This pattern can also be found in the case of the FA-VAR. 

 At the one-step-ahead horizon, varying $\lambda$ between $0.01$ and $0.1$ yields a similar forecasting performance for most models considered. If $\lambda$ is set too large, forecasting accuracy drops markedly. This can be traced back to the fact that a larger penalty leads to an overly sparse model and this, in turn, decreases the predictive variance too much. The resulting predictive distribution is too narrow and this makes capturing outliers increasingly difficult.  

To dig into the sources on why some models work well while others perform poorly when all three variables are jointly considered, we now consider marginal LPLs.  Considering the density forecasting performance for output, we observe that the model which does well for the joint predictive BF also excels (i.e. the medium-size VAR with $\lambda =0.01$). However, considering the large VARs for output alone reveals that they are also highly competitive and close to the single best performing specification (as long as $\lambda$ is set not too large) and that sparse models often outperform the non-sparse competitor. 

Turning to inflation forecasts highlights that they are the main driver of the bad performance of most large-scale models. Irrespective of the choice of $\lambda$ and $\theta_1$, the small-scale VAR outperforms each specification and it seems that using SAVS only reduces predictive performance for inflation in large datasets. This is result is well-known in central bank practice, where these specifications are commonly outperformed by small and medium-sized models that include only a selected set of endogenous variables \citep[e.g.][]{giannone2015prior}.

For interest rates, the story found for inflation is reversed. We find that the best models by large margins feature large information sets and are sparse. The main reason behind this strong performance is similar to the case raised in the case of point forecasts. A setup that sets most coefficients to zero yields a predictive distribution for the interest rate which is strongly centered on zero. During the period of the zero lower bound, the corresponding predictive distribution will feature a mean/median close to zero with a rather small variance and this yields large predictive gains in terms of LPLs.

After discussing all three marginal LPLs we find that large models yield competitive output and very strong interest rate forecasts while they poorly predict inflation. Hence, if little weight is placed on inflation predictions the large VARs appear to be competitive.

\subsection*{Multi-steps-ahead Density Forecasts}
Inspecting multi-steps-ahead density forecasting performance yields similar insights to the one-step-ahead case. Considering relative joint LPLs shows that sparse medium-scale models perform best. The accuracy gains from using SAVS are especially pronounced for this model size. Interestingly, and in contrast to the one-step-ahead case, we find that larger values of $\lambda$ (i.e. $\lambda = 1$ for one-year-ahead and $\lambda = 0.5$ for two-year-ahead forecasts) translate into the largest gains. Again, we find that large VARs are generally beaten by medium-sized models but, as opposed to one-quarter-ahead forecasts, we find that small values of $\lambda$ improve upon the corresponding non-sparse counterpart.

Zooming into variable-specific performance highlights that large and sparse models yield precise density forecasts of GDP that are always better than the non-sparse variant of the model under scrutiny. These gains are often substantial and sometimes significant at the ten percent level. For multi-step-ahead GDP forecasts we even find that predictive performance increases in lockstep with model size. This pattern is quite consistent with one exception. The FA-VAR shows the weakest performance among all models used. This indicates that when the researcher wishes to produce multi-step-ahead forecasts, additional information that might be ignored by introducing a factor structure seems to be important.

In terms of inflation we again see that large models do not perform well, yielding the most inaccurate forecasts across models. For multi-step-ahead inflation predictions we moreover find little evidence that applying SAVS helps to improve forecasts since in most cases, the non-sparse model performs better than the sparse variant. 

Short-term interest rates are again most precisely predicted using large models with sparsification. In both cases (one and two-year-ahead) we find the same ranking according to the MCS procedure, namely that the large VAR with $\theta_1 = 0.075$ and $\lambda = 1$ performs best. Lower values of $\lambda$ yield highly competitive interest rate forecasts with little differences across the different values of $\lambda$.

 Similarly to the discussion of the point forecasts we investigate these statements using MCS. \autoref{tab:MCS_LPS} shows the number of models included in the MCS when we use the negative LPL as a loss function. Starting with an initial set of $33$ models we find that, depending on the variable and forecast horizon, a large number of competitors is eliminated.   The majority of surviving models is comprised of sparse specifications. While the reduction of models in the case of joint LPLs is moderate to large, we find much smaller MCSs when we focus on variable-specific forecasting performance. Especially for output and interest rates, we find that the MCS consists of a handful models. In fact, for two-year-ahead predictions, all except for a single model are removed to form the MCS. In both cases, this specification is the large VAR with $\lambda = 1$ and $\theta_1 = 0.075$. For the remaining cases, the top three of these models can be read of from \autoref{tab:LPS} and always includes several sparsified variants.  For inflation, the MCS corroborate the findings discussed above. The corresponding set of superior models remains elevated and only a relatively small number of models is consequently eliminated.
 
This discussion shows that using SAVS typically improves GDP and interest rate forecasts in large models while it is only of limited help when predicting inflation. One caveat of considering LPLs is that they are computed over the full hold-out period and thus provide a measure of average forecast quality. But it could be the case that some models perform well during selected parts of the hold-out whereas other models excel in other periods. To investigate whether this is the case we now consider density forecasting performance over time and across two distinct periods: the period before the global financial crisis (up to 2008:Q1) and the period afterwards.


\begin{table}[!tbp]
{\tiny
\begin{center}
\scalebox{0.8}{
\begin{tabular}{llcrrrcrrrcrrrcrrr}
\toprule
\multicolumn{1}{l}{\bfseries }&\multicolumn{1}{c}{\bfseries }&\multicolumn{1}{c}{\bfseries }&\multicolumn{3}{c}{\bfseries Joint}&\multicolumn{1}{c}{\bfseries }&\multicolumn{3}{c}{\bfseries Marginal One-quarter-ahead}&\multicolumn{1}{c}{\bfseries }&\multicolumn{3}{c}{\bfseries Marginal One-year-ahead}&\multicolumn{1}{c}{\bfseries }&\multicolumn{3}{c}{\bfseries Marginal Two-year-ahead}\tabularnewline
\cline{4-6} \cline{8-10} \cline{12-14} \cline{16-18}
\multicolumn{1}{l}{}&\multicolumn{1}{c}{}&\multicolumn{1}{c}{}&\multicolumn{1}{c}{One-quarter-ahead}&\multicolumn{1}{c}{One-year-ahead}&\multicolumn{1}{c}{Two-year-ahead}&\multicolumn{1}{c}{}&\multicolumn{1}{c}{GDPC1}&\multicolumn{1}{c}{CPIAUCSL}&\multicolumn{1}{c}{FEDFUNDS}&\multicolumn{1}{c}{}&\multicolumn{1}{c}{GDPC1}&\multicolumn{1}{c}{ CPIAUCSL}&\multicolumn{1}{c}{FEDFUNDS}&\multicolumn{1}{c}{}&\multicolumn{1}{c}{GDPC1}&\multicolumn{1}{c}{ CPIAUCSL}&\multicolumn{1}{c}{FEDFUNDS}\tabularnewline
\midrule
{\scshape }&&&&&&&&&&&&&&&&&\tabularnewline
~~&MIN&&$ 4$&$ 1$&$ 5$&&$ 0$&$ 6$&$0$&&$ 4$&$ 6$&$0$&&$ 2$&$ 6$&$0$\tabularnewline
~~&MIN - $\lambda = 0.01$&&$ 4$&$ 2$&$ 6$&&$ 3$&$ 5$&$0$&&$ 4$&$ 6$&$0$&&$ 4$&$ 6$&$0$\tabularnewline
~~&MIN - $\lambda = 0.1$&&$ 5$&$ 3$&$ 5$&&$ 3$&$ 6$&$1$&&$ 3$&$ 5$&$0$&&$ 2$&$ 6$&$0$\tabularnewline
~~&MIN - $\lambda = 0.5$&&$ 5$&$ 2$&$ 4$&&$ 3$&$ 6$&$1$&&$ 3$&$ 6$&$0$&&$ 5$&$ 6$&$0$\tabularnewline
~~&MIN - $\lambda = 1$&&$ 5$&$ 3$&$ 5$&&$ 3$&$ 6$&$1$&&$ 5$&$ 6$&$1$&&$ 6$&$ 6$&$1$\tabularnewline
~~&SSVS&&$ 0$&$ 0$&$ 2$&&$ 0$&$ 3$&$0$&&$ 0$&$ 2$&$0$&&$ 0$&$ 1$&$0$\tabularnewline
\midrule
{\scshape }&&&&&&&&&&&&&&&&&\tabularnewline
~~&Total&&$23$&$11$&$27$&&$12$&$32$&$3$&&$19$&$31$&$1$&&$19$&$31$&$1$\tabularnewline
\bottomrule
\end{tabular}}
\caption{Number of models included in the superior models sets (SMSs) for density forecasts according to different shrink-and-sparsify specifications. \label{tab:MCS_LPS}
}
\begin{minipage}{\linewidth}
\footnotesize \textbf{Notes}: The SMSs are obtained with the model confidence set (MCS) procedure \citep{hansen2011model} at a $25$ percent significance level across variables and forecast horizons.
The loss function is specified in terms of negative LPLs. For each variable-horizon combination we use $33$ different specifications (including the benchmark). Note that the SSVS prior is only considered for the three smallest information sets. 
\end{minipage}
\end{center}}
\end{table}

Before discussing forecasting performance over time we provide guidance on how to select the penalty parameter $\lambda$ (and thus $\varpi$). As opposed to the simulation exercise, we find that forecast accuracy as measured by LPLs differs across specific values of $\lambda$ and forecast horizons. In general, we can recommend setting $\lambda$ to a rather small value (i.e. to $0.01$ or $0.1$) if interest centers on one-step-ahead predictions. This holds for most model sizes. By contrast, if higher order forecasts are of interest, we can recommend setting $\lambda$ to a larger value (i.e. $0.5$ or $1$). With very few exceptions, these are the values that yield the highest LPLs across models.

\begin{figure}[htbp]
\begin{minipage}{0.49\textwidth}
\centering \textit{Before financial crisis}
\end{minipage}
\begin{minipage}{0.49\textwidth}
\centering \textit{After financial crisis}
\end{minipage}
\begin{minipage}{\textwidth}
\centering
(a) \textit{One-quarter-ahead}
\end{minipage}
\begin{minipage}{0.49\textwidth}
\centering
\includegraphics[scale = 0.5]{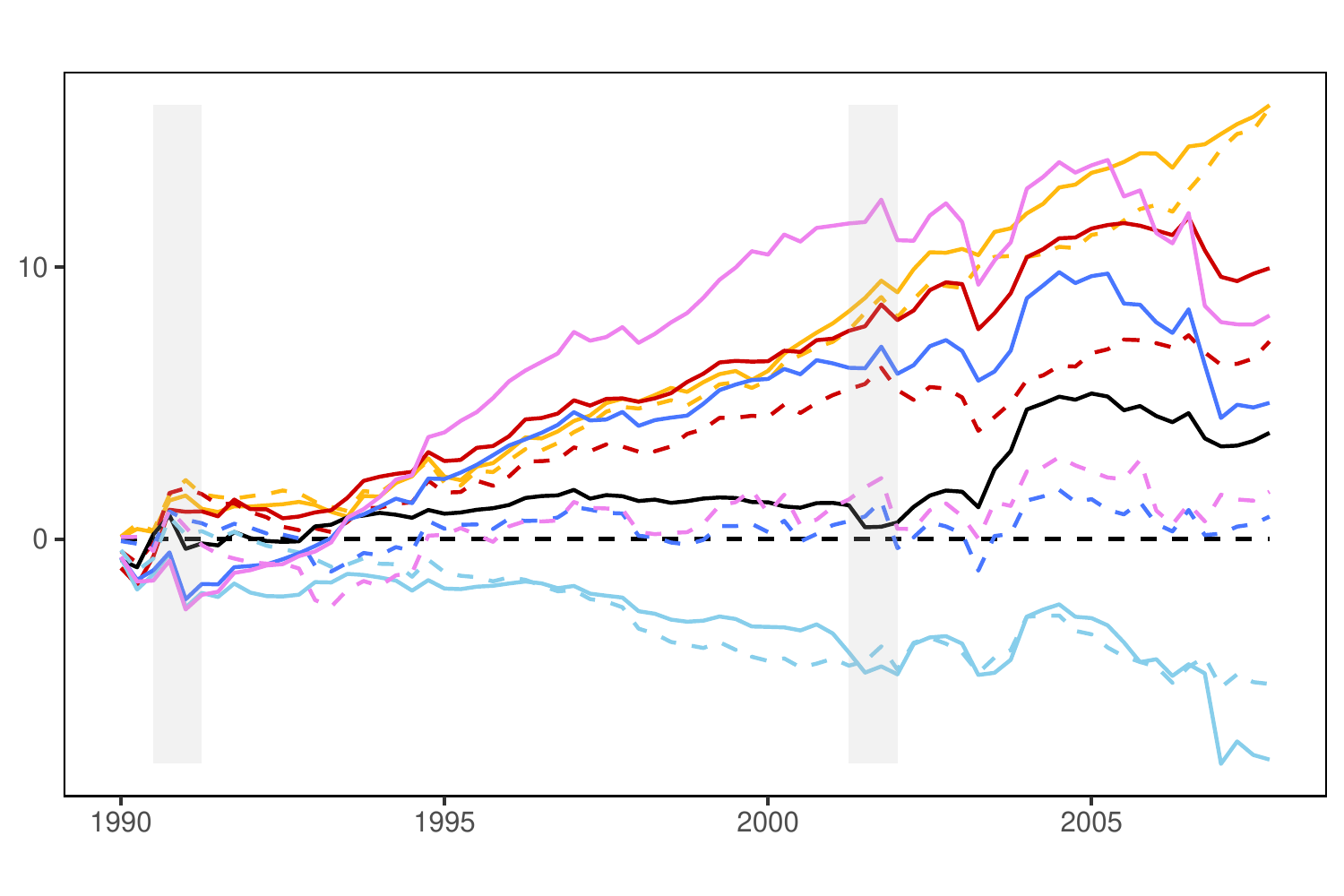}
\end{minipage}
\begin{minipage}{0.49\textwidth}
\centering
\includegraphics[scale = 0.5]{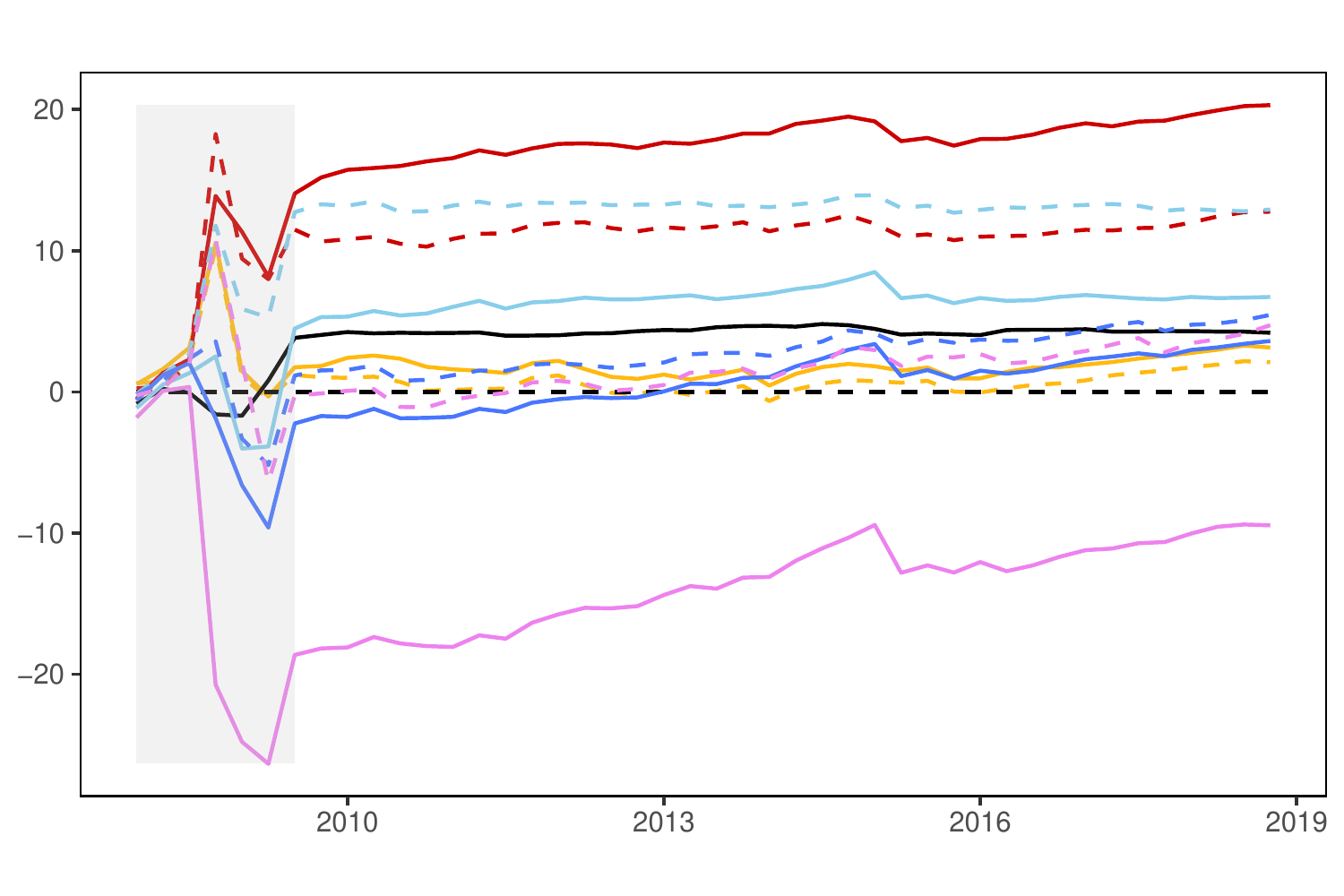}
\end{minipage}
\begin{minipage}{\textwidth}
\centering
(b) \textit{One-year-ahead}
\end{minipage}
\begin{minipage}{0.49\textwidth}
\centering
\includegraphics[scale = 0.5]{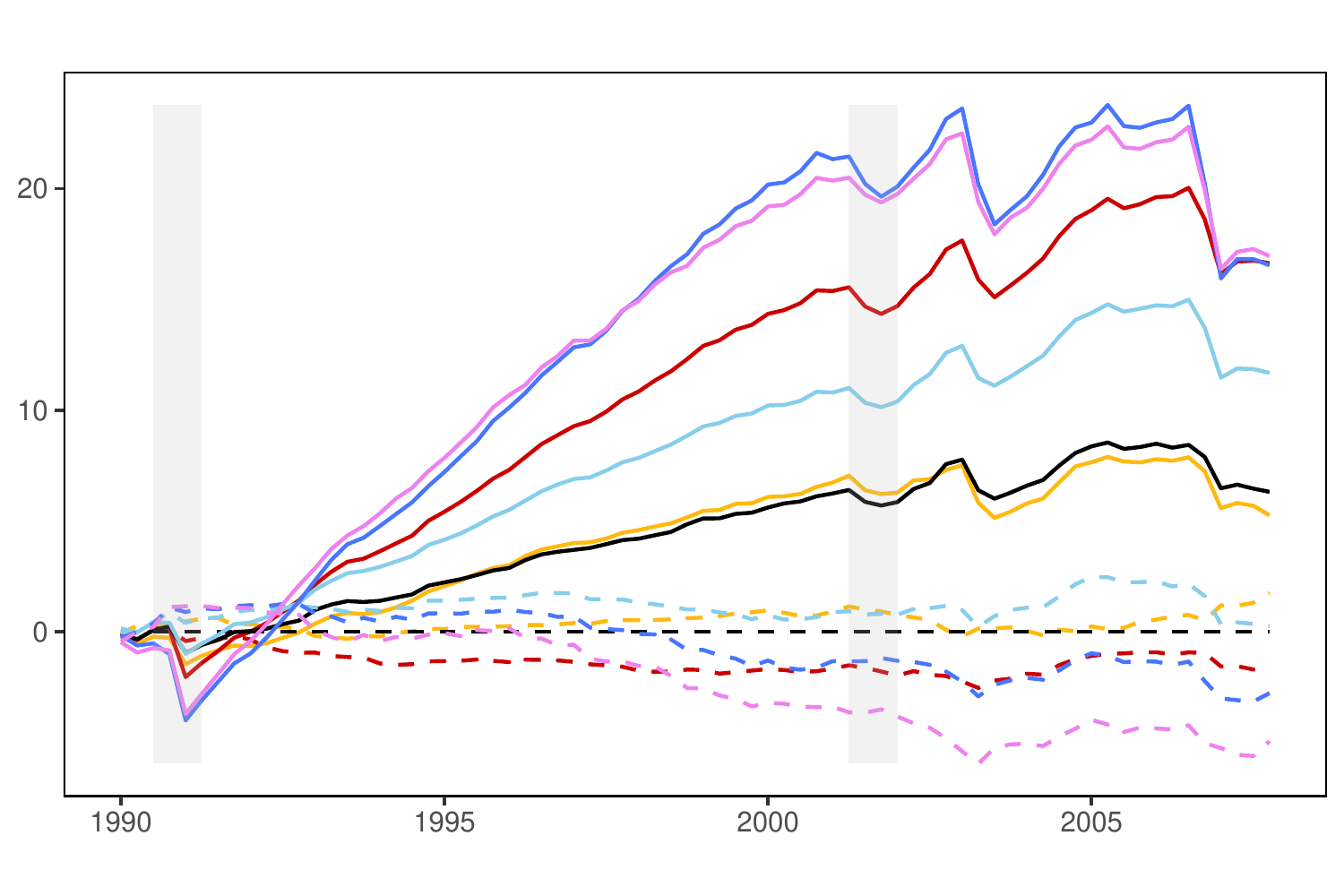}
\end{minipage}
\begin{minipage}{0.49\textwidth}
\centering
\includegraphics[scale = 0.5]{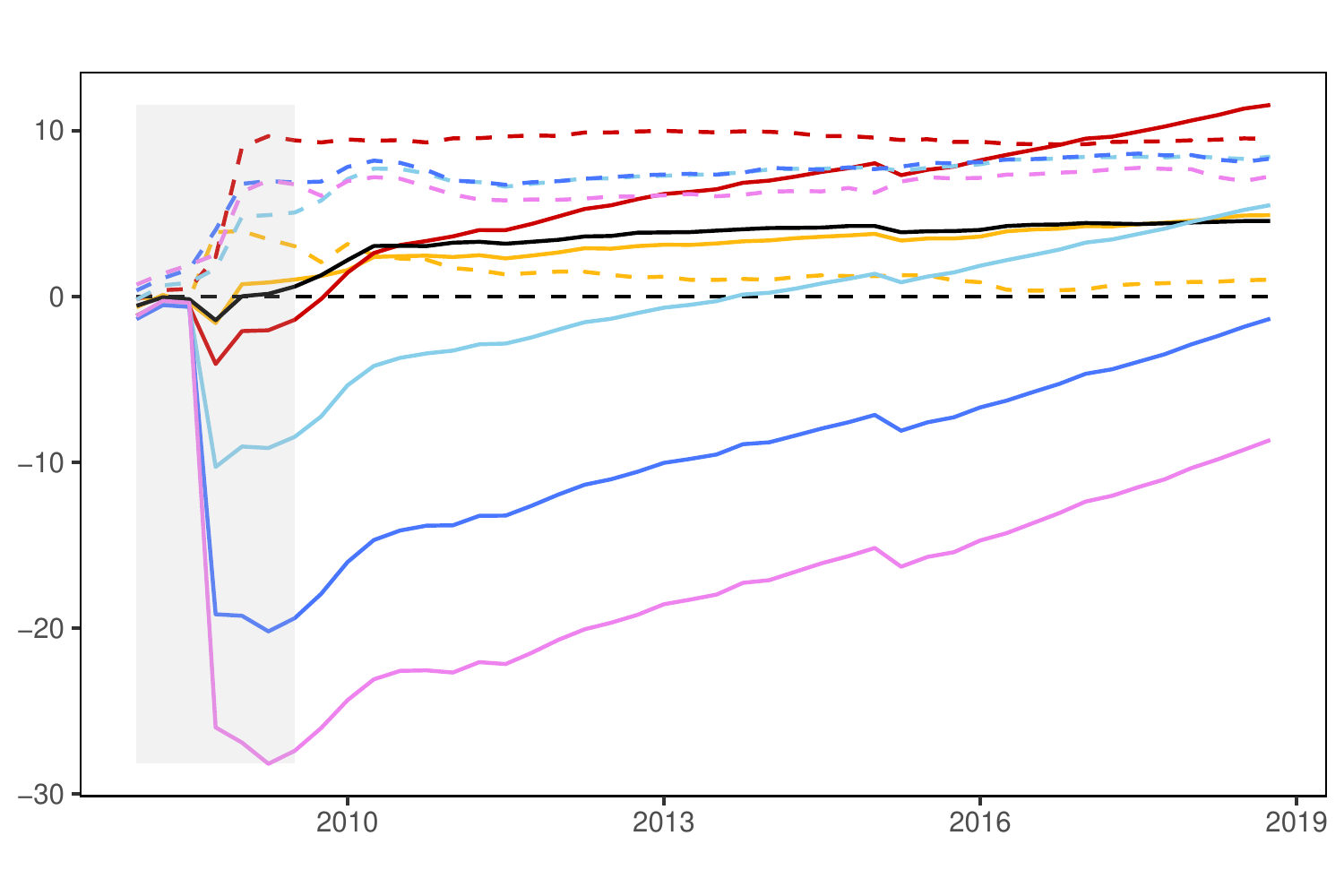}
\end{minipage}
\begin{minipage}{\textwidth}
\centering
(c) \textit{Two-year-ahead}
\end{minipage}
\begin{minipage}{0.49\textwidth}
\centering
\includegraphics[scale = 0.5]{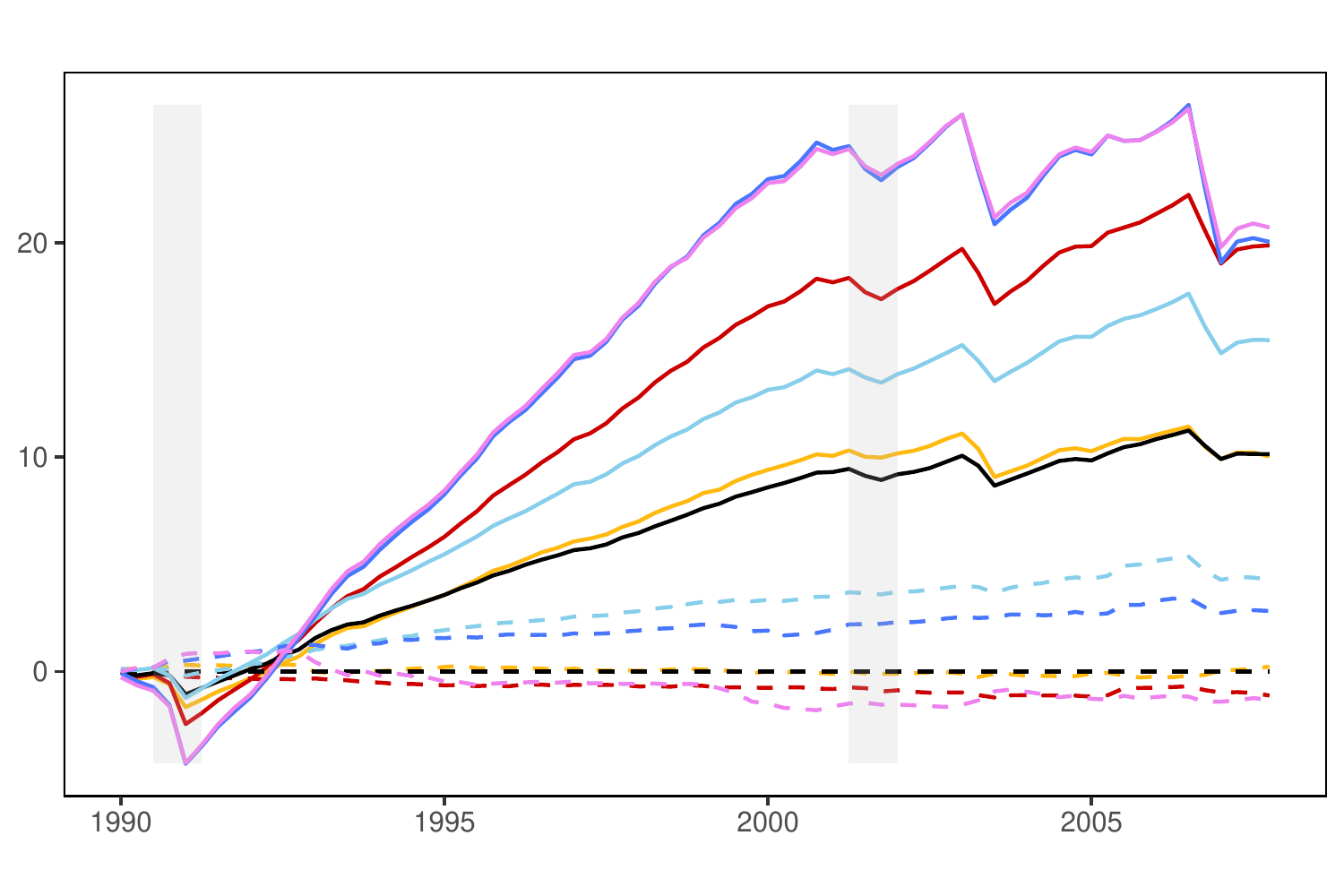}
\end{minipage}
\begin{minipage}{0.49\textwidth}
\centering
\includegraphics[scale = 0.5]{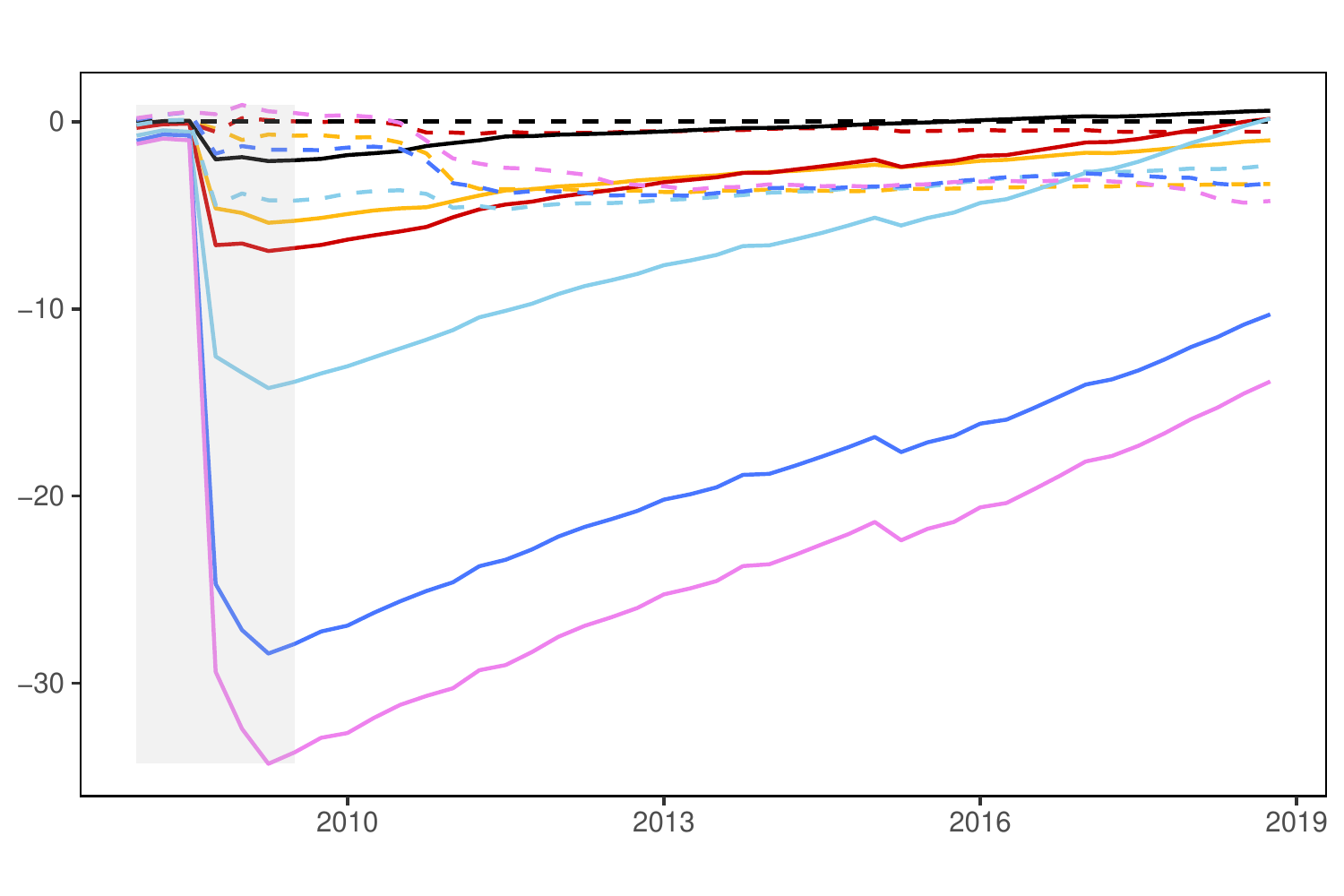}
\end{minipage}

\begin{minipage}{\textwidth}
\centering
\includegraphics[scale = 0.25]{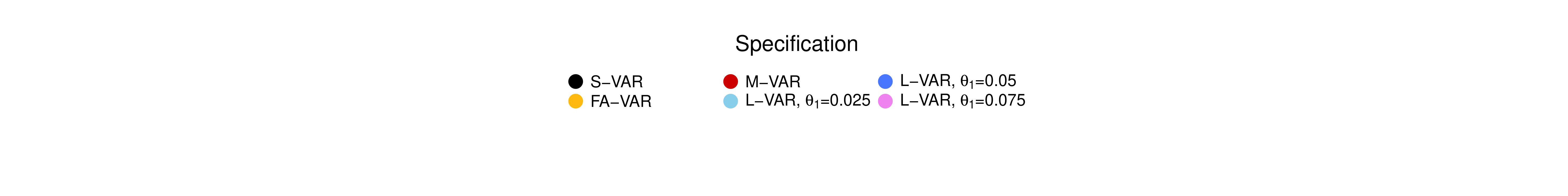}
\end{minipage}
\caption{Cumulative joint log predictive likelihoods for one-quarter-, one-year-, and two-year-ahead predictions benchmarked against the \textit{small-scale} BVAR without SAVS. Dashed lines indicate classic BVARs while solid lines depict the best performing sparsified version within each information set. Gray shaded areas denote NBER recessions.}
	  \label{fig:lpstot}
\end{figure}

\autoref{fig:lpstot} shows the evolution of cumulative joint LPLs relative to the benchmark model for one-quarter-, one-year-, and two-year-ahead predictions over time. The figure includes standard BVARs without SAVS (dashed lines) and BVARs post-processed with an additional SAVS step (solid lines) for all considered information sets. 
We focus on sparse models with the $\lambda$ that maximizes the LPLs at the end of the hold-out within each model class (see \autoref{tab:LPS}). Moreover, for the large-scale BVAR we depict the evolution of BFs for the different values of $\theta_1 \in \{0.025, 0.05, 0.075\}$.

Considering the period before the financial crisis clearly shows that using SAVS often yields more precise forecast distributions. This finding is especially pronounced for higher order forecasts, where we see strong and sustained gains over the period up to the financial crisis. The performance of the sparsified large models with $\theta_1 = \{0.075, 0.05\}$ and the medium-scale BVAR with SAVS appear to be the best specifications during that time span. 

When we focus attention on the global financial crisis we observe a pronounced decline in  model evidence for the large model with SAVS (right panels in \autoref{fig:lpstot}). During the crisis, evidence in favour of sparsification also slightly decreases for the medium-scale model. This can be seen by comparing the solid and dashed red lines in \autoref{fig:lpstot}. While the non-sparse, medium-sized BVAR is outperformed in the run-up to the crisis, model evidence marginally supports the non-sparse variant during the recession. The main reason why predictive accuracy of sparsified models deteriorates in turbulent periods is that the forecast error variance becomes too small and large shocks become increasingly unlikely under the predictive distribution. In such a situation, dense models often feature a larger variance because of many small but non-zero $v_{ij}'$s in \autoref{eq: varPost1}. This makes capturing outliers (or rapid shifts) easier and thus improves LPLs. 

After the financial crisis we see that applying SAVS helps for some models (especially the medium-scale VAR). However, predictive evidence in favor of SAVS declines with the forecast horizon. This pattern is the opposite of the one observed before the financial crisis. But note that this is mostly driven by the dismal performance during the recession. For one- and two-year-ahead forecasts we observe that the slopes of most LPL curves relative to sparse models are steep, indicating a period-by-period outperformance vis-\'{a}-vis to the benchmark model. Nevertheless, while we find that the medium-scale sparse VAR improves upon all benchmarks by the end of the hold-out, this does not carry over to two-year-ahead forecasts.

This discussion highlights that during tranquil periods, which we define to be characterized by small but frequent shocks, applying the SAVS step substantially improves predictive performance. This is even more pronounced if we consider higher order forecasts. In recessions,  by contrast, predictions from sparse models tend to be too conservative. And this is deleterious for density forecast performance.

\begin{landscape}\begin{table}[!tbp]
{\tiny
\begin{center}
\scalebox{0.9}{
\begin{tabular}{llclllclllclllclll}
\toprule
\multicolumn{1}{l}{\bfseries }&\multicolumn{1}{c}{\bfseries Spec.}&\multicolumn{1}{c}{\bfseries }&\multicolumn{3}{c}{\bfseries Joint}&\multicolumn{1}{c}{\bfseries }&\multicolumn{3}{c}{\bfseries Marginal One-quarter-ahead}&\multicolumn{1}{c}{\bfseries }&\multicolumn{3}{c}{\bfseries Marginal One-year-ahead}&\multicolumn{1}{c}{\bfseries }&\multicolumn{3}{c}{\bfseries Marginal Two-year-ahead}\tabularnewline
\cline{2-2} \cline{4-6} \cline{8-10} \cline{12-14} \cline{16-18}
\multicolumn{1}{l}{}&\multicolumn{1}{c}{}&\multicolumn{1}{c}{}&\multicolumn{1}{c}{One-quarter-ahead}&\multicolumn{1}{c}{One-year-ahead}&\multicolumn{1}{c}{Two-year-ahead}&\multicolumn{1}{c}{}&\multicolumn{1}{c}{GDPC1}&\multicolumn{1}{c}{CPIAUCSL}&\multicolumn{1}{c}{FEDFUNDS}&\multicolumn{1}{c}{}&\multicolumn{1}{c}{GDPC1}&\multicolumn{1}{c}{ CPIAUCSL}&\multicolumn{1}{c}{FEDFUNDS}&\multicolumn{1}{c}{}&\multicolumn{1}{c}{GDPC1}&\multicolumn{1}{c}{ CPIAUCSL}&\multicolumn{1}{c}{FEDFUNDS}\tabularnewline
\midrule
{\scshape S}&&&&&&&&&&&&&&&&&\tabularnewline
   ~~&   MIN - $\lambda = 0.01$&   &   5.745**&   4.828*&   4.513&   &   0.634&   0.897&   3.187***&   &   1.570&   0.011&   1.697*&   &   2.596***&   -1.680&   2.722***\tabularnewline
   ~~&   MIN - $\lambda = 0.1$&   &   7.534&   7.745*&   6.118&   &   0.212&   0.198&   6.525**&   &   2.945&   -0.850&   3.558*&   &   4.365***&   -3.968&   5.467***\tabularnewline
   ~~&   MIN - $\lambda = 0.5$&   &   6.817&   9.475&   5.437&   &   0.526&   -2.596&   9.088***&   &   4.457*&   -1.418&   6.169***&   &   5.880***&   -7.634&   8.116***\tabularnewline
   ~~&   MIN - $\lambda = 1$&   &   5.288&   10.737&   5.783&   &   1.515&   -4.985&   9.801***&   &   5.147*&   -1.903&   7.587***&   &   6.561***&   -8.251&   9.464***\tabularnewline
\shadeRow   ~~&   SSVS&   &   -3.705&   -0.176&   -4.098&   &   -3.035***&   4.484 (2)&   -2.205&   &   -4.165***&   \textbf{6.517 (1)}&   -2.751***&   &   -3.609***&   \textbf{4.654 (1)}&   -4.478***\tabularnewline
\midrule
{\scshape FA}&&&&&&&&&&&&&&&&&\tabularnewline
\shadeRow   ~~&   MIN&   &   17.948 (3)&   2.769&   -3.098&   &   11.408**&   4.423 (3)&   6.088&   &   -1.815&   0.329&   1.871***&   &   -3.025&   -0.454&   0.819\tabularnewline
   ~~&   MIN - $\lambda = 0.01$&   &   19.086&   5.844&   2.745&   &   11.634**&   \textbf{4.606 (1)}&   7.538&   &   -0.776&   1.619 (3)&   2.700***&   &   1.460&   -2.279&   2.561***\tabularnewline
   ~~&   MIN - $\lambda = 0.1$&   &   17.368&   5.394&   4.983&   &   11.687**&   1.624&   9.088&   &   -1.320&   0.361&   2.997*&   &   4.656***&   -5.399&   4.700***\tabularnewline
   ~~&   MIN - $\lambda = 0.5$&   &   10.393&   6.816&   7.252&   &   11.893**&   -5.325&   11.055*&   &   0.072&   -1.207&   3.496*&   &   6.809***&   -3.661&   6.073***\tabularnewline
   ~~&   MIN - $\lambda = 1$&   &   5.528&   10.154*&   8.667&   &   12.101**&   -7.035&   11.863**&   &   1.778&   -1.433&   4.315**&   &   7.417**&   -3.304&   6.539***\tabularnewline
\shadeRow   ~~&   SSVS&   &   -46.196&   -28.238&   -35.739&   &   9.724&   -51.723*&   -6.802&   &   -5.828**&   -30.101&   -6.738**&   &   -5.940***&   -31.360&   -10.958***\tabularnewline
\midrule
{\scshape M}&&&&&&&&&&&&&&&&&\tabularnewline
\shadeRow   ~~&   MIN&   &   20.032 &   7.907&   -1.659***&   &   17.376** &   -2.255&   6.753**&   &   0.828&   5.649 (2)&   -0.515&   &   -0.931*&   0.515 (3)&   -0.295\tabularnewline
   ~~&   MIN - $\lambda = 0.01$&   &   26.850 (2)&   12.099**&   8.382&   &   \textbf{20.002*** (1)}&   -3.443&   12.838***&   &   5.318**&   0.214&   6.416***&   &   6.115**&   -4.446&   8.134***\tabularnewline
   ~~&   MIN - $\lambda = 0.1$&   &   \textbf{30.006** (1)}&   23.425*** (2)&   13.825&   &   19.219*** (2)&   -6.633&   18.867***&   &   8.621&   -3.174&   14.480***&   &   9.698*&   -10.551&   16.545***\tabularnewline
   ~~&   MIN - $\lambda = 0.5$&   &   16.531&   22.625 (3)&   \textbf{18.330 (1)}&   &   11.862***&   -13.195&   20.096***&   &   10.468&   -6.613&   21.016***&   &   11.235&   -15.732&   22.923***\tabularnewline
   ~~&   MIN - $\lambda = 1$&   &   10.911&   \textbf{23.489 (1)}&   17.980 (2)&   &   6.597*&   -15.470&   19.296***&   &   10.689&   -8.354&   22.812***&   &   11.501&   -18.285&   24.700***\tabularnewline
\shadeRow   ~~&   SSVS&   &   -68.116&   -50.493*&   -67.513*&   &   2.288&   -68.574*&   -3.879&   &   -9.995**&   -40.090&   -7.634&   &   -17.749**&   -37.498&   -19.086**\tabularnewline
\midrule
{\scshape L ($\theta_1 = 0.025$)}&&&&&&&&&&&&&&&&&\tabularnewline
\shadeRow   ~~&   MIN&   &   7.597&   8.680&   1.966&   &   10.798**&   -4.707&   3.624&   &   4.307*&   0.243&   3.658***&   &   2.764*&   -4.571&   5.158***\tabularnewline
   ~~&   MIN - $\lambda = 0.01$&   &   -1.374&   16.156&   14.614&   &   5.550**&   -15.415&   9.195***&   &   9.270*&   -8.006&   14.482***&   &   10.447*&   -11.916&   16.407***\tabularnewline
   ~~&   MIN - $\lambda = 0.1$&   &   -8.248&   16.134&   15.324&   &   0.544&   -17.598&   8.403**&   &   9.459&   -9.092&   15.961***&   &   10.729*&   -13.387&   17.765***\tabularnewline
   ~~&   MIN - $\lambda = 0.5$&   &   -6.784&   16.336&   15.528&   &   0.900&   -17.445&   8.692**&   &   9.528&   -9.174&   16.198***&   &   10.796*&   -13.520&   17.978***\tabularnewline
   ~~&   MIN - $\lambda = 1$&   &   -6.225&   16.389&   15.621 (3)&   &   1.002&   -17.345&   8.827**&   &   9.537&   -9.180&   16.253***&   &   10.807*&   -13.522&   18.028***\tabularnewline
\midrule
{\scshape L ($\theta_1 = 0.050$)}&&&&&&&&&&&&&&&&&\tabularnewline
\shadeRow   ~~&   MIN&   &   6.284&   5.536&   -0.471&   &   11.666**&   -8.084&   6.990*&   &   3.225&   0.217&   2.864***&   &   -0.132&   -1.901&   4.406***\tabularnewline
   ~~&   MIN - $\lambda = 0.01$&   &   8.616&   14.644&   9.218&   &   13.873***&   -19.797*&   19.791***&   &   11.511 &   -18.799&   23.511***&   &   12.242 &   -24.021&   25.737***\tabularnewline
   ~~&   MIN - $\lambda = 0.1$&   &   -8.814&   1.967&   1.320&   &   4.464&   -33.086**&   22.504***&   &   10.734 (3)&   -34.528&   30.543***&   &   12.070 (3)&   -37.301&   32.828***\tabularnewline
   ~~&   MIN - $\lambda = 0.5$&   &   -14.061&   2.130&   -0.703&   &   1.067&   -38.927*&   22.677***&   &   10.718&   -38.134&   31.721***&   &   12.100&   -39.940&   33.905***\tabularnewline
   ~~&   MIN - $\lambda = 1$&   &   -12.870&   2.495&   -0.578&   &   1.209&   -39.186*&   23.364***&   &   10.841 (2)&   -38.322&   31.969***&   &   12.144 (2)&   -40.097&   34.082***\tabularnewline
\midrule
{\scshape L ($\theta_1 = 0.075$)}&&&&&&&&&&&&&&&&&\tabularnewline
\shadeRow   ~~&   MIN&   &   6.470&   2.303&   -5.543&   &   11.749*&   -9.162&   11.388***&   &   2.937&   0.124&   0.444&   &   -2.815&   -2.295&   1.106\tabularnewline
   ~~&   MIN - $\lambda = 0.01$&   &   -1.536&   8.305&   6.836&   &   17.102*** (3)&   -28.712*&   24.434***&   &   \textbf{12.856* (1)}&   -30.720&   28.291***&   &   \textbf{12.898 (1)}&   -31.633&   31.334***\tabularnewline
   ~~&   MIN - $\lambda = 0.1$&   &   -4.866&   -20.495&   -22.643&   &   9.246**&   -40.233**&   31.976*** (2)&   &   8.744&   -66.632&   39.748*** &   &   10.139&   -69.578&   42.767*** \tabularnewline
   ~~&   MIN - $\lambda = 0.5$&   &   -42.144&   -30.466&   -30.732&   &   -1.292&   -69.624*&   31.766*** (3)&   &   7.426&   -76.417&   42.522*** &   &   8.980&   -83.052&   45.338*** \tabularnewline
   ~~&   MIN - $\lambda = 1$&   &   -46.109&   -30.932&   -31.000&   &   -2.662&   -71.845*&   \textbf{32.671*** (1)}&   &   7.565&   -77.508&   \textbf{43.035*** (1)}&   &   8.984&   -84.153&   \textbf{45.706*** (1)}\tabularnewline
\bottomrule
\end{tabular}}
\caption{Log predictive likelihoods relative to the \textit{small-scale} BVAR with a Minnesota prior. \label{tab:LPS}}
\begin{minipage}{\linewidth}
\footnotesize \textbf{Notes}: Bold numbers indicate highest Bayes factors for each horizon and target variable (and therefore best performing models over the full hold-out sample in terms of density forecasts), while the numbers in parenthesis refer to the ranking of the three best specifications according to the MCS \cite{hansen2011model} procedure (including the benchmark). Note for one-year- and two-year-ahead FEDFUNDS density forecasts the superior model set consists of a single best model. Gray shaded rows denote non-sparsified models. Asterisks indicate statistical significance for each model relative to the benchmark at the $1$ ($^{***}$), $5$ ($^{**}$) and $10$ ($^{*}$) percent significance levels. 
\end{minipage}\end{center}}
\end{table}\end{landscape}

\subsection{Assessing Model Calibration using Probability Integral Transforms}
In this section we investigate how sparsification impacts the calibration of predictive densities. Therefore  we follow \cite{giordani2010forecasting} and \cite{clark2011real} in analyzing normalized forecast errors that are obtained by applying the probability integral transform (PIT).\footnote{For more details on how the normalized forecast errors are computed, see \cite{clark2011real}.}  If a given model is well calibrated, these normalized forecast errors should be standard normally distributed. Deviations from the standard normal distribution provide information along which dimension the model might be miss-specified.

In this discussion we focus on one-year-ahead predictive densities for the reasons outlined above. The normalized forecast errors for one-quarter- and two-year-ahead forecasts look similar, displaying no discernible differences in qualitative terms.\footnote{The normalized forecast errors for the other forecast horizons are available from the first author upon request.}



\begin{figure}[htbp]
\begin{minipage}{0.33\textwidth}
\centering \textit{GDPC1}
\end{minipage}
\begin{minipage}{0.33\textwidth}
\centering \textit{CPIAUCSL}
\end{minipage}
\begin{minipage}{0.33\textwidth}
\centering \textit{FEDFUNDS}
\end{minipage}
\begin{minipage}{\textwidth}
\centering (a) \textit{S-VAR}
\end{minipage}
\begin{minipage}{0.33\textwidth}
\centering
\includegraphics[width=1\textwidth]{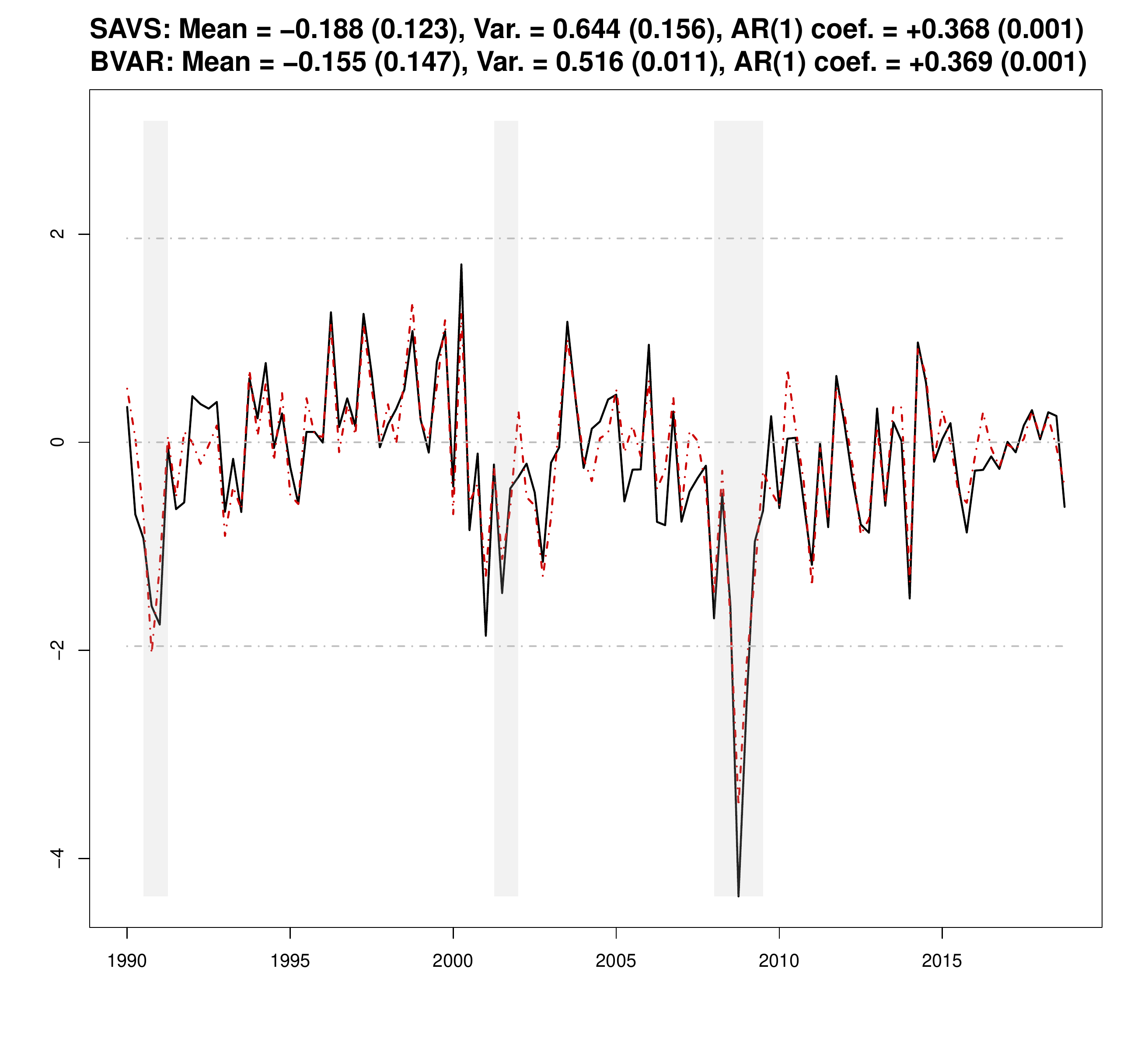}
\end{minipage}
\begin{minipage}{0.33\textwidth}
\centering
\includegraphics[width=1\textwidth]{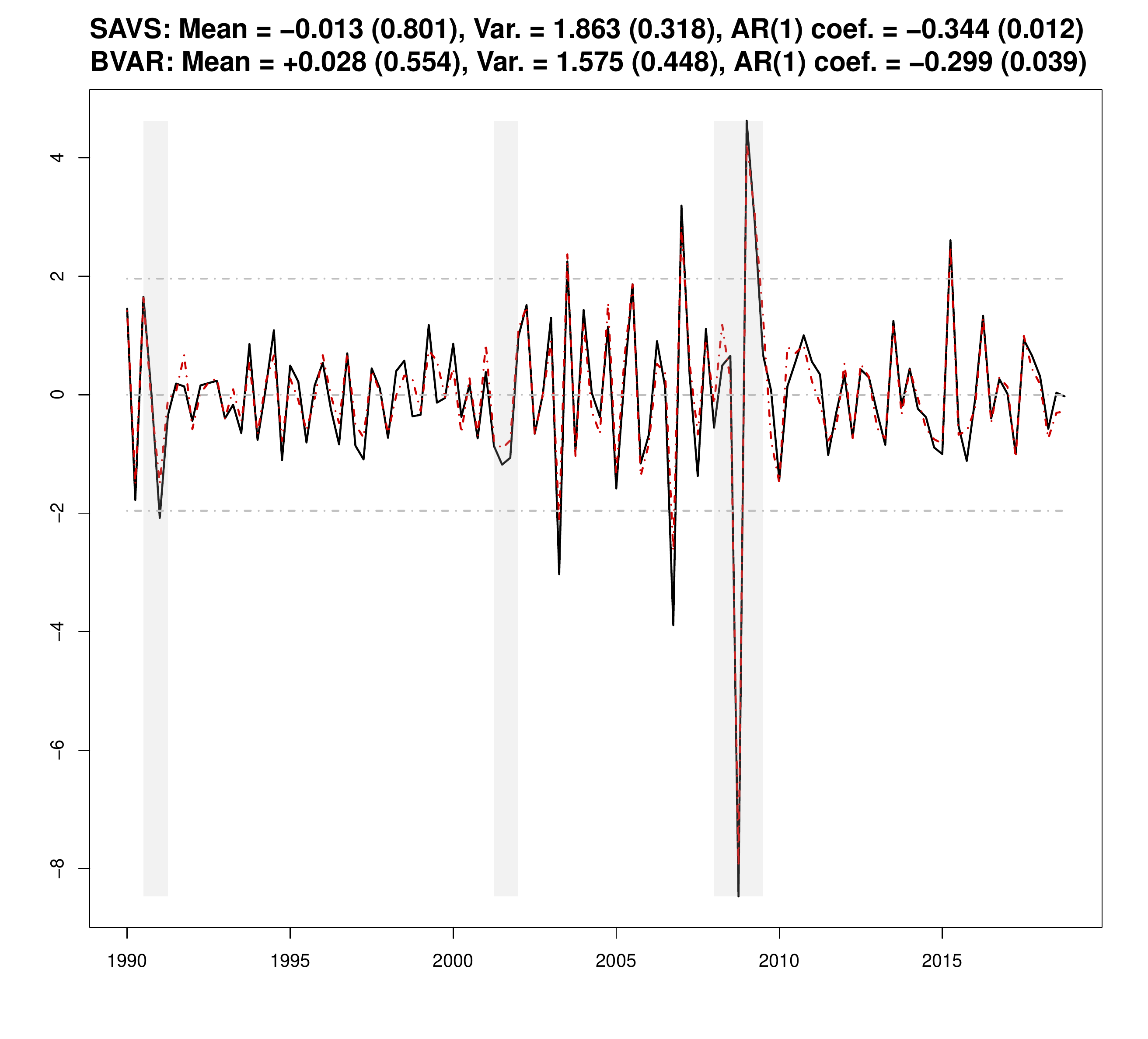}
\end{minipage}
\begin{minipage}{0.33\textwidth}
\centering
\includegraphics[width=1\textwidth]{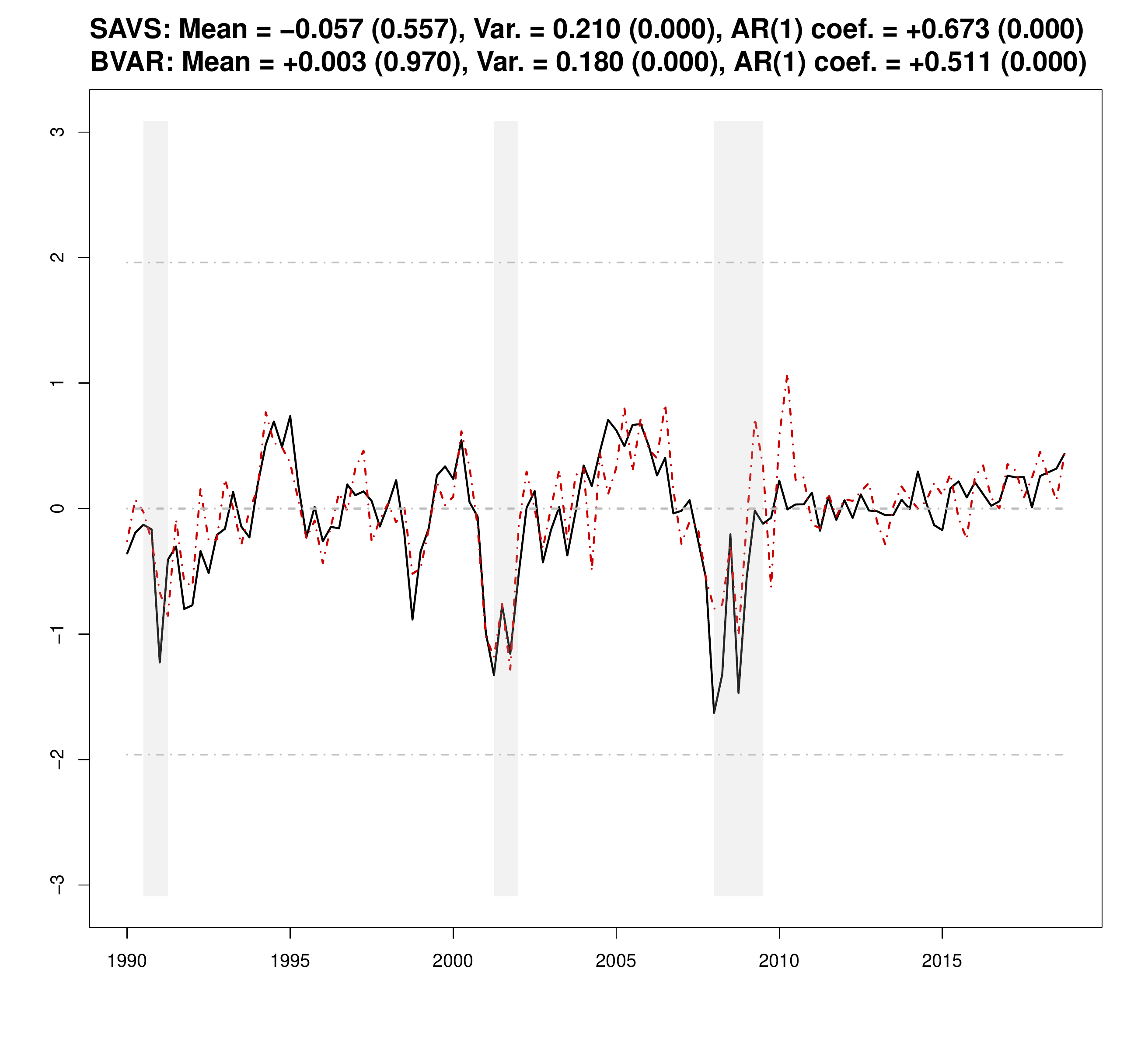}
\end{minipage}
\begin{minipage}{\textwidth}
\centering (b) \textit{FA-VAR}
\end{minipage}
\begin{minipage}{0.33\textwidth}
\centering
\includegraphics[width=1\textwidth]{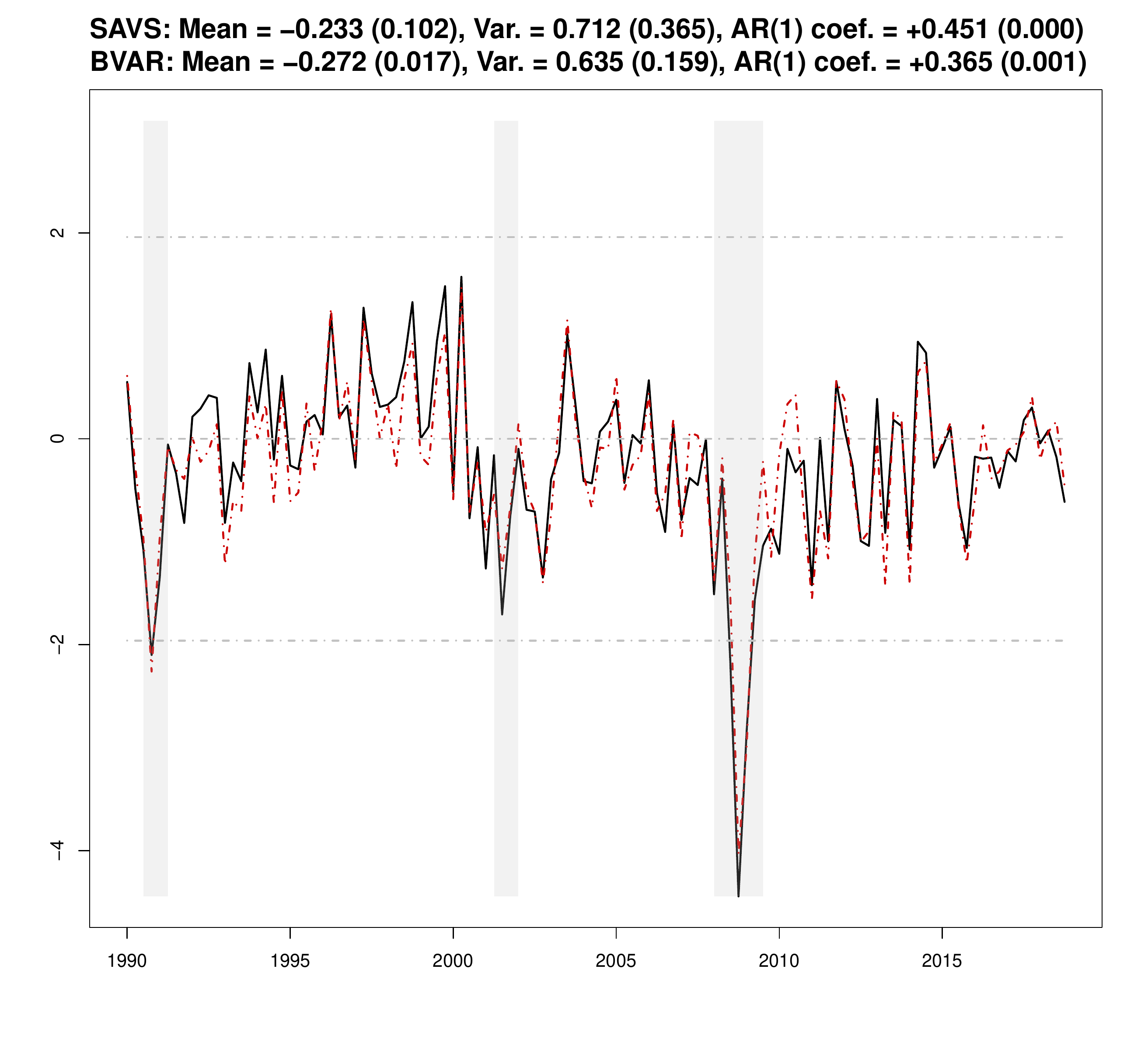}
\end{minipage}
\begin{minipage}{0.33\textwidth}
\centering
\includegraphics[width=1\textwidth]{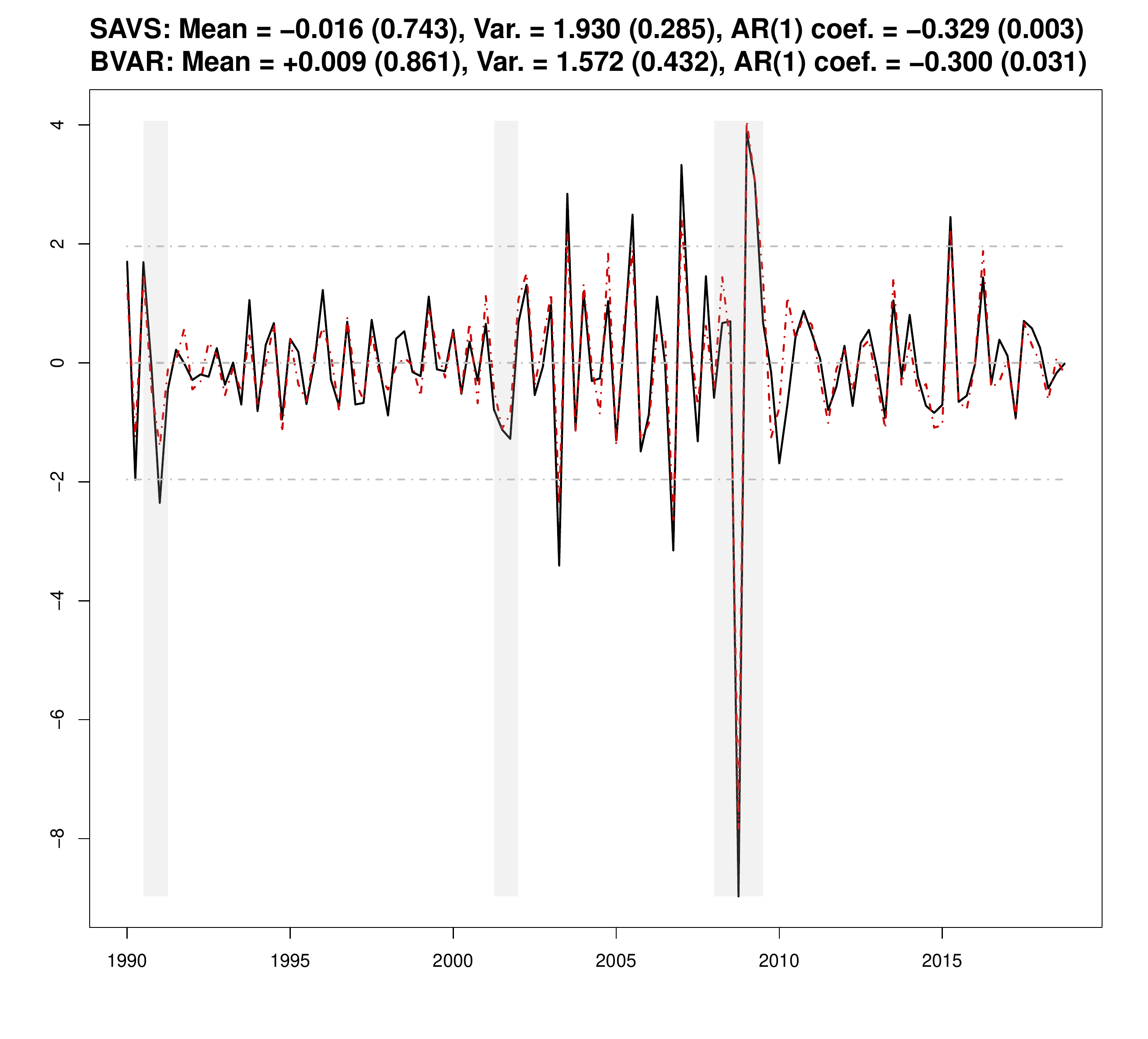}
\end{minipage}
\begin{minipage}{0.33\textwidth}
\centering
\includegraphics[width=1\textwidth]{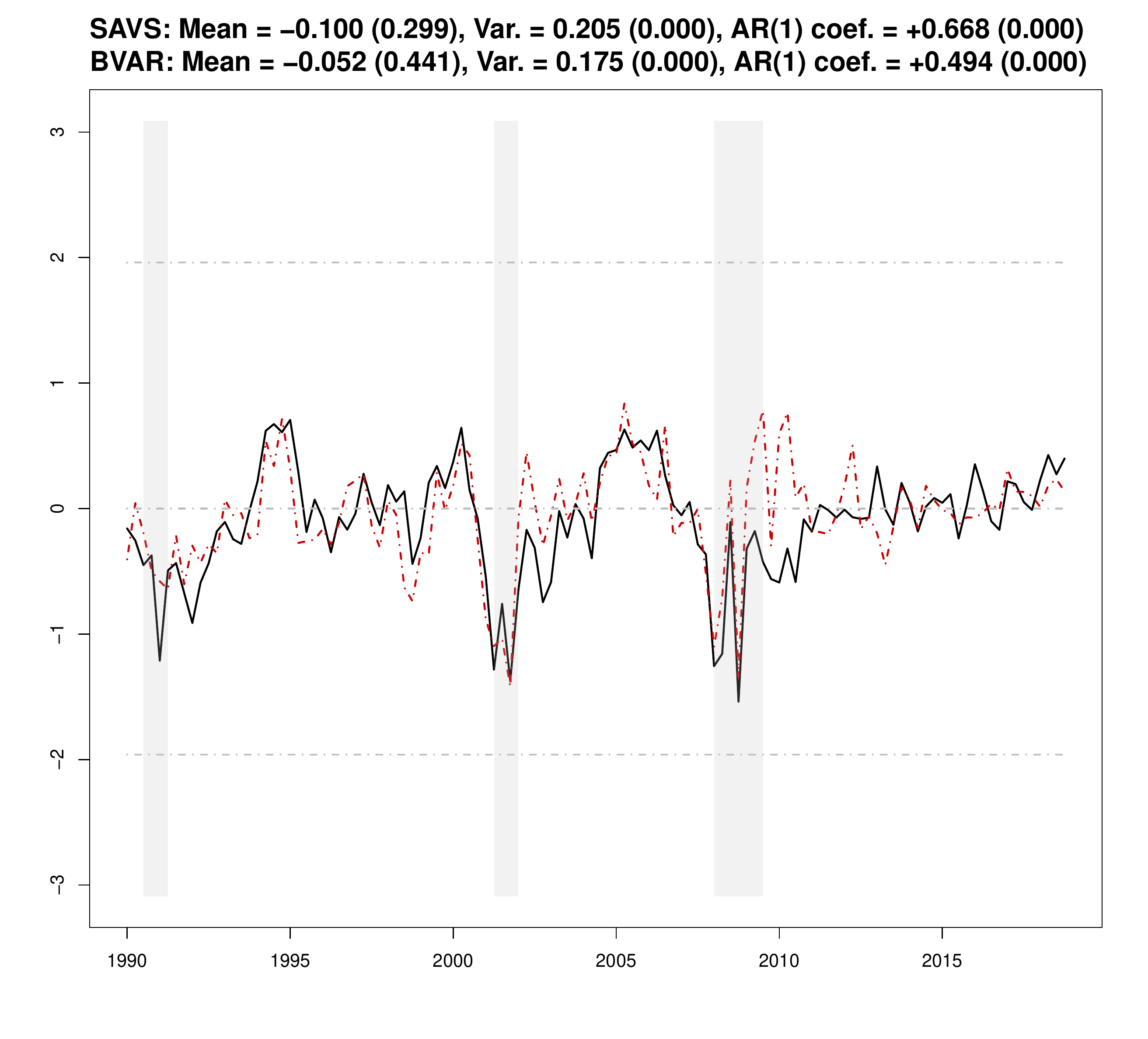}
\end{minipage}
\begin{minipage}{\textwidth}
\centering (c) \textit{M-VAR}
\end{minipage}
\begin{minipage}{0.33\textwidth}
\centering
\includegraphics[width=1\textwidth]{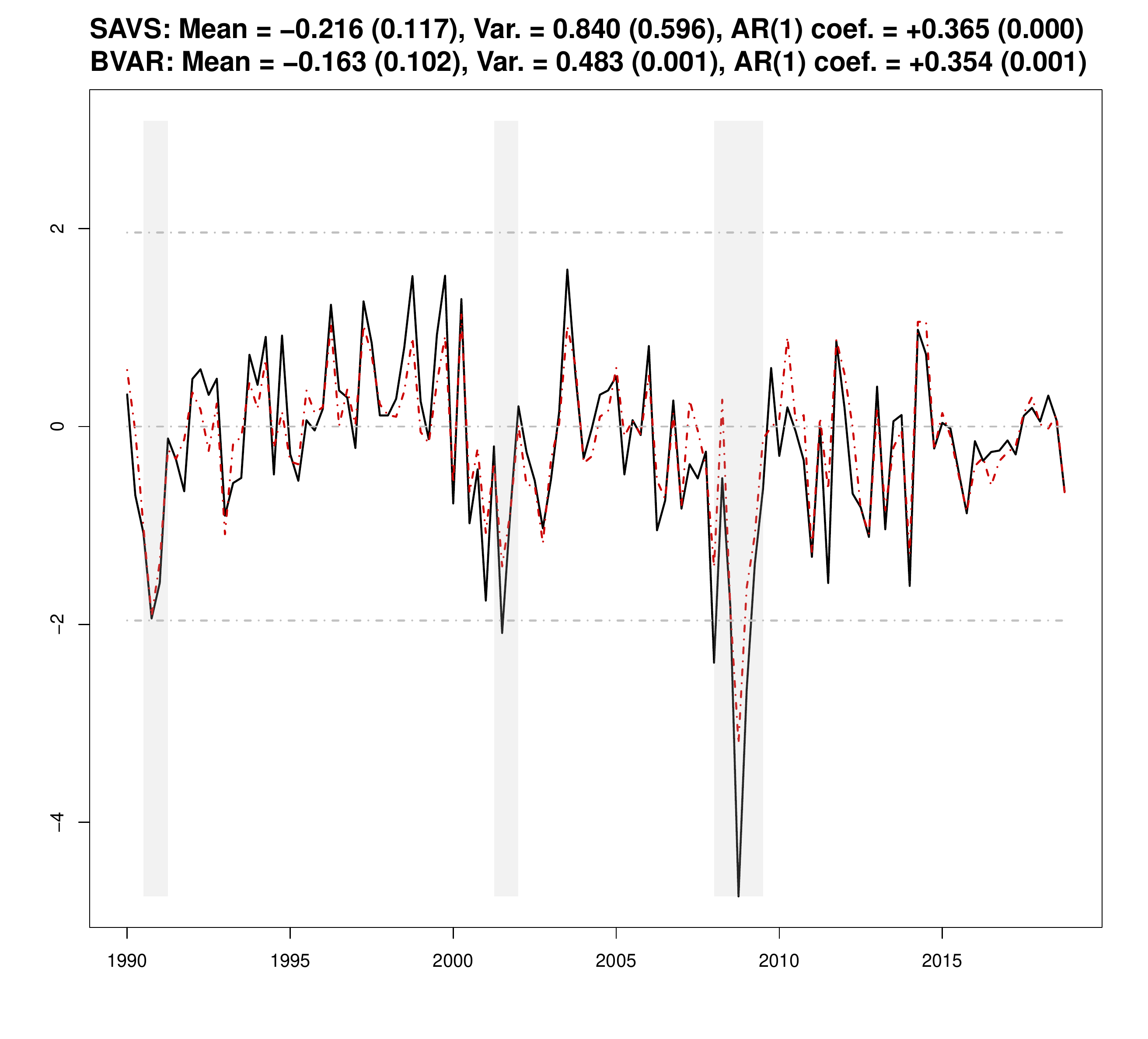}
\end{minipage}
\begin{minipage}{0.33\textwidth}
\centering
\includegraphics[width=1\textwidth]{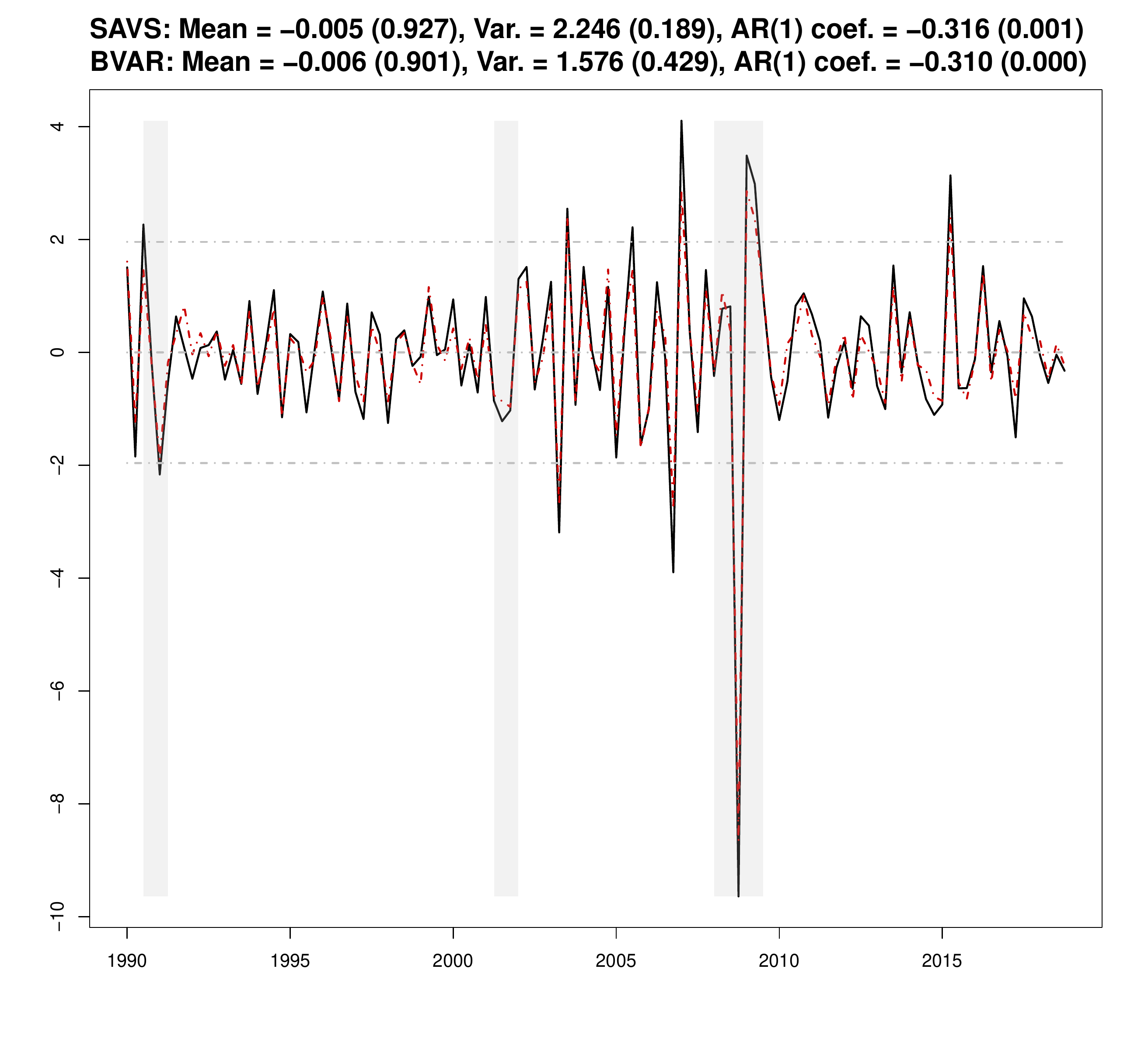}
\end{minipage}
\begin{minipage}{0.33\textwidth}
\centering
\includegraphics[width=1\textwidth]{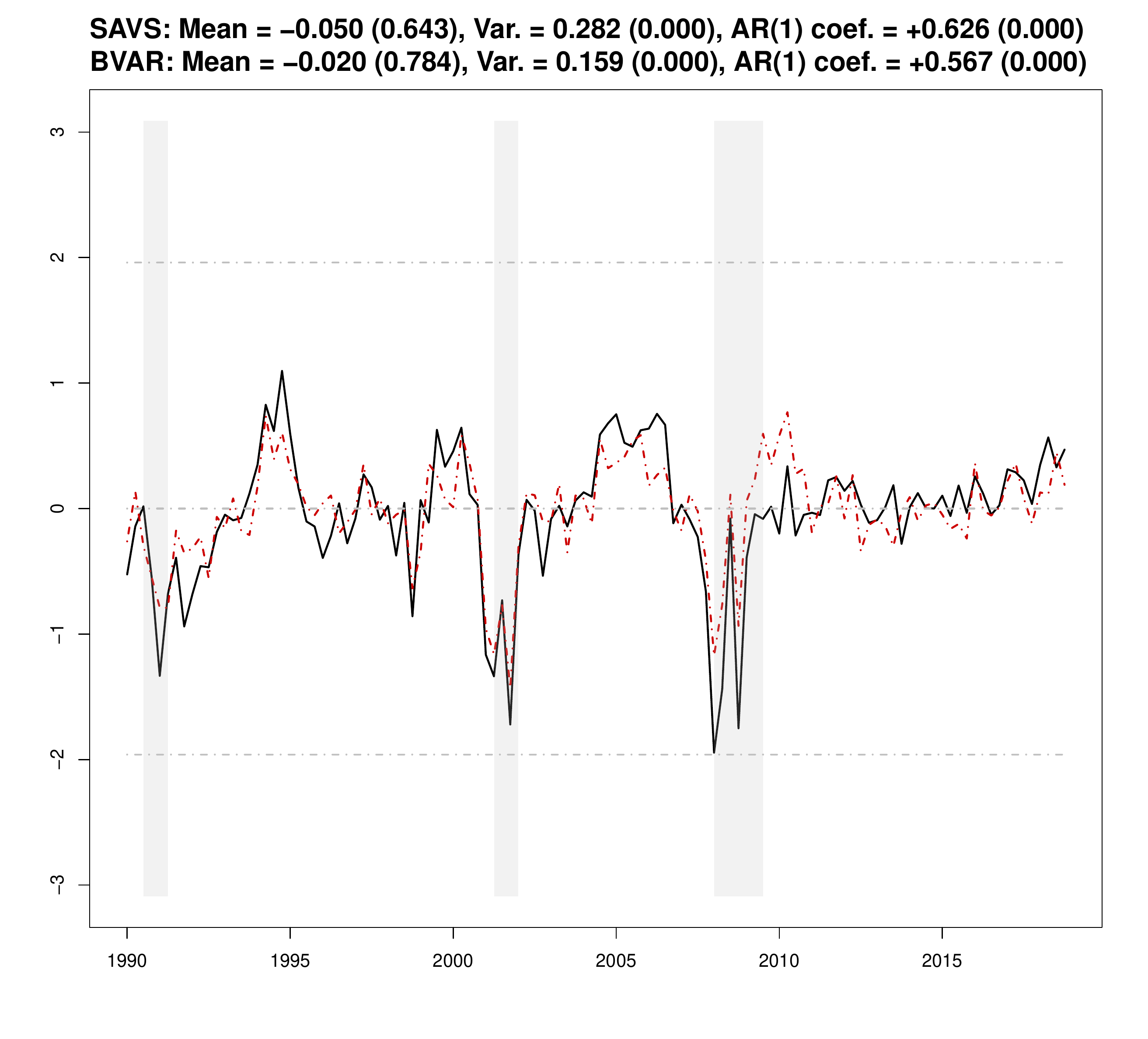}
\end{minipage}

\caption{Normalized one-year-ahead forecast errors of small and moderate sized models. The black solid lines represent the sparsified versions (SAVS) with $\lambda = 1$ while the red dash-dotted lines depict classic BVARs. The gray dash-dotted horizontal lines indicate the $95\%$ interval of the standard normal distribution and the gray shaded areas denote NBER recessions. Moreover, the legends show the corresponding test statistics of normalized errors. Here we follow \citep{clark2011real} and show the mean, the variance (Var.), and the autoregressive coefficient (AR($1$) coef.) of normalized errors. In parenthesis we depict the corresponding $p$-values. The null-hypotheses, a zero mean, a variance of one, and no autocorrelated errors, are tested separately.}
	  \label{fig:pitmar}
\end{figure}
\begin{figure}[htbp]
\begin{minipage}{0.33\textwidth}
\centering \textit{GDPC1}
\end{minipage}
\begin{minipage}{0.33\textwidth}
\centering \textit{CPIAUCSL}
\end{minipage}
\begin{minipage}{0.33\textwidth}
\centering \textit{FEDFUNDS}
\end{minipage}
\begin{minipage}{\textwidth}
\centering (a) \textit{L-VAR ($\theta_1 = 0.025$)}
\end{minipage}
\begin{minipage}{0.33\textwidth}
\centering
\includegraphics[width=1\textwidth]{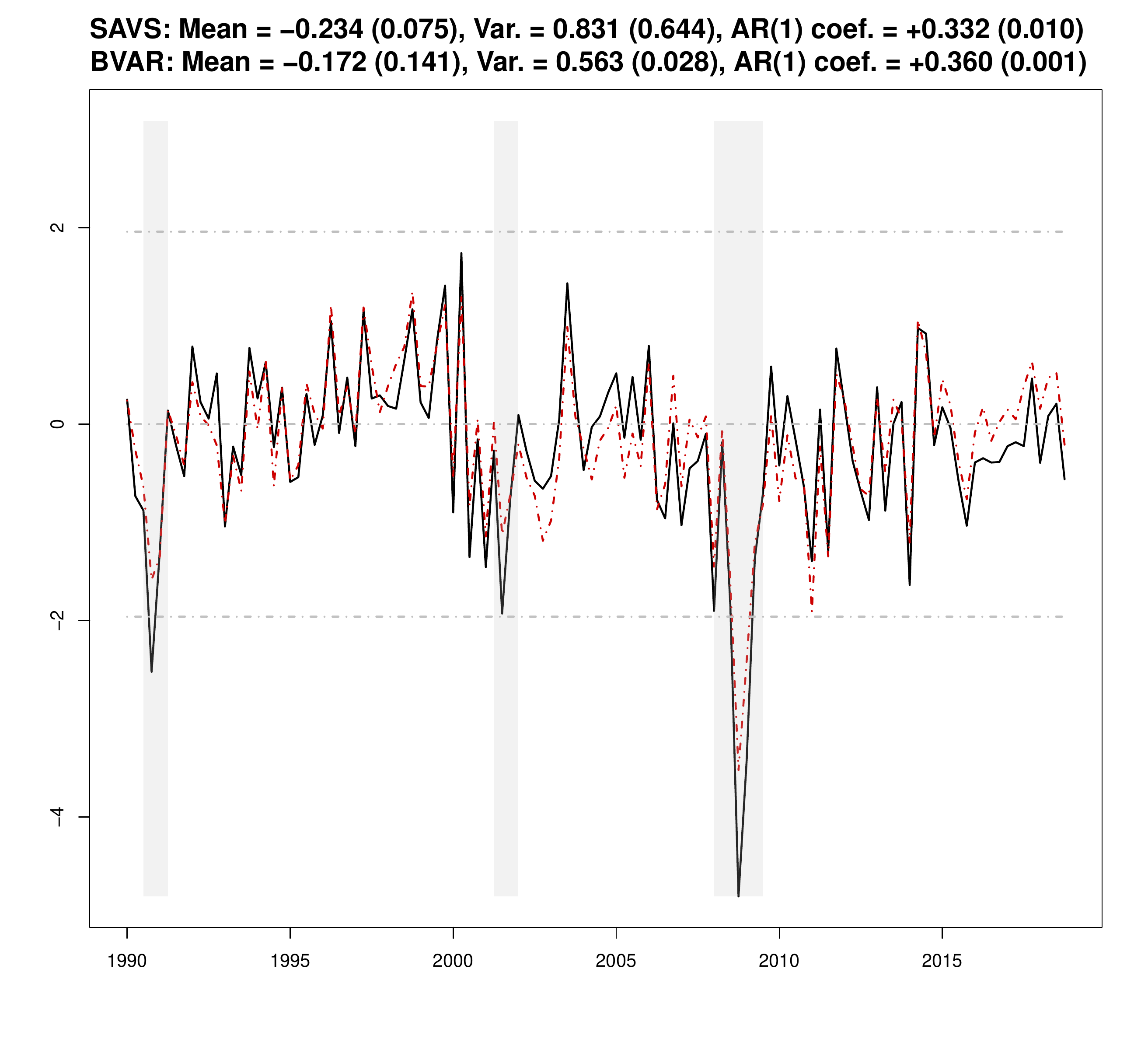}
\end{minipage}
\begin{minipage}{0.33\textwidth}
\centering
\includegraphics[width=1\textwidth]{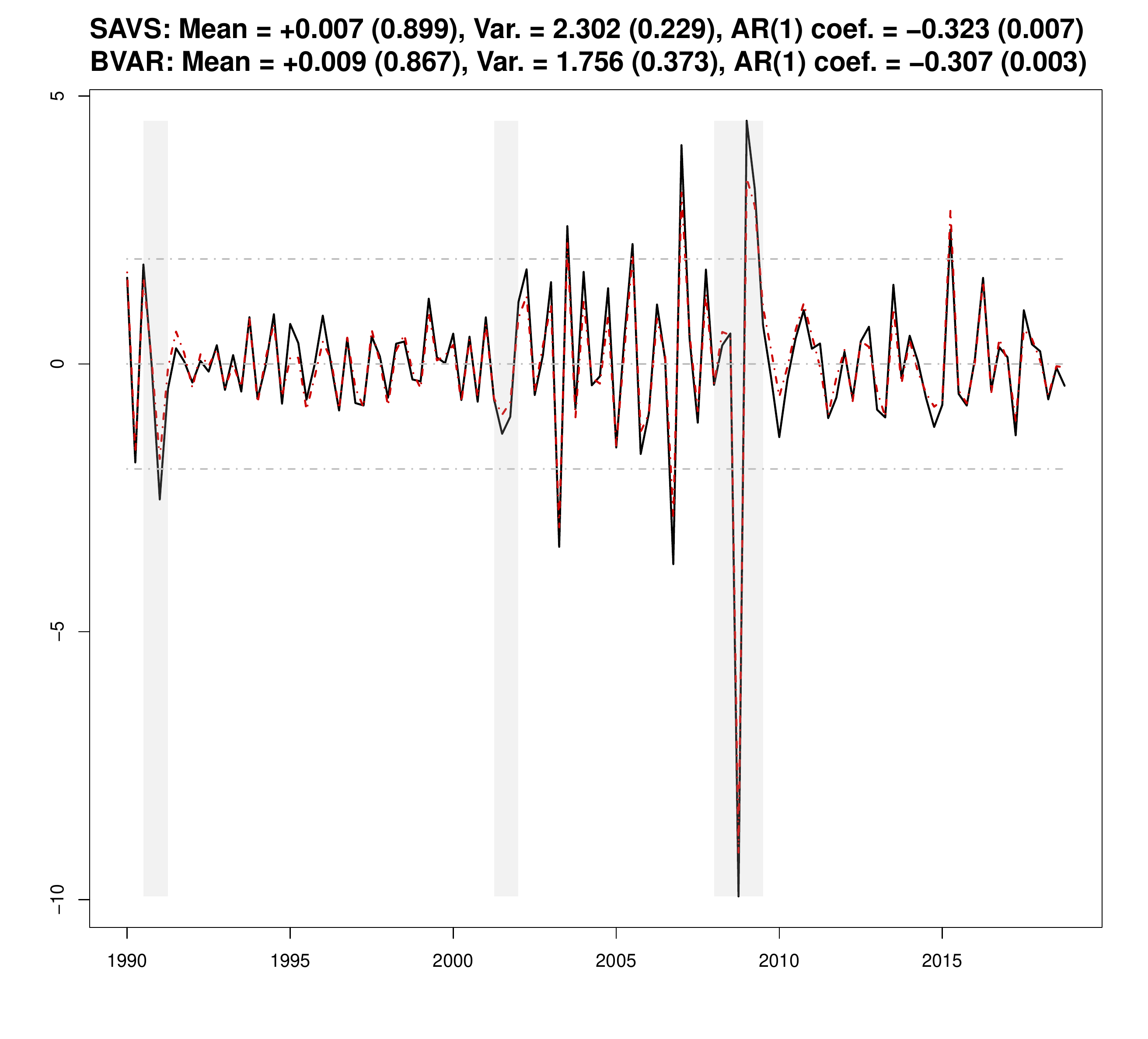}
\end{minipage}
\begin{minipage}{0.33\textwidth}
\centering
\includegraphics[width=1\textwidth]{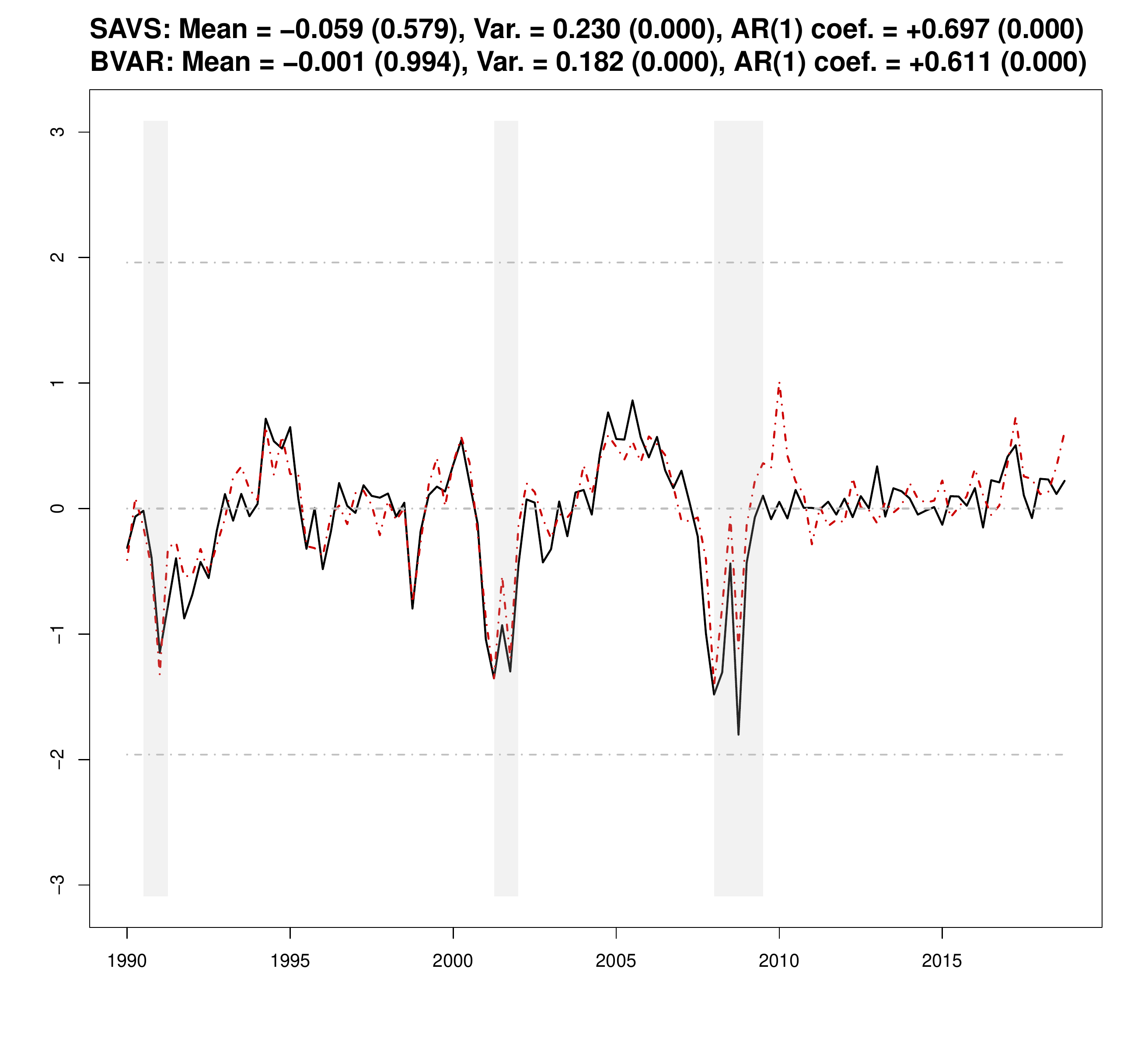}
\end{minipage}
\begin{minipage}{\textwidth}
\centering (b) \textit{L-VAR ($\theta_1 = 0.05$)}
\end{minipage}
\begin{minipage}{0.33\textwidth}
\centering
\includegraphics[width=1\textwidth]{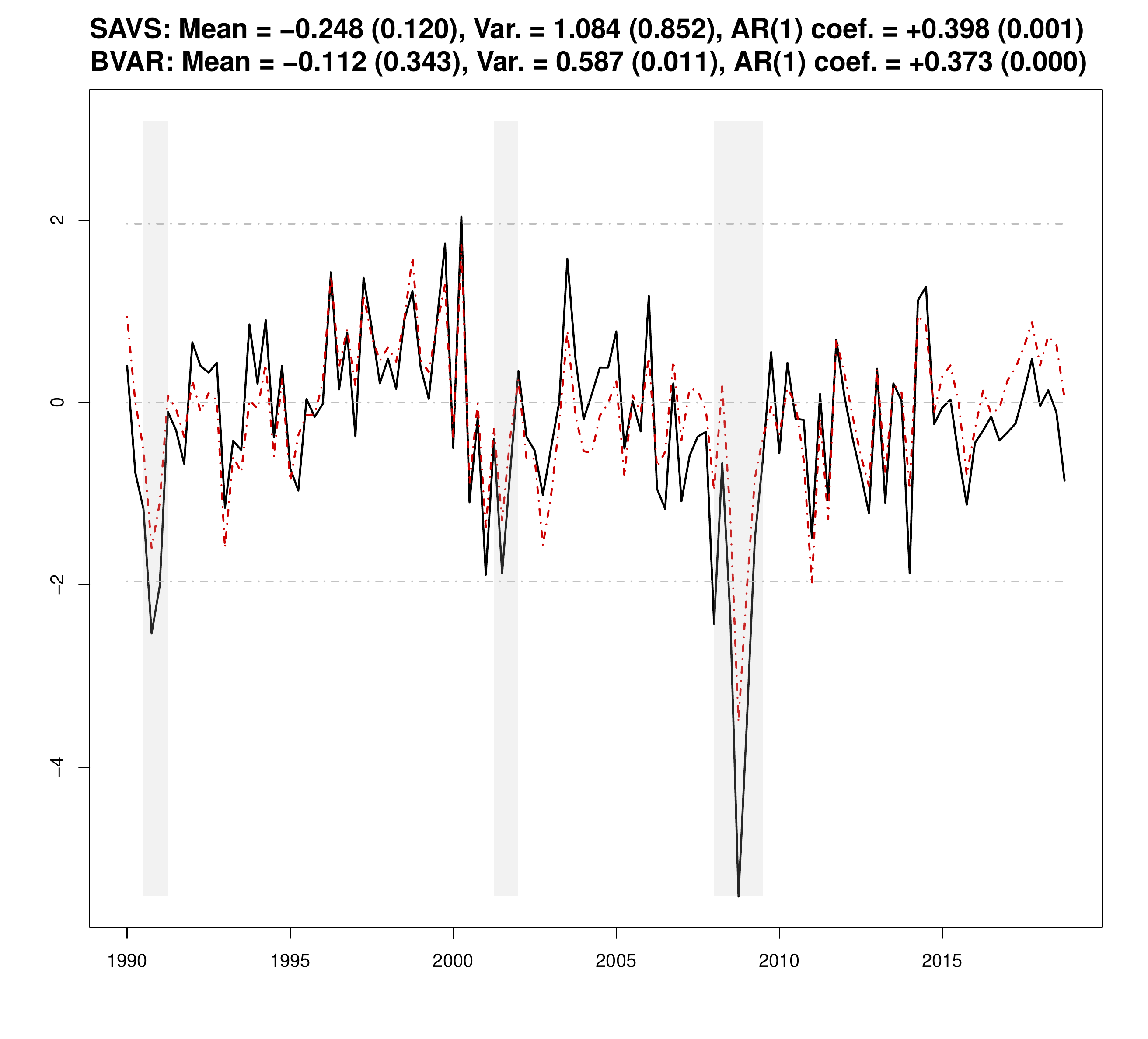}
\end{minipage}
\begin{minipage}{0.33\textwidth}
\centering
\includegraphics[width=1\textwidth]{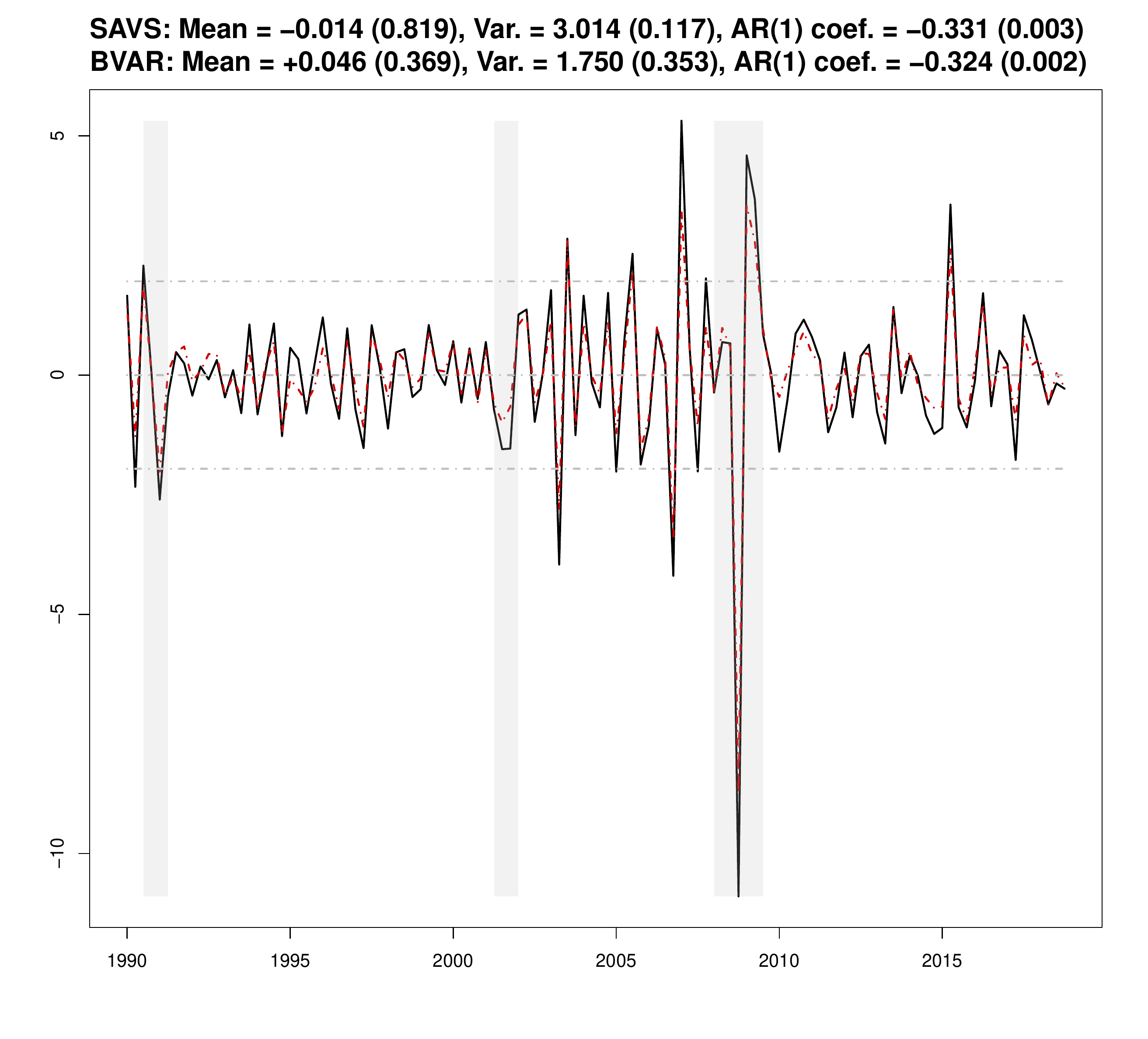}
\end{minipage}
\begin{minipage}{0.33\textwidth}
\centering
\includegraphics[width=1\textwidth]{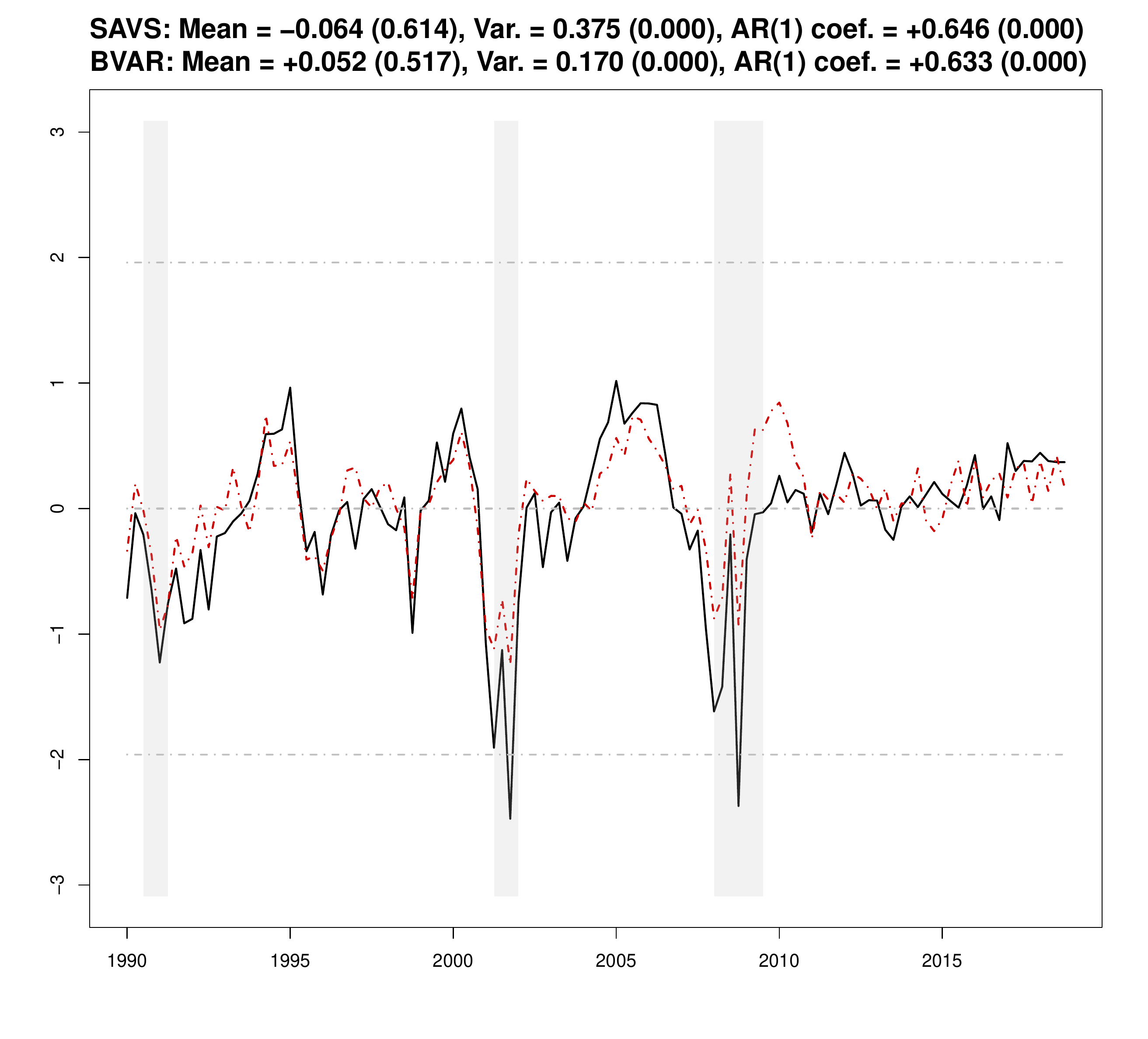}
\end{minipage}
\begin{minipage}{\textwidth}
\centering (c) \textit{L-VAR ($\theta_1 = 0.075$)}
\end{minipage}
\begin{minipage}{0.33\textwidth}
\centering
\includegraphics[width=1\textwidth]{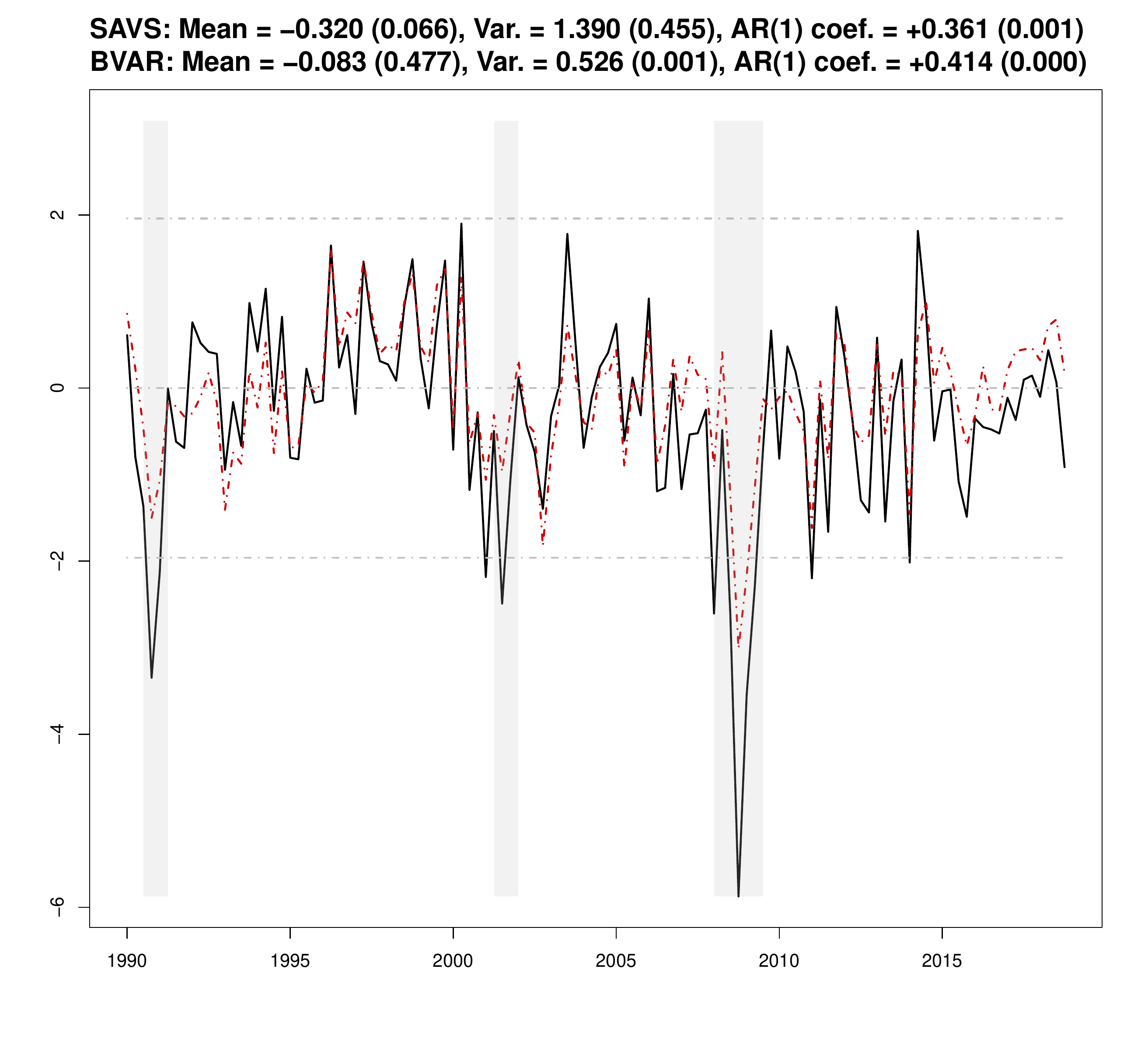}
\end{minipage}
\begin{minipage}{0.33\textwidth}
\centering
\includegraphics[width=1\textwidth]{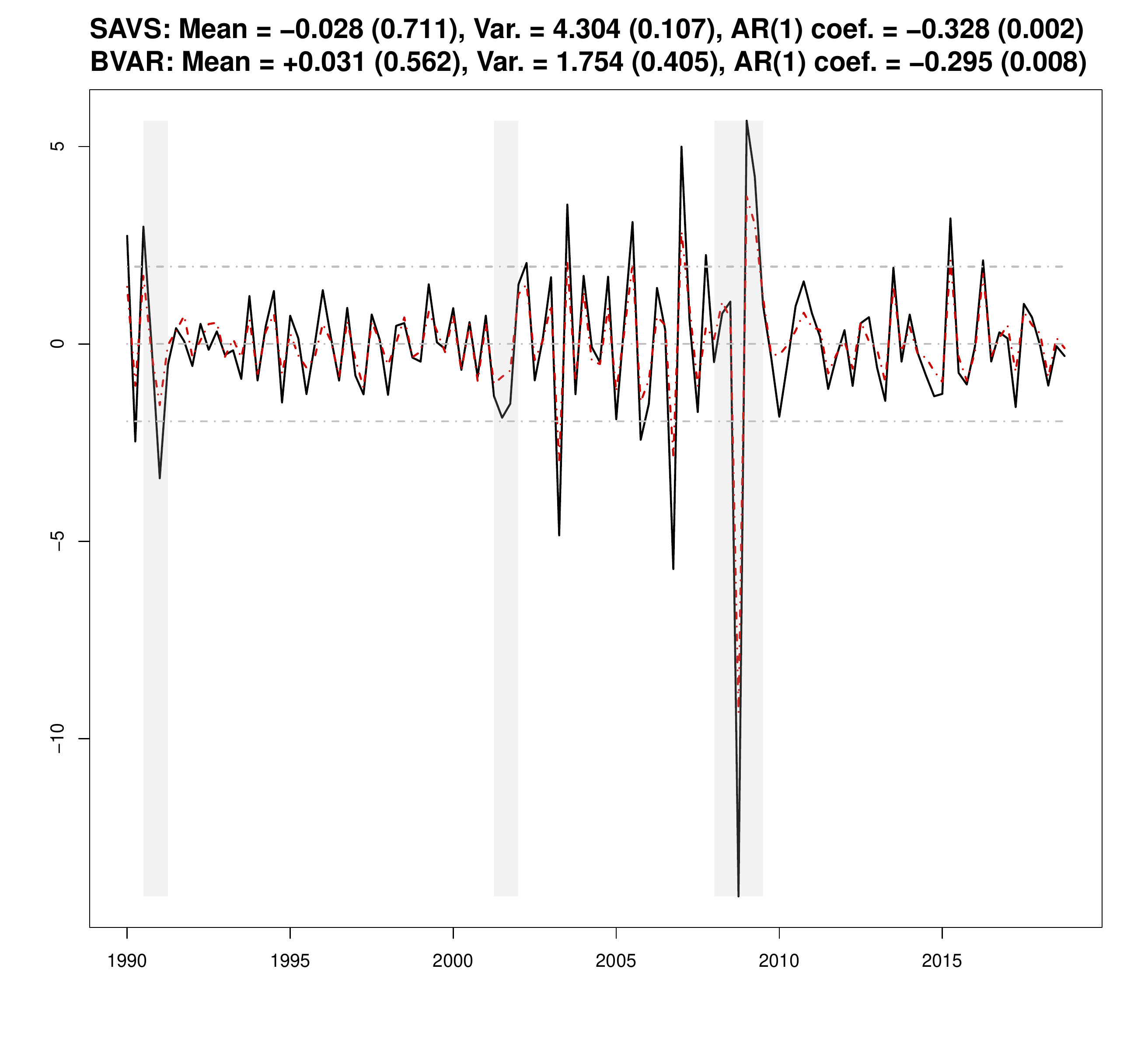}
\end{minipage}
\begin{minipage}{0.33\textwidth}
\centering
\includegraphics[width=1\textwidth]{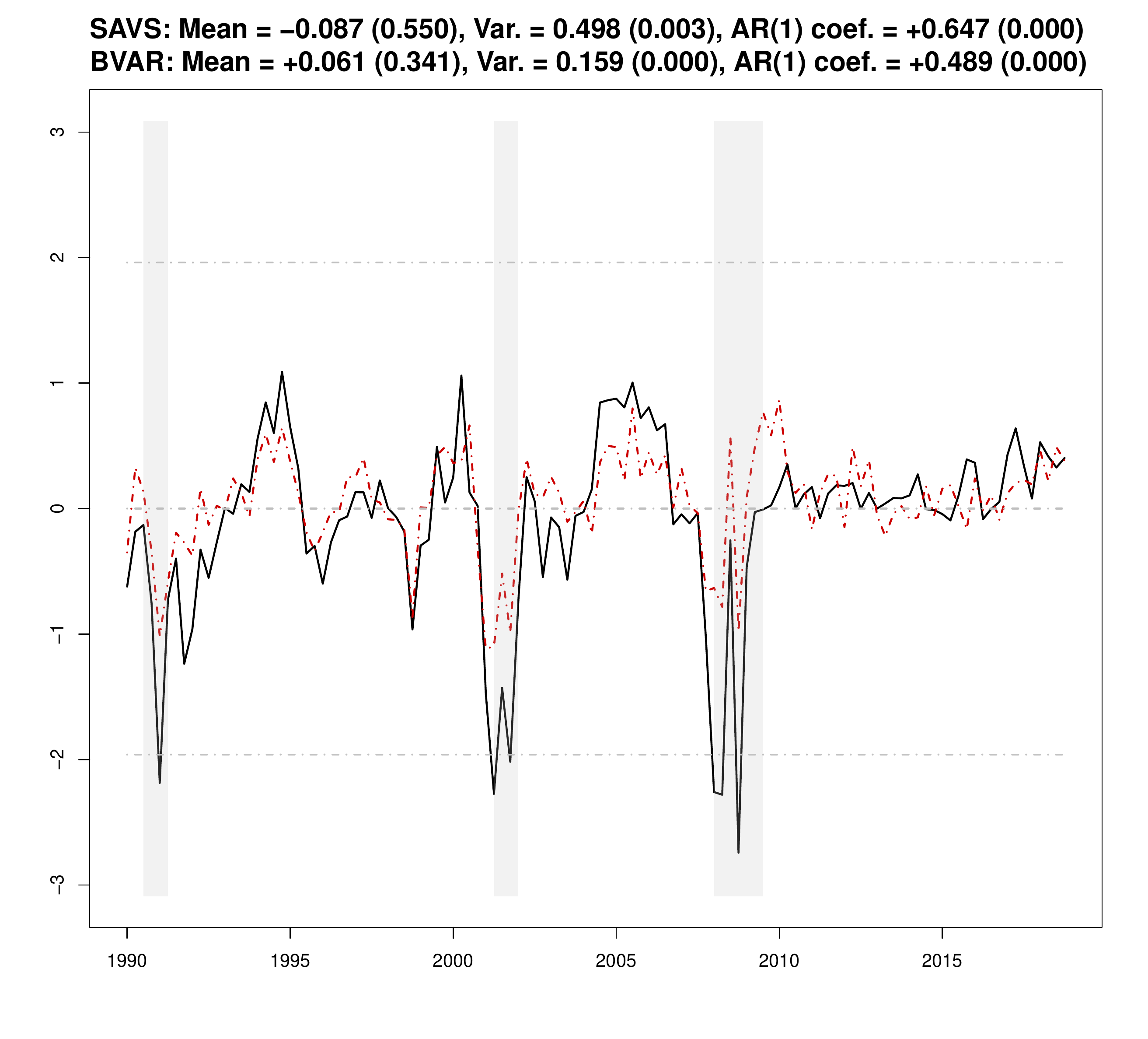}
\end{minipage}
\caption{Normalized one-year-ahead forecast errors of large-scale models. The black solid lines represent the sparsified versions (SAVS) with $\lambda = 1$ while the red dash-dotted lines depict classic BVARs. The gray dash-dotted horizontal lines indicate the $95\%$ interval of the standard normal distribution and the gray shaded areas denote NBER recessions. Moreover, the legends show the corresponding test statistics of normalized errors. We follow \citep{clark2011real} and show the mean, the variance (Var.), and the autoregressive coefficient (AR($1$) coef.) of normalized errors. In parenthesis we depict the corresponding $p$-values. The null-hypotheses, a zero mean, a variance of one, and no autocorrelated errors, are tested separately.}\label{fig:pitmar_large}
\end{figure}

In \autoref{fig:pitmar} and \autoref{fig:pitmar_large} we compare the one-year-ahead normalized forecast errors for the setup proposed in \cite{bhattacharya2018signal} (with $\lambda = 1$) to the non-sparsified BVAR model across information sets and target variables. Since this purely visual analysis might be misleading, we add a legend to each panel that provides information on whether departures from standard normality are statistically significant. The  null hypothesis is that the normalized forecast errors are zero mean, feature a variance of one and no serial correlation \citep{berkowitz2001testing}.\footnote{Following \cite{clark2011real}, the $p$-value of a zero mean is computed with a Newey-West variance with five lags, the $p$-value of a unit variance is obtained by regressing the squared normalized forecast errors on an intercept and using a Newey-West variance with three lags, and the $p$-value of no autocorrelation is computed with an AR($1$) model, featuring an unconditional mean and heteroskedasticity-robust standard errors.}

Considering each of the three focus variables separately provides some interesting insights. Before focusing on sparsified models, we analyze the non-sparsified counterparts because these are the models that are then  sparsified.

For the standard BVAR forecasts, the mean is very often not significantly different from zero. With regards to the variance we find that  normalized forecast errors display variances well below one for both GDP and the interest rate, while displaying a particularly high variance for inflation forecasts. A variance of forecast errors below one indicates that for many periods in the hold-out, the predictive distribution is too spread out.  As discussed above, this may be at least in part attributed to the imprecisely estimated small parameters. By contrast,  a variance above one would indicate that the predictive density tends to be too tight.  Moreover, we observe a high autocorrelation of forecast errors for the interest rate. This feature is also commonly found in the literature \citep[see, for example][]{clark2011real}.
   
 
If we apply the additional SAVS step the properties of the normalized forecast errors improve. In the case of GDP, sparsifying the BVAR typically increases the variances at a small cost of a slightly more negative mean (which is not significantly different from zero in almost all instances).  In particular, sparsified large-scale models attain a variance close to one (see, e.g.,  \autoref{fig:pitmar} L-VAR with $\theta_1 = 0.05$). The variances of the normalized forecast errors are much closer to unity (and in fact do not differ from one in a statistical significant manner). 

In the case of inflation, and especially during the financial crisis, sparsification does not seem to help.
The non-sparsified BVAR estimates tend to produce  already too tight predictive distributions. These become even narrower after applying sparsification. It is precisely this feature that hurts forecasting performance after the large drop in inflation in $2008$:Q$4$. In $2008$:Q$4$, models are generally overly confident and, in the following period ($2009$:Q$1$), somewhat too pessimistic due to the autoregressive nature of the models. Additionally, note that forecast errors are particularly high for large-scale models.

For the interest rate, BVAR models produce normalized forecast errors with variances well below unity. Using large VARs coupled with a relative loose prior and adding SAVS improves this. More precisely, for the sparsified large VAR with $\theta_1=0.075$ we find that the variance increases from $0.16$ to approximately $0.5$. While this is still below the unit variance we would expect under a well calibrated model we consider this a substantial improvement at the costs of slightly higher autocorrelation in the errors.

In general, the PITs (for the one-year-ahead forecasts) suggest that using SAVS has the potential to improve model calibration for some models and most variables. In the case of output and interest rates, we can improve model calibration by adding SAVS. A recommendation for practitioners might be that if the variance of normalized forecast errors of a BVAR is well below one, SAVS can substantially improve forecast density calibration (see, e.g., output and interest rate forecasts of large VARs). Conversely, if the predictive variance of a BVAR is already too tight, characterized by normalized forecast errors that feature a variance above one, SAVS tends to hurt predictive accuracy (see the case of inflation).
        
\section{Conclusions}\label{sec:concl}
This paper proposes methods to shrink-and-sparsify VAR models with conjugate priors. The main feature of our SAVS approach is that we post-process each draw from the joint posterior by solving an optimization problem to search for a sparse coefficient vector. Without destroying the conjugacy of the model, this approach allows for different predictors across the equations in the VAR. And, instead of pushing coefficients close to zero, our approach introduces exact zeros, removing the lower bound on accuracy one can achieve under a popular shrinkage prior in the Minnesota tradition. Since the error covariance matrix in large VARs also features a large number of coefficients, we adapt techniques from the literature on graphical models to obtain a sparse estimate of the variance-covariance matrix of the system. Using a synthetic and real-data application, we illustrate the merits of combining shrinkage and sparsification in large multivariate models.

\clearpage

\small{\scfont\setstretch{0.85}
\addcontentsline{toc}{section}{References}
\bibliographystyle{fischer}
\bibliography{sparse.bib}}

\clearpage

\begin{appendices}\crefalias{section}{appsec}
\setcounter{equation}{0}
\renewcommand\theequation{A.\arabic{equation}}
\normalsize

\clearpage

\section{Data description}
\label{app:data}
Here, we provide detailed information on the transformation applied for each variable, as we transform the data to stationarity, according to the suggestions of \cite{mccracken2016fred}. With stationary data the prior is centered on zero, assuming a white noise process for each variable a priori. Moreover, we standardise the data by demeaning each variable and dividing through the standard deviation. Due to the scale-variance of PCs the data is also standardised before extracting the factors.

\begin{table}[!htp]
{\tiny
\begin{center}
\scalebox{0.8}{
\begin{tabular}{lllrlll}
\toprule
\multicolumn{1}{l}{\ }&\multicolumn{1}{c}{\ FRED.Mnemonic}&\multicolumn{1}{c}{\ Description}&\multicolumn{1}{c}{\ Trans I(0)}&\multicolumn{1}{c}{\ SMALL}&\multicolumn{1}{c}{\ MEDIUM}&\multicolumn{1}{c}{\ LARGE}\tabularnewline
\midrule
{\scshape Slow}&&&&&&\tabularnewline
~~&GDPC1&Real Gross Domestic Product&$5$&x&x&x\tabularnewline
~~&PCECC96&Real Personal Consumption Expenditures&$5$&&x&x\tabularnewline
~~&PCDGx&Real personal consumption expenditures:  Durable goods &$5$&&&x\tabularnewline
~~&PCESVx&Real Personal Consumption Expenditures:  Services &$5$&&&x\tabularnewline
~~&PCNDx&Real Personal Consumption Expenditures:  Nondurable Goods &$5$&&&x\tabularnewline
~~&GPDIC1&Real Gross Private Domestic Investment&$5$&&&x\tabularnewline
~~&FPIx&Real private fixed investment &$5$&&x&x\tabularnewline
~~&Y033RC1Q027SBEAx&Real Gross Private Domestic Investment:  Fixed Investment:  Nonresidential Equipment&$5$&&&x\tabularnewline
~~&PNFIx&Real private fixed investment:  Nonresidential &$5$&&&x\tabularnewline
~~&PRFIx&Real private fixed investment:  Residential &$5$&&&x\tabularnewline
~~&A014RE1Q156NBEA&Shares of gross domestic product:  Gross private domestic investment: Change
in private inventories&$1$&&&x\tabularnewline
~~&GCEC1&Real Government Consumption Expenditures and Gross Investment&$5$&&x&x\tabularnewline
~~&A823RL1Q225SBEA&Real Government Consumption Expenditures and Gross Investment:  Federal&$1$&&&x\tabularnewline
~~&FGRECPTx&Real Federal Government Current Receipts &$5$&&&x\tabularnewline
~~&SLCEx&Real government state and local consumption expenditures &$5$&&&x\tabularnewline
~~&EXPGSC1&Real Exports of Goods and Services&$5$&&&x\tabularnewline
~~&IMPGSC1&Real Imports of Goods and Services&$5$&&&x\tabularnewline
~~&DPIC96&Real Disposable Personal Income&$5$&&&x\tabularnewline
~~&OUTNFB&Nonfarm Business Sector:  Real Output&$5$&&&x\tabularnewline
~~&OUTBS&Business Sector:  Real Output&$5$&&&x\tabularnewline
~~&INDPRO&IP:Total index Industrial Production Index (Index 2012=100)&$5$&&x&x\tabularnewline
~~&IPFINAL&IP:Final products Industrial Production: Final Products (Market Group) (Index 2012=100)&$5$&&&x\tabularnewline
~~&IPCONGD&IP:Consumer goods Industrial Production: Consumer Goods (Index 2012=100)&$5$&&&x\tabularnewline
~~&IPMAT&Materials (Index 2012=100)&$5$&&&x\tabularnewline
~~&IPDMAT&Durable Materials (Index 2012=100)&$5$&&&x\tabularnewline
~~&IPNMAT&Nondurable Materials (Index 2012=100)&$5$&&&x\tabularnewline
~~&IPDCONGD&Durable Consumer Goods (Index 2012=100)&$5$&&&x\tabularnewline
~~&IPB51110SQ&Durable Goods:  Automotive products (Index 2012=100)&$5$&&&x\tabularnewline
~~&IPNCONGD&Nondurable Consumer Goods (Index 2012=100)&$5$&&&x\tabularnewline
~~&IPBUSEQ&Business Equipment (Index 2012=100)&$5$&&&x\tabularnewline
~~&IPB51220SQ&Consumer energy products (Index 2012=100)&$5$&&&x\tabularnewline
~~&CUMFNS&Capacity Utilization:  Manufacturing (SIC) (Percent of Capacity)&$1$&&&x\tabularnewline
~~&IPMANSICS&Industrial Production:  Manufacturing (SIC) (Index 2012=100)&$5$&&&x\tabularnewline
~~&IPB51222S&Industrial Production:  Residential Utilities (Index 2012=100)&$5$&&&x\tabularnewline
~~&IPFUELS&Industrial Production:  Fuels (Index 2012=100)&$5$&&&x\tabularnewline
~~&PAYEMS& Emp:Nonfarm All Employees: Total nonfarm (Thousands of Persons)&$5$&&&x\tabularnewline
~~&USPRIV& All Employees: Total Private Industries (Thousands of Persons)&$5$&&&x\tabularnewline
~~&MANEMP& All Employees: Manufacturing (Thousands of Persons)&$5$&&&x\tabularnewline
~~&SRVPRD&All Employees:  Service-Providing Industries (Thousands of Persons)&$5$&&&x\tabularnewline
~~&USGOOD&All Employees:  Goods-Producing Industries (Thousands of Persons)&$5$&&&x\tabularnewline
~~&DMANEMP&All Employees:  Durable goods (Thousands of Persons)&$5$&&&x\tabularnewline
~~&NDMANEMP&All Employees:  Nondurable goods (Thousands of Persons)&$5$&&&x\tabularnewline
~~&USCONS&All Employees:  Construction (Thousands of Persons)&$5$&&&x\tabularnewline
~~&USEHS&All Employees:  Education \& Health Services (Thousands of Persons)&$5$&&&x\tabularnewline
~~&USFIRE&All Employees:  Financial Activities (Thousands of Persons)&$5$&&&x\tabularnewline
~~&USINFO&All Employees:  Information Services (Thousands of Persons)&$5$&&&x\tabularnewline
~~&USPBS&All Employees:  Professional \& Business Services (Thousands of Persons)&$5$&&&x\tabularnewline
~~&USLAH&All Employees:  Leisure \& Hospitality (Thousands of Persons)&$5$&&&x\tabularnewline
~~&USSERV&All Employees:  Other Services (Thousands of Persons)&$5$&&&x\tabularnewline
~~&USMINE&All Employees:  Mining and logging (Thousands of Persons)&$5$&&&x\tabularnewline
~~&USTPU&All Employees:  Trade, Transportation \& Utilities (Thousands of Persons)&$5$&&&x\tabularnewline
~~&USGOVT&All Employees:  Government (Thousands of Persons)&$5$&&&x\tabularnewline
~~&USTRADE&All Employees:  Retail Trade (Thousands of Persons)&$5$&&&x\tabularnewline
~~&USWTRADE&All Employees:  Wholesale Trade (Thousands of Persons)&$5$&&&x\tabularnewline
~~&CES9091000001&All Employees:  Government:  Federal (Thousands of Persons)&$5$&&&x\tabularnewline
~~&CES9092000001&All Employees:  Government:  State Government (Thousands of Persons)&$5$&&&x\tabularnewline
~~&CES9093000001&All Employees:  Government:  Local Government (Thousands of Persons)&$5$&&&x\tabularnewline
~~&CE16OV&Civilian Employment (Thousands of Persons)&$5$&&x&x\tabularnewline
~~&CIVPART&Civilian Labor Force Participation Rate (Percent)&$2$&&&x\tabularnewline
~~&UNRATE&Civilian Unemployment Rate (Percent)&$2$&&x&x\tabularnewline
~~&UNRATESTx&Unemployment Rate less than 27 weeks (Percent)&$2$&&&x\tabularnewline
~~&UNRATELTx&Unemployment Rate for more than 27 weeks (Percent)&$2$&&&x\tabularnewline
~~&LNS14000012&Unemployment Rate - 16 to 19 years (Percent)&$2$&&&x\tabularnewline
~~&LNS14000025&Unemployment Rate - 20 years and over, Men (Percent)&$2$&&&x\tabularnewline
~~&LNS14000026&Unemployment Rate - 20 years and over, Women (Percent)&$2$&&&x\tabularnewline
~~&UEMPLT5&Number of Civilians Unemployed - Less Than 5 Weeks (Thousands of Persons)&$5$&&&x\tabularnewline
~~&UEMP5TO14&Number of Civilians Unemployed for 5 to 14 Weeks (Thousands of Persons)&$5$&&&x\tabularnewline
~~&UEMP15T26&Number of Civilians Unemployed for 15 to 26 Weeks (Thousands of Persons)&$5$&&&x\tabularnewline
~~&UEMP27OV&Number of Civilians Unemployed for 27 Weeks and Over (Thousands of Persons)&$5$&&&x\tabularnewline
~~&AWHMAN&Average Weekly Hours of Production and Nonsupervisory Employees:  Manufacturing (Hours)&$1$&&&x\tabularnewline
~~&AWOTMAN&Average Weekly Overtime Hours of Production and Nonsupervisory Employees: Manufacturing (Hours)&$2$&&&x\tabularnewline
~~&HWIx&Help-Wanted Index&$1$&&&x\tabularnewline
~~&CES0600000007&Average Weekly Hours of Production and Nonsupervisory Employees:  Goods-Producing&$2$&&x&x\tabularnewline
~~&CLAIMSx&Initial Claims&$5$&&&x\tabularnewline
~~&HOUST&Housing Starts: Total: New Privately Owned Housing Units Started&$5$&&x&x\tabularnewline
\bottomrule
\end{tabular}}
\caption{Data description\label{tab:data-descr1}}\end{center}}
\end{table}

\begin{table}[!tbp]
{\tiny
\begin{center}
\scalebox{0.8}{
\begin{tabular}{lllrlll}
\toprule
\multicolumn{1}{l}{\ }&\multicolumn{1}{c}{\ FRED.Mnemonic}&\multicolumn{1}{c}{\ Description}&\multicolumn{1}{c}{\ Trans I(0)}&\multicolumn{1}{c}{\ SMALL}&\multicolumn{1}{c}{\ MEDIUM}&\multicolumn{1}{c}{\ LARGE}\tabularnewline
\midrule
{\scshape Slow}&&&&&&\tabularnewline
~~&HOUST5F&Privately Owned Housing Starts: 5-Unit Structures or More&$5$&&&x\tabularnewline
~~&PERMIT&New Private Housing Units Authorized by Building Permits&$5$&&x&x\tabularnewline
~~&HOUSTMW&Housing Starts in Midwest Census Region (Thousands of Units)&$5$&&&x\tabularnewline
~~&HOUSTNE&Housing Starts in Northeast Census Region (Thousands of Units)&$5$&&&x\tabularnewline
~~&HOUSTS&Housing Starts in South Census Region (Thousands of Units)&$5$&&&x\tabularnewline
~~&HOUSTW&Housing Starts in West Census Region (Thousands of Units)&$5$&&&x\tabularnewline
~~&RSAFSx&Real Retail and Food Services Sales (Millions of Chained 2012 Dollars)&$5$&&&x\tabularnewline
~~&AMDMNOx&Real Manufacturers' New Orders:  Durable Goods (Millions of 2012 Dollars)&$5$&&&x\tabularnewline
~~&AMDMUOx&Real Value of Manufacturers' Unfilled Orders for Durable Goods Industries&$5$&&&x\tabularnewline
~~&BUSINVx&Total Business Inventories (Millions of Dollars)&$5$&&&x\tabularnewline
~~&ISRATIOx&Total Business:  Inventories to Sales Ratio&$2$&&&x\tabularnewline
~~&PCECTPI&Personal Consumption Expenditures: Chain-type Price Index &$6$&&x&x\tabularnewline
~~&PCEPILFE&Personal Consumption Expenditures Excluding Food and Energy&$6$&&&x\tabularnewline
~~&GDPCTPI&Gross Domestic Product: Chain-type Price Index&$5$&x&x&x\tabularnewline
~~&GPDICTPI&Gross Private Domestic Investment: Chain-type Price Index &$6$&&&x\tabularnewline
~~&IPDBS&Business Sector:  Implicit Price Deflator (Index 2012=100)&$6$&&&x\tabularnewline
~~&DGDSRG3Q086SBEA&Personal consumption expenditures:  Goods &$6$&&&x\tabularnewline
~~&DDURRG3Q086SBEA&Personal consumption expenditures:  Durable goods &$6$&&&x\tabularnewline
~~&DSERRG3Q086SBEA&Personal consumption expenditures:  Services &$6$&&&x\tabularnewline
~~&DNDGRG3Q086SBEA&Personal consumption expenditures:  Nondurable goods&$6$&&&x\tabularnewline
~~&DHCERG3Q086SBEA&Personal consumption expenditures:  Services:  Household consumption expenditures&$6$&&&x\tabularnewline
~~&DMOTRG3Q086SBEA&Personal consumption expenditures:  Durable goods:  Motor vehicles and parts&$6$&&&x\tabularnewline
~~&DFDHRG3Q086SBEA&Personal consumption expenditures:  Durable goods:  Furnishings and durable household equipment&$6$&&&x\tabularnewline
~~&DREQRG3Q086SBEA&Personal consumption expenditures:  Durable goods:  Recreational goods and vehicles&$6$&&&x\tabularnewline
~~&DODGRG3Q086SBEA&Personal consumption expenditures:  Durable goods:  Other durable goods&$6$&&&x\tabularnewline
~~&DFXARG3Q086SBEA&Personal consumption expenditures:  Nondurable goods:  Food and beverages purchased for off-premises consumption&$6$&&&x\tabularnewline
~~&DCLORG3Q086SBEA&Personal consumption expenditures:  Nondurable goods:  Clothing and footwear&$6$&&&x\tabularnewline
~~&DGOERG3Q086SBEA&Personal consumption expenditures:  Nondurable goods:  Gasoline and other energy goods&$6$&&&x\tabularnewline
~~&DONGRG3Q086SBEA&Personal consumption expenditures:  Nondurable goods:  Other nondurable goods&$6$&&&x\tabularnewline
~~&DHUTRG3Q086SBEA&Personal consumption expenditures:  Services:  Housing and utilities&$6$&&&x\tabularnewline
~~&DHLCRG3Q086SBEA&Personal consumption expenditures:  Services:  Health care&$6$&&&x\tabularnewline
~~&DTRSRG3Q086SBEA&Personal consumption expenditures:  Transportation services&$6$&&&x\tabularnewline
~~&DRCARG3Q086SBEA&Personal consumption expenditures: Recreation services&$6$&&&x\tabularnewline
~~&DFSARG3Q086SBEA&Personal consumption expenditures:  Services:  Food services and accomodations&$6$&&&x\tabularnewline
~~&DIFSRG3Q086SBEA&Personal consumption expenditures:  Financial services and insurance&$6$&&&x\tabularnewline
~~&DOTSRG3Q086SBEA&Personal consumption expenditures:  Other services &$6$&&&x\tabularnewline
~~&CPIAUCSL&Consumer Price Index for All Urban Consumers:  All Items&$6$&&x&x\tabularnewline
~~&CPILFESL&Consumer Price Index for All Urban Consumers:  All Items Less Food \& Energy&$6$&&&x\tabularnewline
~~&WPSFD49207&Producer Price Index by Commodity for Finished Goods &$6$&&&x\tabularnewline
~~&PPIACO&Producer Price Index for All Commodities &$6$&&&x\tabularnewline
~~&WPSFD49502&Producer Price Index by Commodity for Finished Consumer Goods &$6$&&&x\tabularnewline
~~&WPSFD4111&Producer Price Index by Commodity for Finished Consumer Foods&$6$&&&x\tabularnewline
~~&PPIIDC&Producer Price Index by Commodity Industrial Commodities &$6$&&&x\tabularnewline
~~&WPSID61&Producer Price Index by Commodity Intermediate Materials:  Supplies \& Components&$6$&&&x\tabularnewline
~~&WPU0561&Producer Price Index by Commodity for Fuels and Related Products and Power&$5$&&&x\tabularnewline
~~&OILPRICEx&Real Crude Oil Prices: West Texas Intermediate (WTI) - Cushing, Oklahoma&$5$&&&x\tabularnewline
~~&WPSID62&Producer Price Index: Crude Materials for Further Processing &$6$&&&x\tabularnewline
~~&PPICMM&Producer Price Index: Commodities:  Metals and metal products:  Primary nonferrous metals&$6$&&&x\tabularnewline
~~&CPIAPPSL&Consumer Price Index for All Urban Consumers:  Apparel&$6$&&&x\tabularnewline
~~&CPITRNSL&Consumer Price Index for All Urban Consumers:  Transportation&$6$&&&x\tabularnewline
~~&CPIMEDSL&Consumer Price Index for All Urban Consumers:  Medical Care&$6$&&&x\tabularnewline
~~&CUSR0000SAC&Consumer Price Index for All Urban Consumers:  Commodities&$6$&&&x\tabularnewline
~~&CES2000000008x&Real Average Hourly Earnings of Production and Nonsupervisory Employees: Construction&$5$&&&x\tabularnewline
~~&CES3000000008x&Real Average Hourly Earnings of Production and Nonsupervisory Employees: Manufacturing&$5$&&&x\tabularnewline
~~&COMPRNFB&Nonfarm Business Sector:  Real Compensation Per Hour (Index 2012=100)&$5$&&&x\tabularnewline
~~&CES0600000008&Average Hourly Earnings of Production and Nonsupervisory Employees:&$6$&&x&x\tabularnewline
\midrule
{\scshape Policy rate}&&&&&&\tabularnewline
~~&FEDFUNDS&Effective Federal Funds Rate (Percent)&$2$&x&x&x\tabularnewline
\midrule
{\scshape Fast}&&&&&&\tabularnewline
~~&TB3MS&3-Month Treasury Bill: Secondary Market Rate (Percent)&$2$&&&x\tabularnewline
~~&TB6MS&6-Month Treasury Bill: Secondary Market Rate (Percent)&$2$&&&x\tabularnewline
~~&GS1&1-Year Treasury Constant Maturity Rate (Percent)&$2$&&x&x\tabularnewline
~~&GS10&10-Year Treasury Constant Maturity Rate (Percent)&$2$&&x&x\tabularnewline
~~&AAA&Moody's Seasoned Aaa Corporate Bond Yield (Percent)&$2$&&&x\tabularnewline
~~&BAA&Moody's Seasoned Baa Corporate Bond Yield (Percent)&$2$&&&x\tabularnewline
~~&BAA10YM&Moody's Seasoned Baa Corporate Bond Yield Relative to Yield on 10-Year Treasury&$1$&&&x\tabularnewline
~~&TB6M3Mx&6-Month Treasury Bill Minus 3-Month Treasury Bill, secondary market (Percent)&$1$&&&x\tabularnewline
~~&GS1TB3Mx&1-Year Treasury Constant Maturity Minus 3-Month Treasury Bill, secondary market&$1$&&&x\tabularnewline
~~&GS10TB3Mx&10-Year Treasury Constant Maturity Minus 3-Month Treasury Bill, secondary market&$1$&&&x\tabularnewline
~~&CPF3MTB3Mx&3-Month Commercial Paper Minus 3-Month Treasury Bill, secondary market&$1$&&&x\tabularnewline
~~&GS5&5-Year Treasury Constant Maturity Rate&$2$&&&x\tabularnewline
~~&TB3SMFFM&3-Month Treasury Constant Maturity Minus Federal Funds Rate&$1$&&&x\tabularnewline
~~&T5YFFM&5-Year Treasury Constant Maturity Minus Federal Funds Rate&$1$&&&x\tabularnewline
~~&AAAFFM&Moody's Seasoned Aaa Corporate Bond Minus Federal Funds Rate&$1$&&&x\tabularnewline
~~&BUSLOANSx&Real Commercial and Industrial Loans, All Commercial Banks&$5$&&&x\tabularnewline
~~&CONSUMERx&Real Consumer Loans at All Commercial Banks &$5$&&&x\tabularnewline
~~&NONREVSLx&Total Real Nonrevolving Credit Owned and Securitized, Outstanding&$5$&&&x\tabularnewline
~~&REALLNx&Real Real Estate Loans, All Commercial Banks&$5$&&&x\tabularnewline
~~&TOTALSLx&Total Consumer Credit Outstanding&$5$&&&x\tabularnewline
~~&TOTRESNS&Total Reserves of Depository Institutions &$6$&&x&x\tabularnewline
~~&NONBORRES&Reserves Of Depository Institutions, Nonborrowed&$7$&&x&x\tabularnewline
~~&DTCOLNVHFNM&Consumer Motor Vehicle Loans Outstanding Owned by Finance Companies&$6$&&&x\tabularnewline
~~&DTCTHFNM&Total Consumer Loans and Leases Outstanding Owned and Securitized by Finance Companies &$6$&&&x\tabularnewline
~~&INVEST&Securities in Bank Credit at All Commercial Banks &$6$&&&x\tabularnewline
~~&TABSHNOx&Real Total Assets of Households and Nonprofit Organizations&$5$&&&x\tabularnewline
~~&EXSZUSx&Switzerland / U.S. Foreign Exchange Rate&$5$&&&x\tabularnewline
~~&EXJPUSx&Japan / U.S. Foreign Exchange Rate&$5$&&&x\tabularnewline
~~&EXUSUKx&U.S. / U.K. Foreign Exchange Rate&$5$&&&x\tabularnewline
~~&EXCAUSx&Canada / U.S. Foreign Exchange Rate&$5$&&&x\tabularnewline
~~&S.P.500&S\&P's Common Stock Price Index:  Composite&$5$&&x&x\tabularnewline
~~&S.P..indust&S\&P's Common Stock Price Index:  Industrials&$5$&&&x\tabularnewline
~~&S.P.div.yield&S\&P's Composite Common Stock:  Dividend Yield&$2$&&&x\tabularnewline
\bottomrule
\end{tabular}}
\caption{Data description (cont.)\label{tab:data-descr2}}\end{center}}
\end{table}

\end{appendices}

\end{document}